\documentclass[12pt]{article}
\usepackage[utf8]{inputenc}

\usepackage{amsmath}
\usepackage{color}
\usepackage{amsfonts}
\usepackage{amssymb}
\usepackage{graphicx}
\usepackage{geometry}
\usepackage{amssymb,epsfig,subfigure}
\usepackage{hyperref}
\usepackage{comment}
\usepackage{tabularx}
\usepackage{bm}
\usepackage{euscript}
\usepackage{graphicx}
\usepackage{color}
\usepackage{amsfonts}
\usepackage{exscale}
\usepackage{amsbsy}
\usepackage{subfigure}
\usepackage{textcomp}
\usepackage{comment}
\usepackage{hyperref}
\usepackage{slashed}
\usepackage{authblk}
\usepackage{mathtools}
\usepackage{tabularx}
\usepackage{bm}
\usepackage{euscript}
\usepackage{graphicx}
\usepackage{color}
\usepackage{amsfonts}
\usepackage{exscale}
\usepackage{amsbsy}
\usepackage{subfigure}
\usepackage{textcomp}
\usepackage{comment}
\usepackage{hyperref}

\usepackage[dvipsnames]{xcolor}

\def\be{\begin{equation}}
\def\ee{\end{equation}}
\def\bea{\begin{eqnarray}}
\def\eea{\end{eqnarray}}
\def\ba#1\ea{\begin{align}#1\end{align}}
\def\bg#1\eg{\begin{gather}#1\end{gather}}

\def\bmd#1\emd{\begin{multlined}#1\end{multlined}}

\newcommand{\nn}{\nonumber}

\newcommand{\psb}{\overline{\psi}}

\newcommand{\wb}{\overline{w_x}}
\newcommand{\gmb}{\overline{g_m}}
\newcommand{\gob}{\overline{g_0}}
\newcommand{\gjb}{\overline{g_j}}
\newcommand{\seb}{\overline{\sigma_e}}

\newcommand{\Dob}{\overline{\Delta_0}} 
\newcommand{\Djb}{\overline{\Delta_j}}

\newcommand{\longDeno}{\sqrt{ 1+y^2 } (g_1^2 +\kappa^2)(\seb+|y|) + g_1 \, \wb \, |y| }

\newcommand{\longNum}{
4 \; \epsilon \; N_f \; \pi^2 \,  (v^2 +z^2)^2
\bigg[ 
\sqrt{v^2+ z^2} \big( g_1^2 +\kappa^2 \,    \big) \sigma_e
+|z| \, \Big(  \, (g_1^2+\kappa^2    )  \, \sqrt{v^2+ z^2}
 +     \, g_1 \, w_x     \; \Big)
\bigg] 
}

\usepackage[numbers,sort&compress]{natbib}

\def\simge{
    \mathrel{\rlap{\raise 0.511ex 
        \hbox{$>$}}{\lower 0.511ex \hbox{$\sim$}}}}

\def\simle{
    \mathrel{\rlap{\raise 0.511ex 
        \hbox{$<$}}{\lower 0.511ex \hbox{$\sim$}}}}

\makeatletter
\renewcommand\section{\@startsection {section}{1}{\z@}%
                                 {-3.5ex \@plus -1ex \@minus -.2ex}
                                   {2.3ex \@plus.2ex}%
                                   {\normalfont\large\bfseries}}
\renewcommand\subsection{\@startsection{subsection}{2}{\z@}%
                                   {-3.25ex\@plus -1ex \@minus -.2ex}%
                                     {1.5ex \@plus .2ex}%
                                     {\normalfont\bfseries}}
\renewcommand\subsubsection{\@startsection{subsubsection}{3}{\z@}%
                                   {-3.25ex\@plus -1ex \@minus -.2ex}%
                                     {1.5ex \@plus .2ex}%
                                     {\normalfont\itshape}}
\makeatother

\def\pplogo{\vbox{\kern-\headheight\kern -29pt
\halign{##&##\hfil\cr&{\ppnumber}\cr\rule{0pt}{2.5ex}&\ppdate\cr}}}
\makeatletter
\def\ps@firstpage{\ps@empty \def\@oddhead{\hss\pplogo}%
  \let\@evenhead\@oddhead 
}
\def\maketitle{\par
 \begingroup
 \def\thefootnote{\fnsymbol{footnote}}
 \def\@makefnmark{\hbox{$^{\@thefnmark}$\hss}}
 \if@twocolumn
 \twocolumn[\@maketitle]
 \else \newpage
 \global\@topnum\z@ \@maketitle \fi\thispagestyle{firstpage}\@thanks
 \endgroup
 \setcounter{footnote}{0}
 \let\maketitle\relax
 \let\@maketitle\relax
 \gdef\@thanks{}\gdef\@author{}\gdef\@title{}\let\thanks\relax}
\makeatother

\numberwithin{equation}{section}

\begin{document}

\setcounter{page}0
\def\ppnumber{\vbox{\baselineskip14pt
}}

\def\ppdate{
} \date{\today}

\title{\bf Scaling and Diffusion of Dirac Composite Fermions 
\vskip 0.5cm}
\author{Chao-Jung Lee}
\affil[1]{\small \it Department of Physics,
California Institute of Technology, Pasadena, CA 91125, USA}
\author{Michael Mulligan}
\affil[2]{\small \it Department of Physics and Astronomy, University of California,
Riverside, CA 92511, USA}

\bigskip

\maketitle

\begin{abstract}
We study the effects of quenched disorder and a dissipative Coulomb interaction on an anyon gas in a periodic potential undergoing a quantum phase transition.
We use a $(2+1)$d low-energy effective description that involves $N_f = 1$ Dirac fermion coupled to a $U(1)$ Chern-Simons gauge field at level $(\theta - 1/2)$.
When $\theta = 1/2$ the anyons are free Dirac fermions that exhibit an integer quantum Hall transition; when $\theta = 1$ the anyons are bosons undergoing a superconductor-insulator transition in the universality class of the 3d XY model.
Using the large $N_f$ approximation we perform a renormalization group analysis.
The dissipative Coulomb interaction allows for two classes of IR stable fixed points: those with a finite, nonzero Coulomb coupling and dynamical critical exponent $z = 1$ and those with an effectively infinite Coulomb coupling and $1 < z  < 2$.
We find the Coulomb interaction to be an irrelevant perturbation of the clean fixed point for any $\theta$.
At $\theta = 1/2$ the clean fixed point is stable to charge-conjugation preserving (random mass) disorder, while a line of diffusive fixed points obtains when the product of charge-conjugation and time-reversal symmetries is preserved.
At $\theta = 1$ we find a finite disorder fixed point with unbroken charge-conjugation symmetry whether or not the Coulomb interaction is present.
Other cases result in runaway flows.
We comment on the relation of our results to other theoretical studies and the relevancy to experiment.
\end{abstract}
\bigskip

\newpage

\tableofcontents

\newpage

\vskip 1cm

\section{Introduction}

Delocalization transitions determine the phase diagrams of various electronic systems \cite{RevModPhys.57.287, RevModPhys.66.261, lagendijk2009fifty}. 
In three spatial dimensions, such transitions can occur between a diffusive metal and a localized insulator.
In two dimensions (and fewer), localization generally relegates $T=0$ metallic states to isolated critical points.
The integer quantum Hall transition (IQHT) and the superconductor-insulator transition (SIT) are prototypical examples of such two-dimensional diffusive quantum critical points, having been well characterized by extensive experimental and numerical work over the past 30 years (see \cite{SondhiGirvinCariniShahar, PhysRevLett.94.206807, PhysRevLett.102.216801, PhysRevB.81.033305} and references therein).
Nevertheless, our understanding of these quantum states remains incomplete.

Theories of noninteracting electrons have provided valuable insight to the IQHT \cite{RevModPhys.67.357}.
As the critical point is approached by tuning the external magnetic field or electron density to criticality $\delta \rightarrow 0$, the localization length is found to diverge as $|\delta|^{- \nu}$ with $\nu = 2.593(5)$ \cite{PhysRevB.80.041304}, while $\nu_{\rm expt} \approx 2.38$ experimentally \cite{PhysRevLett.94.206807, PhysRevLett.102.216801}.
On the other hand, a diverging timescale $\xi_t \sim \xi^{- \nu z}$ is also expected near the quantum critical point. 
Theories of noninteracting electrons yield a dynamical critical exponent $z = 2$ \cite{RevModPhys.67.357, PhysRevLett.61.593, PhysRevLett.72.713}; $z_{\rm expt} \approx 1$ \cite{SondhiGirvinCariniShahar, PhysRevLett.102.216801} (although see \cite{2019arXiv190300489S}).\footnote{For the magnetic field-tuned SIT, $\nu_{\rm expt} \approx 4/3$ or $\nu_{\rm expt} \approx 7/3$ and $z_{\rm expt} \approx 1$ experimentally \cite{masonthesis}.} 
The challenge is to develop a framework that 
combines the effects of electron interactions with those of disorder \cite{PhysRevLett.82.5100}.

Duality is a powerful tool for understanding the behavior of strongly interacting systems.
Recent work has uncovered a duality web that relates various (2+1)d relativistic quantum field theories (see \cite{2018arXiv181005174S} and references therein).
Included in this set are simple, toy models for integer quantum Hall and superconductor-insulator transitions.
In this paper, we study the combined effects of quenched disorder and a dissipative Coulomb interaction on the critical properties of two such models. 
The hope is to abstract lessons that may be valid more generally.
As we discuss, these theories have a rich set of random critical behaviors.


For the first member of the duality web, consider a system of spinless electrons hopping on a square lattice with a half-unit of magnetic flux penetrating each plaquette \cite{PhysRevB.50.7526} (\cite{PhysRevLett.61.2015} may alternatively be considered). 
An IQHT obtains as the ratio of the (staggered) chemical potential to next-neighbor hopping is varied. 
The critical properties of the transition are controlled by a free Dirac fermion $\Psi$ with Lagrangian,\footnote{Additional details for the Lagrangians appearing in this section are given in \S\ref{setup}.}
\begin{align}
\label{dirac}
{\cal L}_{\rm Dirac} = \overline{\Psi} i \slashed D_A \Psi - M \overline{\Psi} \Psi + {1 \over 2} {1 \over 4 \pi} A d A,
\end{align}
where $A_\mu$ is a non-dynamical $U(1)$ gauge field and the Chern-Simons term $A d A = \epsilon^{\mu \nu \rho} A_\mu \partial_\nu A_\rho$.
The mass $M$ vanishes at criticality.
In the presence of an external magnetic field, \eqref{dirac} describes the particle-hole symmetric limit of the half-filled zeroth/lowest Landau level of Dirac/nonrelativistic electrons
\cite{Son2015}.
In this paper, we consider vanishing magnetic field.
A dual effective theory to \eqref{dirac} consists of a Dirac fermion $\psi$ coupled to a dynamical (emergent) $U(1)$ gauge field $a_\mu$,
\begin{align}
\label{diraccf}
{\cal L}_{\rm F} = \overline{\psi} i \slashed D_a \psi - m \overline{\psi} \psi - {1 \over 2} {1 \over 2 \pi} a d A  + {1 \over 2} {1 \over 4 \pi} A d A - {1 \over 4} f_{\mu \nu}^2,
\end{align}
where the mass $m \propto M$ and the field strength $f_{\mu \nu} = \partial_\mu a_\nu - \partial_\nu a_\mu$.\footnote{We use ``condensed matter" notation when writing these Lagrangians; see \cite{Seiberg:2016gmd} for a precise explanation of the meaning of, e.g., Chern-Simons terms with half-integer levels.}
\eqref{diraccf} was first introduced as a dual description of the half-filled Landau level \cite{Son2015} or the gapless surface state of a time-reversal invariant topological insulator \cite{WangSenthilfirst2015, MetlitskiVishwanath2016} (when the $AdA$ term is absent) with $\psi$ being the Dirac composite fermion; its inclusion in the duality web was explained in \cite{Seiberg:2016gmd, PhysRevX.6.031043, 2017JHEP...05..159M}.
When the external magnetic field is zero, the Dirac composite fermion chemical potential sits at the Dirac point.

For the second member of the duality web, consider a collection of repulsive bosons in a periodic potential \cite{PhysRevB.40.546}. 
For commensurate filling, the system exhibits a superfluid to Mott insulator transition with a charge-conjugation symmetry as the ratio of the boson hopping strength to repulsion is tuned.
The long wavelength critical properties are described by the 3d XY model,
\begin{align}
\label{bose}
{\cal L}_{\rm XY} = |D_A \Phi|^2 - M |\Phi|^2 - |\Phi|^4.
\end{align}
(Broken charge-conjugation symmetry generally results in a term proportional to $\Phi^\ast i \partial_t \Phi$.)
In mean-field theory, the $M < 0$ region is a superfluid, while the $M > 0$ region is an insulator; we'll view \eqref{bose} as describing a SIT.
A dual effective theory \cite{ChenFisherWu1993, BarkeshliMcGreevy2012continuous, PhysRevB.95.045118} to \eqref{bose} is 
\begin{align}
\label{bosecf}
{\cal L}_{\rm B} = \overline{\psi} i \slashed D_a \psi - m \overline{\psi} \psi + {1 \over 2} {1 \over 4 \pi} a d a - {1 \over 2 \pi} a d A + {1 \over 4 \pi} A d A - {1 \over 4} f_{\mu \nu}^2.
\end{align}
The statistics of the particles that \eqref{diraccf} and \eqref{bosecf} describe is controlled by the coefficient of the $ada$ term.

Quenched disorder can have a profound effect on the nature of the above critical points and lead to new universality classes.
Ref.~\cite{PhysRevB.50.7526} considered the effects of quenched randomness on the free Dirac fermion fixed point in \eqref{dirac}.
While for generic disorder the theory flows to strong coupling, if only a random vector potential ${\bf A}({\bf x})$ is present the theory features a line of diffusive fixed points characterized by a continuously variable dynamical exponent $z$; the clean fixed point is stable to random mass disorder $M(\bf {x})$.
Sachdev and Ye \cite{PhysRevLett.80.5409, PhysRevB.60.8290} generalized this study to fractional quantum Hall transitions in the presence of an unscreened Coulomb interaction using a model closely related to \eqref{bosecf}.
Recently, Goswami, Goldman, and Raghu \cite{PhysRevB.95.235145} and Thomson and Sachdev \cite{PhysRevB.95.235146} considered the effects of randomness on \eqref{diraccf} with $2N_f$ fermion flavors.
We use the large $N_f$ expansion and the dimensional reduction renormalization group (RG) scheme\footnote{This scheme is valid for theories with or without Chern-Simons terms and is closely related to the approach in \cite{PhysRevLett.80.5409, PhysRevB.60.8290, PhysRevB.95.235146}.} to reexamine these works and extend them to include the effects of ``topological disorder" (\S\ref{randomness}) and a dissipative Coulomb interaction, generally finding agreement with this prior work that found interacting, diffusive fixed points.
Related work studying the effects of quenched randomness on theories of Dirac fermions coupled to a fluctuating boson include \cite{PhysRevB.73.155104, PhysRevB.98.195142}.

In contrast to the fermion models, only random mass disorder $M({\bf x})$ has resulted in accessible diffusive fixed points of the XY model.
Early work \cite{DOROGOVTSEV1980169, PhysRevB.26.154, Lawrie_1984, PhysRevB.77.214516} studying the $O(2N_f)$ generalization of \eqref{bose} used a double-$\epsilon$ expansion to find an interacting, finite disorder fixed point.
However, the nature of the renormalization group flow in the vicinity of the fixed point is peculiar, exhibiting an anomalously long ``time" to achieve criticality.
Recently, this problem was reexamined within a large $N_f$ expansion by Goldman, Thomson, Nie, and Bi \cite{2019arXiv190909167G}, where it was argued that the anomalous renormalization group trajectories \cite{DOROGOVTSEV1980169, PhysRevB.26.154, Lawrie_1984, PhysRevB.77.214516} are a relic of the double-$\epsilon$ expansion.
Furthermore, \cite{2019arXiv190909167G} find remarkable agreement with the critical exponents of the dirty XY model calculated by numerical simulation \cite{PhysRevLett.92.015703, PhysRevLett.114.255701, PhysRevLett.108.055701, PhysRevB.94.134501}.
We consider this analysis from the perspective of the ``fermionic dual" of the XY model in \ref{bosecf}, providing qualitative confirmation of the renormalization group flow found in \cite{2019arXiv190909167G}.
To ${\cal O}(1/N_f)$, we find a finite-disorder fixed point with critical exponents,
\begin{align}
\nu^{-1} = 1  \quad \text{and} \quad z = 1 + {1.411 \over N_f};
\end{align}
$\nu = 1$ and $z = 1 + {.54/N_f}$ is reported in \cite{2019arXiv190909167G}.
We also consider other types of disorder that is sourced by the random gauge field $A_\mu({\bf x})$.

The important influence of a Coulomb interaction on the critical properties of the above transitions was stressed long ago \cite{Fisher1990}, where it was argued that an unscreened Coulomb interaction generically results in a dynamical critical exponent $z =1$.
In addition, the observed IQHT and SIT appear to be sensitive to the precise nature of the Coulomb interaction (\cite{Kapitulnik2001, 2018PhRvL.120p7002W} and references therein).
For example, a capacitively-coupled screening plane has been found to affect the metallic behavior in thin films \cite{PhysRevB.65.220505}, lifting an anomalous low-temperature metallic regime that intervenes a direct magnetic field-tuned SIT.
To investigate such effects, we consider a Coulomb interaction that is screened by a diffusive two-dimensional Fermi gas \cite{PhysRevB.69.054507}.
The dissipative Coulomb interaction that results allows for two types of fixed points: those with a finite, nonzero Coulomb coupling and $z=1$ and those with an effectively infinite Coulomb interaction and $z \neq 1$ \cite{PhysRevB.75.235423}.
For the ``fermionic dual" of the XY model with random mass disorder, we find critical exponents, 
\begin{align}
\nu^{-1} = 1 \quad \text{and} \quad 1 \leq z < 2,
\end{align}
with $z$ saturating the lower bound for the unscreened Coulomb interaction and varying continuously with an effective dissipation parameter for $z > 1$.
In our approach, we're unable to access the ``infinite $z$" fixed point found in the study of the dissipative XY model in \cite{PhysRevB.79.184501}.
Our result differs from that of Vishwanath, Moore, and Senthil \cite{PhysRevB.69.054507} who studied the effects a dissipative Coulomb interaction on the dirty XY model using the double-$\epsilon$ expansion and found a line of fixed points with $z= 1$ and continuously varying $\nu$.
We also consider the effects of other types of disorder on the theories in \eqref{bosecf} (and \eqref{diraccf}) when a dissipative Coulomb interaction is present.


\section{Setup}
\label{setup}

In this section, we introduce the effective model that realizes an IQHT/SIT and whose critical properties we'll analyze in \S\ref{rganalysis}.

Consider the $(2+1)$d theory of $N_f$ Dirac fermions $\psi_I$ coupled to a $U(1)$ Chern-Simons gauge field $a_\mu$ at level $(\theta - 1/2)$\footnote{The notation in \eqref{unite} is as follows: 
$\bar \psi = \psi^\dagger \gamma^0$; $\slashed D_a = (\partial_\mu - i a_\mu)\gamma^\mu$ with $\mu \in \{0,1,2\} = \{t, x, y\}$; 
and a Chern-Simons term $AdA = \epsilon^{\mu \nu \rho} A_\mu \partial_\nu A_\rho$.
For the purpose of discussing the symmetries of \eqref{unite} later in this section, we choose Minkowski signature $\eta^{\mu \nu} = {\rm diag}(+1, -1, -1)$ and $\gamma$-matrices $(\gamma^0, \gamma^1, \gamma^2) = (\sigma^3, i \sigma^1, i \sigma^2)$ where $\sigma^j$ are the Pauli $\sigma$-matrices; in the renormalization group analysis in \S\ref{rganalysis}, we'll work in Euclidean signature.}
\begin{align}
\label{unite}
{\cal L}^{(1)} = \sum_{I = 1}^{N_f} \overline{\psi}_I ( i \slashed D_a - m ) \psi_I - {1 \over 2} {1 \over 4 \pi} a d a + {\theta \over 4 \pi} (a - A) d (a - A) - {1 \over 4} f_{\mu \nu}^2.
\end{align}
When $N_f = 2 \theta = 1$, we recover \eqref{diraccf}, the dual of a free Dirac fermion; when $N_f = \theta = 1$, we find the dual \eqref{bosecf} to the 3d XY model.
Reminiscent of conventional flux attachment \cite{jainCF, Fradkinbook}, $\theta^{-1}$ quantifies the number of attached flux quanta; for general $\theta$, ${\cal L}^{(1)}$ is the model for an anyon gas introduced by Chen, Fisher, and Wu \cite{ChenFisherWu1993}.
We refer to $\psi_I$ as a Dirac composite fermion.
$A_\mu$ is a nondynamical $U(1)$ gauge field that we identify with electromagnetism.\footnote{In \S\ref{dissipativecoulomb} we give $A_0$ dynamics to discuss the Coulomb interaction.}

In \S\ref{mfphasediagram} and \S\ref{symmetry}, where we discuss the phase diagram and symmetry of \eqref{unite}, we take $N_f = 1$.
Otherwise, $N_f$ is an arbitrary parameter that allows for analytic control as $N_f \rightarrow \infty$.

\subsection{Mean-Field Phase Diagram at $N_f = 1$}
\label{mfphasediagram}

For a given $\theta$, the mean-field phase diagram of \eqref{unite} at $N_f = 1$ is parameterized by the Dirac composite fermion mass $m$.
At energies less than $|m|$, we may integrate out\footnote{By ``integrate out," we refer to path integral relations of the form: $\int {\cal D}\phi \ e^{i \int ({1 \over 2} \phi K \phi + \phi J)} \propto e^{i \int (- {1 \over 2} J K^{-1} J)}$, where $\phi$ and $J$ are real fields and $K$ is some kernel, e.g., a kinetic term for $\phi$. 
Thus, we equate the Lagrangians ${1 \over 2} \phi K \phi + \phi J = - {1 \over 2} J K^{-1} J$ upon integrating out $\phi$.
Such identities follow directly from the $\phi$ equation of motion when $\phi$ appears quadratically in the Lagrangian.}
the Dirac composite fermion to obtain the effective Lagrangian,
\begin{align}
{\cal L}_{\rm eff} ={{\rm sign}(m) - 1 + 2 \theta \over 2} {1 \over 4 \pi} a d a - {\theta \over 2 \pi} a d A+ {\theta \over 4 \pi} A d A - {1 \over 4} f_{\mu \nu}^2.
\end{align}
Higher-order terms in $a_\mu$ can be ignored as $|m| \rightarrow \infty$.
The Maxwell term $f_{\mu \nu}^2$ can also be dropped in this long wavelength analysis.

\subsubsection*{$\theta = 1/2$}

Setting $\theta = 1/2$, there are two phases.
For $m > 0$ we find the effective Lagrangian for an insulator at zero temperature,
\begin{align}
{\cal L}_{\rm INS} = {1 \over 2} \Big({1 \over 4 \pi} a d a - {1 \over 2 \pi} a d A+ {1 \over 4 \pi} A d A \Big) = 0,
\end{align}
where the second equality follows from integrating out $a_\mu$.
For $m < 0$ we find the long wavelength Lagrangian for an integer Hall state,
\begin{align}
{\cal L}_{\rm IQH} = {1 \over 2} \Big(- {1 \over 4 \pi} a d a - {1 \over 2 \pi} a d A+ {1 \over 4 \pi} A d A \Big) = {1 \over 4 \pi} A d A.
\end{align}

\subsubsection*{$\theta = 1$}

Next set $\theta = 1$. 
We again find the insulator when $m > 0$,
\begin{align}
{\cal L}_{\rm INS} ={1 \over 4 \pi} a d a - {1 \over 2 \pi} a d A+ {1 \over 4 \pi} A d A = 0.
\end{align}
To identify the $m < 0$ phase, it's helpful to include the charge $e_\ast = q$ (measured in units of the electric charge $e$) carried by the boson $\Phi$ in \eqref{bose} by substituting $A_\mu \rightarrow q A_\mu$:
\begin{align}
\label{superconductor}
{\cal L}_{\rm SC} = - {q \over 2 \pi} a d A+ {q^2 \over 4 \pi} A d A.
\end{align}
\eqref{superconductor} describes a $\mathbb{Z}/q$ gauge theory, the long wavelength description of a superconductor with charge-$q$ condensate \cite{2004AnPhy.313..497H, KapustinSeibergqfttqft2014}.

\subsection{Discrete Symmetry at $N_f = 1$}
\label{symmetry}

The types of randomness that can be added to \eqref{unite} are characterized by charge-conjugation ${\cal C}$ and time-reversal ${\cal T}$ symmetries.
(Parity, i.e., spatial reflection, is necessarily broken in the presence of quenched disorder.)
These symmetries are defined with respect to the electron and boson Lagrangians in Eqs~\eqref{dirac} and \eqref{bose}.
We discuss their implementation \cite{Son2015, Seiberg:2016gmd, PhysRevX.7.041016, 2018arXiv180906886H, PhysRevLett.120.016602} in the dual Lagrangian \eqref{unite} at $N_f = 1$ at criticality $m = 0$.

\subsubsection*{$\theta = 1/2$}

The free Dirac Lagrangian in \eqref{dirac} is invariant under charge-conjugation ${\cal C}$,
\begin{align}
\Psi \mapsto \sigma^1 \Psi^\ast, \quad A_\mu \mapsto - A_\mu.
\end{align}
The presence of the Chern-Simons term for $A_\mu$ reflects the violation of time-reversal ${\cal T}$:
\begin{align}
\label{Tdirac}
t \mapsto - t, \quad \Psi \mapsto - i \sigma^2 \Psi, \quad (A_0, A_i) \mapsto (A_0, - A_i),
\end{align}
which is anti-unitary ($i \mapsto - i$).
On the surface of a time-reversal invariant topological insulator, this Chern-Simons term is absent and so ${\cal T}$ can be preserved.

The dual Lagrangian \eqref{unite} at $\theta = 1/2$ is also invariant under ${\cal C}$:
\begin{align}
\label{chargediraccf}
\psi \mapsto \sigma^1 \psi, \quad a_\mu \mapsto - a_\mu, \quad A_\mu \mapsto - A_\mu.
\end{align}
Identifying the electromagnetic currents across the duality between \eqref{dirac} and \eqref{unite}, ${\delta {\cal L_{\rm Dirac}} \over \delta A_\mu} = {\delta {\cal L}^{(1)} \over \delta A_\mu}$, we equate
\begin{align}
\label{electroconsequence}
\bar \Psi \gamma^\mu \Psi = {1 \over 4 \pi} \epsilon^{\mu \nu \rho} \partial_\nu a_\rho.
\end{align}
Similarly, the $a_0$ equation of motion relates
\begin{align}
\label{a0EOM}
{1 \over 4 \pi} \epsilon^{\mu \nu \rho} \partial_\nu A_\rho = \bar \psi \gamma^\mu \psi.
\end{align}
Eqs.~\eqref{electroconsequence} and \eqref{a0EOM} imply that in \eqref{unite}, ${\cal C T}$:
\begin{align}
\label{Tdiraccf}
t \mapsto - t, \quad \psi \mapsto - i \sigma^2 \psi^\ast, \quad (a_0, a_i) \mapsto (a_0, - a_i), \quad (A_0, A_i) \mapsto (- A_0, A_i).
\end{align}
Thus, ${\cal T}$ and ${\cal CT}$ are exchanged across the duality: the ${\cal T}$ transformations on $\Psi$ and $A_\mu$ is identical to the ${\cal CT}$ transformations on $\psi$ and $a_\mu$, and vice versa.
In the absence of the Chern-Simons term for $A_\mu$, \eqref{unite} is time-reversal invariant.

While the dual Lagrangians in \eqref{dirac} and \eqref{unite} violate time-reversal invariance as (2+1)d theories, they do preserve a ``non-local" particle-hole (PH) transformation.
To define this, consider the following transformations of a general Lagrangian $L(A)$ which has a $U(1)$ symmetry current that is coupled to a non-dynamical field $A_\mu$  \cite{WittenSL2Z2003}:
\begin{align}
\label{modularT}
& {\bf T}: L(A) \mapsto L(A) + {1 \over 4 \pi} A d A; \\
\label{modularS}
& {\bf S}: L(A) \mapsto L(c) + {1 \over 2 \pi} c d A.
\end{align}
${\bf T}$ shifts the Hall conductivity by a unit; ${\bf S}$ converts $A_\mu$ into a dynamical $U(1)$ gauge field $c_\mu$ and adds a BF term, which couples the field strength $dc$ to a new external field $A_\mu$.
\eqref{modularT} and \eqref{modularS} implement modular transformations on the conductivity tensor of the $U(1)$ symmetry current coupling to $A_\mu$.
The PH transformation is defined as ${\cal T}$ followed by the modular ${\bf T}$ transformation \eqref{modularT}.
Notice that the Dirac masses $\bar \Psi \Psi$ and $\bar \psi \psi$ are odd under PH symmetry and even under ${\cal C}$.
The ${\bf S}$ transformation will play a role in our discussion of the SIT theory.

\subsubsection*{$\theta = 1$}

The XY model in \eqref{bose} is invariant under charge-conjugation ${\cal C}$,
\begin{align}
\Phi \mapsto \Phi^\ast, \quad A_\mu \mapsto - A_\mu,
\end{align}
and time-reversal ${\cal T}$,
\begin{align}
t \mapsto - t, \quad \Phi \mapsto \Phi, \quad (A_0, A_i) \mapsto (A_0, - A_i).
\end{align}
The dual Lagrangian in \eqref{unite} at $\theta = 1$ is only invariant under ${\cal C}$ defined in \eqref{chargediraccf}; it isn't invariant under ${\cal T}$,
\begin{align}
\label{sittimereversal}
t \mapsto - t, \quad \psi \mapsto - \sigma^3 \psi^\ast, \quad (a_0, a_i) \mapsto (- a_0, a_i), \quad (A_0, A_i) \mapsto (A_0, - A_i),
\end{align}
with $i \mapsto - i$.
Instead, time-reversal is an emergent symmetry of the long wavelength physics \cite{Seiberg:2016gmd, PhysRevX.7.041016}.
In addition, \eqref{unite} is invariant under a ``non-local" particle-vortex (PV) transformation:
\begin{align}
t \mapsto - t, \quad \psi \mapsto - i \sigma^2 \psi, \quad (a_0, a_i) \mapsto (a_0, - a_i), \quad (A_0, A_i) \mapsto (- A_0, A_i),
\end{align}
followed by the modular ${\bf S}$ transformation \eqref{modularS}.
The PV transformation is analogous to the PH transformation of the previous section \cite{MulliganRaghuCFsatSIT}; it maps the 3d XY model to its scalar quantum electrodynamics dual \cite{Peskin:1977kp, DasguptaHalperin1981}, and vice versa.

Duality maps $|\Phi|^2 \leftrightarrow \bar \psi \psi$.
While it's clear that the Dirac mass is even under ${\cal C}$, it's less obvious that perturbation by $\bar \psi \psi$ is time-reversal invariant.
This can be understood in the following sense:
Perturbation of \eqref{unite}  by $\bar \psi \psi$ and its time-reversal, obtained using \eqref{sittimereversal}, by $- \bar \psi \psi$ result in identical phases.

\subsubsection*{Symmetry Assignment Summary}

Table \ref{symmetrytable} summarizes the transformations of the operators that appear in \eqref{unite} under charge-conjugation ${\cal C}$ and time-reversal ${\cal T}$ symmetries.
We use these transformation assignments to characterize the types of randomness that may be added to ${\cal L}^{(1)}$ for general $N_f$.
\begin{table} [h!]
\centering
 \begin{tabular}{| c | c | c |}
        \hline    ~~ 
  &    $ \mathcal{C} $ &  $\mathcal{T} $   \\ \hline 
  $  \overline{\psi} \psi$          &  + & -  \\ \hline    
 $\overline{\psi} \gamma_0  \psi $  &  - & -   \\ \hline 
 $\overline{\psi} \gamma_j  \psi $  &  - & +   \\ \hline 
 $a_0$  & -  & -       \\ \hline 
 $a_j$  & -  & +    \\ \hline 
$b=\partial_x a_y -\partial_y a_x$          &  - &  +    \\ \hline  
$e_j  = \partial_0 a_j - \partial_j a_0$  &  - &  -    \\ \hline   
        \end{tabular}
        \caption{Charge-conjugation ${\cal C}$ and time-reversal ${\cal T}$ symmetry assignments of various operators.}
        \label{symmetrytable}
\end{table}
 
\subsection{Dissipative Coulomb Interaction}
\label{dissipativecoulomb}

\subsubsection*{Dualizing the Coulomb Interaction}

The Coulomb interaction between fermions/bosons carrying charge $e_\ast$ arises from the exchange of a dynamical $(3+1)d$ electromagnetic scalar potential $A_0$. 
In Fourier space, we consider the action that couples a (2+1)d charge density $J_0(k_0, k)$ to the scalar potential $A_0(k_0, \vec{k})$:
\begin{align}
- {1 \over 2 e_\ast^2}\int d^4 k\ A_0(k_0, \vec{k}) \Big( \vec{k}^2 \Big) A_0(- k_0, - \vec{k}) - \int d^3 k J_0(k_0, k) \int dk_3 A_0(-k_0, - \vec{k}),
\end{align} 
where $k = (k_1, k_2)$ and $\vec{k} = (k_1, k_2, k_3)$.
$J_0(k_0, k)$ is the Fourier transform of $\bar \Psi \gamma^0 \Psi(x)$ for the free Dirac fermion \eqref{dirac} or $i \Phi^\ast \partial_0 \Phi(x) - i (\partial_0 \Phi^\ast) \Phi(x)$ for the XY model \eqref{bose}. 
The absence of an $A_0 k_0^2 A_0$ term means that $A_0$ mediates an instantaneous interaction for particles moving at speeds much less than the photon velocity.
Integrating out the $A_0$ field we find the unscreened Coulomb interaction,
\begin{align}
S_{\rm unscreened} = - {\pi \over 2} \int d^3k\ J_0(k_0, k) {e_\ast^2 \over |k|} J_0(- k_0, -k),
\end{align}
between $(2+1)d$ particles.
It's convenient to interpret $S_{\rm unscreened}$ as arising from the exchange of a purely $(2+1)$d gauge field $\tilde{A}_0$ with kinetic term and coupling to $J_0$ as
\begin{align}
S_{\tilde A_0} = - \int d^3 k\ \Big( \tilde{A}_0(k_0, k) {|k| \over \pi e_\ast^2} \tilde{A}_0(-k_0, - k) + J_0(k_0, k) \tilde{A}_0(- k_0, - k) \Big).
\end{align}

The electromagnetic charge density $J_0(x)$ dualizes in \eqref{unite} according to
\begin{align}
\label{dualcharge}
J_0(x) = {\delta {\cal L}^{(1)} \over \delta A_0} = - {\theta \over 2 \pi} \epsilon_{ij} \partial_i a_j,
\end{align}
for vanishing $A_j$.
Decomposing the gauge field $a_i(k_0, k) = i {k_i \over |k|} a_L(k_0, k) - i {k_j \over |k|} \epsilon_{j i} a_T(k_0, k)$ in terms of its longitudinal and transverse components, the (unscreened) Coulomb interaction becomes a kinetic term for $a_T$ \cite{KMTW2015}:
\begin{align}
S_{\rm unscreened} = - {e_\ast^2 \theta^2 \over 8 \pi} \int d^3k\ a_T(k_0, k) |k| a_T(- k_0, - k).
\end{align}
A similar transformation of the Coulomb interaction occurs in nonrelativistic composite fermion theories \cite{halperinleeread}.
Notice that the unscreened Coulomb interaction results in a kinetic term that dominates a possible Maxwell coupling for $a_\mu$ at long wavelengths.

\subsubsection*{Dissipation}


To model dissipation following \cite{PhysRevB.69.054507}, we consider an auxiliary system consisting of a parallel two-dimensional electron gas (2DEG) that is coupled to \eqref{unite} through the Coulomb interaction, specifically, through $\tilde A_0$.
The spatial separation between the system \eqref{unite} and electron gas is assumed negligible. 
The electron Green's function is assumed to take a diffusive form,
\begin{align}
G_{2D}^{-1}(i k_0, k) = i k_0 - ({k^2 \over 2m} - \epsilon_F) + {i \over 2 \tau} {\rm sign}(k_0).
\end{align}

The dissipative effects arising from the coupling to the two-dimensional electron gas are encoded in a correction to the $\tilde A_0$ kinetic term in $S_{\tilde A_0}$
\cite{altlandsimonsbook, colemanbook},
\begin{align}
\delta S_{\tilde A_0} = \int d^3k\ \tilde{A}_0(k_0, k) {\sigma_e k^2 \over |k_0| + D_e k^2} \tilde{A}_0(- k_0, - k),
\end{align}
where the Drude conductivity $\tilde \sigma_e = q_e^2 N D_e$ with $N$ the density of states at Fermi energy $\epsilon_F$ of the two-dimensional electron gas and $D_e$ its diffusivity.
Higher-order corrections due to the two-dimensional electron gas will be ignored.
Including $\delta S_{\tilde A_0}$ we obtain the dissipation-corrected density-density \eqref{dualcharge} interaction upon integrating out $\tilde A_0$:
\begin{align}
\label{disscoulomb}
S^{(2)} = - {e_\ast^2 \theta^2 \over 8 \pi} \int d^3 k\ a_T(k_0, k) \Big({k^2 \over |k| + f(k_0, k)} \Big) a_T(- k_0, - k),
\end{align}
where
\begin{align}
f(k_0, k) = {\sigma_e k^2 \over |k_0| + D_e k^2}, \quad \sigma_e = e_\ast^2 \tilde \sigma_e.
\end{align}
We recover the dual of an unscreened Coulomb interaction when $q_e = 0$, as expected, or as $|k|/|k_0| \rightarrow 0$.
The Coulomb interaction is shortranged as $D_e \rightarrow \infty$ at finite density of states $N$ or when $|k_0|/|k| \rightarrow 0$; in either of these limits, we find a Maxwell-like kinetic term for $a_T$ (albeit with inverted charge $1/e_\ast$). 

\subsection{Quenched Randomness}
\label{randomness}

We consider the effects of quenched disorder that's induced by random $A_\mu({\bf x})$ and $M({\bf x})$.
In this discussion, we assume the Coulomb interaction has been included via \eqref{disscoulomb} and $A_\mu({\bf x})$ is a non-dynamical quenched random variable.
Since $m({\bf x}) \propto M({\bf x})$, these perturbations readily map across the duality to
\begin{align}
\label{directdisordermapping}
\delta {\cal L} = - m({\bf x}) \overline{\psi} \psi(x)  - {\theta \over 2 \pi} A({\bf x}) d a(x),
\end{align}
where ${\bf x} = (x_1, x_2)$ and $x = (x_0, x_1, x_2)$.
The second term in Eq.~\eqref{directdisordermapping} is ``topological disorder," i.e., a random source to the field strength or ``topological" current $da$.
We have dropped a possible term proportional to $\epsilon_{ij} A_0({\bf x}) \partial_i A_j({\bf x})$ arising from the Chern-Simons term for $A$ in \eqref{unite}.

Interactions generate additional operators with random couplings, consistent with the symmetries of $A_\mu({\bf x})$ and $m({\bf x})$.
The Harris criterion \cite{Harris_1974} (for Gaussian-correlated randomness) implies the relevant terms at low energies correspond to operators with scaling dimensions $\Delta \leq z + 1$.
At large $N_f$ \cite{Borokhov:2002ib, 2016JHEP...08..069C, ChenFisherWu1993} the most generic random terms to include are \cite{PhysRevB.95.235146}
\begin{align}
\label{generaldisorderLagrangian}
{\cal L}_{\rm dis} = m({\bf x}) \bar \psi \psi(x) + i \tilde a_0({\bf x}) \bar \psi \gamma^0 \psi(x) + i \tilde a_j({\bf x}) \bar \psi \gamma^j \psi(x) - A_0({\bf x}) b(x) + \epsilon_{j k} A_j({\bf x}) e_k(x),
\end{align}
where $b = \epsilon_{ij} \partial_i a_j$ and $e_k = \partial_0 a_k - \partial_k a_0$.
The random couplings are assumed to be {\it independent} Gaussian-correlated quenched random variables with zero mean:
\begin{align}
\label{randomvariables}
\langle m({\bf x}) m({\bf x}') \rangle_{\rm dis} & = g_m \delta^{(2)}({\bf x} - {\bf x}'), \cr
\langle \tilde a_0({\bf x}) \tilde a_0({\bf x}') \rangle_{\rm dis} & = g_0 \delta^{(2)}({\bf x} - {\bf x}'), \cr
\langle \tilde a_k({\bf x}) \tilde a_k ({\bf x}') \rangle_{\rm dis} & = g_j \delta^{(2)}({\bf x} - {\bf x}'), \quad k \in \{x, y\}, \cr
\langle A_0({\bf x}) A_0({\bf x}') \rangle_{\rm dis} & = \Delta_0 \delta^{(2)}({\bf x} - {\bf x}'), \cr
\langle A_k({\bf x}) A_k({\bf x}') \rangle_{\rm dis} & = \Delta_j \delta^{(2)}({\bf x} - {\bf x}'), \quad k \in \{x, y\},
\end{align}
where $\langle \ \cdot \ \rangle_{\rm dis}$ indicates a disorder average and there is no sum over $k$. 
The disorder variances $g_m, g_0, g_j, \Delta_0, \Delta_j$ are positive constants. 

We study the effects of the randomness in \eqref{generaldisorderLagrangian} using the replica trick, which enables the calculation of the disorder-averaged free energy and all observables that derive from it.
To this end, we introduce $n_r$ replicas $\psi_{I, \ell}$ and $a_{\mu, \ell}$ with $\ell \in \{1, \ldots, n_r \}$ and consider the replicated partition function,
\begin{align}
Z^{n_r} = \prod_\ell \Big( \int {\cal D} \psi_\ell {\cal D} \bar \psi_\ell {\cal D} a_\ell \Big) e^{i \sum_\ell \Big(S^{(1)}[\psi_\ell, a_\ell] + S^{(2)}[a_\ell] + S_{\rm dis}[\psi_\ell, a_\ell] \Big)},
\end{align}
where $S^{(1)}[\psi_\ell, a_\ell] = \int d^3 x {\cal L}^{(1)}(\psi_\ell, a_\ell)$ with ${\cal L}^{(1)}$ given in \eqref{unite}, $S^{(2)}[a_\ell]$ is given in \eqref{disscoulomb}, and $S_{\rm dis}[\psi_\ell, a_\ell] = \int d^3 x {\cal L}_{\rm dis}(\psi_\ell, a_\ell$ with ${\cal L}_{\rm dis}$ given in \eqref{generaldisorderLagrangian}.
Using the identity,
\begin{align}
\log Z = \lim_{n_r \rightarrow 0} {Z^{n_r} - 1 \over n_r},
\end{align}
the disorder-averaged free energy, proportional to $\langle \log Z \rangle_{\rm dis}$, is found upon disorder-averaging.
Using \eqref{randomvariables}:
\begin{align}
\langle Z^{n_r} \rangle_{\rm dis} = \prod_\ell \Big( \int {\cal D} \psi_\ell {\cal D} \bar \psi_\ell {\cal D} a_\ell \Big) e^{i \sum_\ell \Big(S^{(1)}[\psi_\ell, a_\ell] + S^{(2)}[a_\ell] + i S^{(3)}[\psi_\ell, a_\ell] \Big)},
\end{align}
where 
\begin{align}
\label{minkowskidisorder}
S^{(3)}[\psi_\ell, a_\ell] & = - {1 \over 2} \sum_k \int dt dt' d^2x \Big[g_m \Big(\bar \psi_\ell \psi_\ell \Big)(t) \Big(\bar \psi_k \psi_k\Big)(t') + g_0 \Big(\bar \psi_\ell \gamma^0 \psi_\ell \Big) (t) \Big(\bar \psi_k \gamma^0 \psi_k\Big)(t') \cr 
& + g_j \Big(\bar \psi_\ell \gamma^j \psi_\ell \Big) (t) \Big( \bar \psi_k \gamma^j \psi_k \Big) (t') + \Delta_0 b_\ell(t) b_k(t') + \Delta_j {\bf e}_\ell(t) \cdot {\bf e}_k(t') \Big],
\end{align}
$\Big(\bar \psi_\ell^{(I)} \psi_\ell^{(I)}\Big)(t) \Big(\bar \psi_k^{(J)} \psi_k^{(J)}\Big)(t') \equiv \bar \psi_\ell^{(I)}(x, t) \psi_\ell^{(I)}(x, t) \bar \psi_k^{(J)}(x, t') \psi_k^{(J)}(x, t')$ and similarly for the other terms appearing in $S_E^{(3)}$.

\section{Renormalization Group Analysis}
\label{rganalysis}

We now study the critical properties of the model introduced in \S\ref{setup}.
Details of our calculations are presented in Appendix \ref{barerenormalizedappendix}.

\subsection{Large $N_f$ Expansion and Renormalization Group Scheme}

The Euclidean effective action in $D+1$ dimensions is
\begin{align}
S_E = S^{(1)}_E + S^{(2)}_E  + S^{(3)}_E
\end{align}
where\footnote{Replica indices $\ell, k \in \{1, \ldots, n_r \}$ and flavor indices $I, J \in \{1, \ldots, N_f \}$ with repeated indices summed. In Euclidean signature $(+, +, +)$, the coordinates $(\tau, x_j) = (i t, x_j)$, Fourier space variables $(\omega, k_j) = (- i k_0, k_j)$, and the $\gamma$-matrices $(\gamma_0, \gamma_1, \gamma_2) = (\sigma^3, \sigma^1, \sigma^2)$.}:
\begin{align}
\label{action1}
S^{(1)}_E & = \int d\tau d^D x \Big[ \bar{\psi}^{(I)}_\ell \Big( \gamma_\tau  (\partial_\tau + i {g \over \sqrt{N_f}} a_{\tau, \ell})
+ v \gamma_j  (\partial_j+ i {g \over \sqrt{N_f}} a_{j, \ell}) 
 \Big) \psi^{(I)}_\ell + m \bar \psi^{(I)}_\ell \psi^{(I)}_\ell \cr
 & + {i \kappa \over 2} a_\ell d a_\ell \Big], \\
 \label{action2}
S^{(2)}_E & = \int d\omega d^Dk\ {w_x \over 2} a_{\ell, T}(\omega, k) {k^2 \over |k| + f(\omega, k)} a_{T, \ell}(- \omega, - k), \\
\label{action3}
S^{(3)}_E  & = - {1 \over 2} \int d\tau d\tau' d^Dx \Big[g_m \Big(\bar \psi_\ell^{(I)} \psi_\ell^{(I)}\Big)(\tau) \Big(\bar \psi_k^{(J)} \psi_k^{(J)}\Big)(\tau') + g_0 \Big(\bar \psi^{(I)}_\ell \gamma^0 \psi^{(I)}_\ell \Big) (\tau) \Big(\bar \psi^{(J)}_k \gamma^0 \psi^{(J)}_k\Big)(\tau') \cr 
& + g_j \Big(\bar \psi^{(I)}_\ell \gamma^j \psi^{(I)}_\ell \Big) (\tau) \Big( \bar \psi^{(J)}_k \gamma^j \psi^{(J)}_k \Big) (\tau') + \Delta_0 b_\ell(\tau) b_k(\tau') + \Delta_j {\bf e}_\ell(\tau) \cdot {\bf e}_k(\tau') \Big].
\end{align}
We've set the longitudinal component of $a_i$ to zero (Coulomb gauge): $a_i(\omega, k) = i {k_j \over |k|} \epsilon_{ji} a_T(\omega, k)$.
The gauge coupling is $g/\sqrt{N_f}$ with $g$ fixed and $N_f \rightarrow \infty$ \cite{Coleman}, $v$ is the Dirac composite fermion velocity, and the Dirac mass $m$ vanishes at criticality.
The disorder variances $g_m, g_0, g_j, \Delta_0, \Delta_j$ are assumed to scale as $1/N_f$.
The Chern-Simons level is controlled by $\kappa = {2 \theta - 1 \over 4 \pi}$: $\kappa = 0$ gives an IQHT and $\kappa = 1/4\pi$ gives a SIT.
$w_x = {e_\ast^2 \over 4 \pi}$ parameterizes the strength of the dissipative Coulomb interaction and $f(\omega, k) = {\sigma_e k^2 \over |\omega| + D_e k^2}$.
The non-dynamical (electromagnetic) field $A_\mu = 0$.
In the remainder, we'll often leave replica and flavor indices, as well as the spacetime dependence of fields implicit.

We regularize UV divergent integrals that appear in our renormalization group analysis of $S_E$ using {\it dimensional reduction} \cite{SIEGEL1979193, Chen:1992ee, ChenFisherWu1993}.
This is the standard approach (e.g., \cite{Avdeev:1991za, Chen:1992ee, ChenFisherWu1993, GMPTWY2012, AharonyGurAriYacoby2012} and references therein) used in the study of theories of Chern-Simons gauge fields coupled to matter and, in contrast to dimensional regularization, has been shown to preserve gauge invariance at least to 2-loop order in the perturbative analysis of $S_E^{(1)}$ \cite{Chen:1992ee}.
We assume without proof that this regularization procedure maintains gauge invariance in our large $N_f$ study of $S_E$, which involves 3-loop integrals.
We consider a slight variation of the conventional dimensional reduction approach.
First, all vector, tensor, and spinor algebra is performed in $3$d; in particular, the antisymmetric symbol $\epsilon^{\mu \nu \rho}$ obeys the usual $3$d identities.
Second, loop integrals are analytically continued to general (Euclidean) spatial dimension $D \leq 2$:
\begin{align}
\int {d\omega d^2k \over (2 \pi)^3} \rightarrow \mu^{\epsilon} \int {d \omega d^D k \over (2 \pi)^{D + 1}},
\end{align}
where $\epsilon = 2 - D$ and $\mu$ is the renormalization group scale.\footnote{Typically in dimensional reduction the spacetime dimension is analytically continued.}
Simple poles proportional to $2/\epsilon$ are identified with logarithmic divergences proportional to $\log(\Lambda^2/\mu^2)$ in a theory with momentum cutoff $\Lambda$; power-law divergences are set to zero.

\begin{figure}[h!]
  \centering
\includegraphics[width=1.0\linewidth]{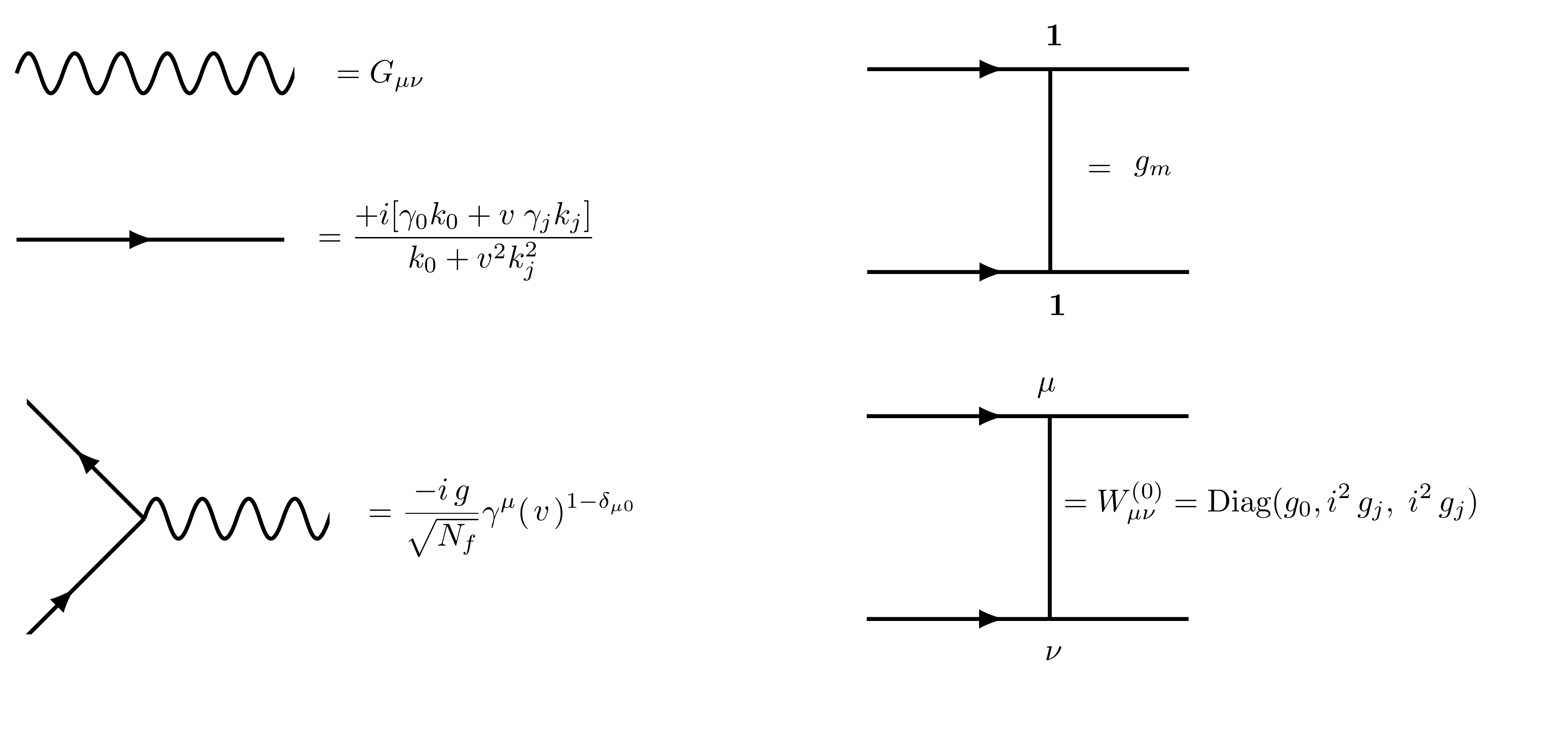} \qquad 
\caption{Feynman rules of $S_E$. 
The wavy line denotes the effective gauge field propagator and the directed solid line indicates the fermion propagator with $m = 0$.
Disorder is represented by a solid line without an arrow and specified by its disorder variance ($g_m, g_0, g_j$). 
Screening of the disorder ($g_0, g_j$) and topological disorder ($\Delta_0, \Delta_j$) are discussed in the Appendix \ref{disorderscreening}.}
\label{Feyn-rules}
\end{figure}
\begin{figure}[h!]
  \centering
\includegraphics[width=1\linewidth]{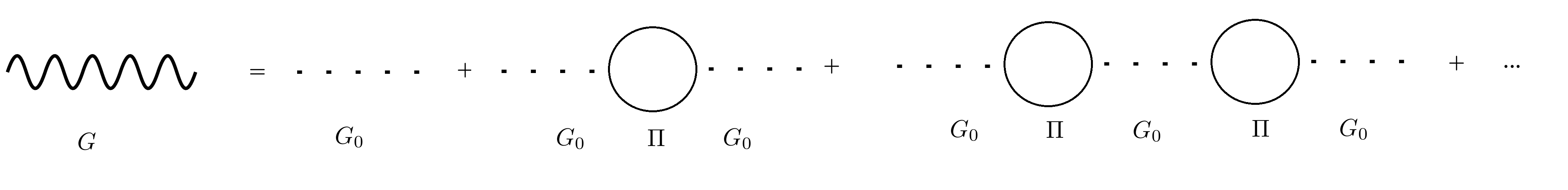} \qquad 
\caption{The effective gauge field propagator $G$. 
The dotted line $G_0$ represents the bare gauge field propagator and $\Pi_{\mu \nu}$ is the 1-loop gauge field self-energy. 
Each term in this geometric series of diagrams produces an ${\cal O}(N_f^0)$ correction to the gauge field propagator, since each fermion loop contributes a factor of $N_f$ and the two vertices associated to each loop contribute an additional factor of $g^2/N_f$.
}
\label{RPA_series}
\end{figure}

The large $N_f$ Feynman rules that derive from $S_E$ at $m=0$ are given in Fig.~\ref{Feyn-rules}. 
We've summed once and for all the geometric series of fermion bubble diagrams in Fig.~\ref{RPA_series} and replaced the bare gauge field propagator by the effective propagator,
\begin{align}
G_{m n} = \begin{pmatrix} {g^2 \over 16} {k^2 \over \sqrt{\omega^2 + v ^2 k^2}} & i \kappa |k| \cr i \kappa |k| & w_x {k^2 \over |k| + f(\omega, k)} + {g^2 \over 16} \sqrt{\omega^2 + v^2 k^2} \end{pmatrix}_{m n}^{-1},
\end{align}
where $m, n \in \{0, T \}$ correspond to the zeroth and transverse components of $a_\mu$.
At large $N_f$ this resummation is equivalent to the random phase approximation.
The same effect also leads to a screening of the $g_0$ and $g_j$ disorders (see Appendix \ref{disorderscreening}) \cite{PhysRevB.60.8290, PhysRevB.95.235145}.
Aside from a few exceptions that we'll discuss, we've found disorder screening to be a subleading effect in our analysis. 

We use minimal subtraction \cite{tHooft:1973mfk, Weinberg:1996kr} to renormalize $S_E$.
In this scheme, simple poles in $\epsilon$ appear in counterterms $b_{\lambda_a}(\vec \lambda^R)/\epsilon$ that relate bare ($B$) and renormalized ($R$) couplings:
\begin{align}
\label{barerenorm}
\lambda^B_a \mu^{- \Delta_a(\epsilon)} = \lambda^R_a(\mu, \epsilon) + {b_{\lambda_a}(\vec \lambda^R(\mu, \epsilon)) \over \epsilon},
\end{align}
where the vector of coupling constants (either $B$ or $R$)
\begin{align}
\vec \lambda = \Big({g^2 \over N_f}, v, m, \kappa, w_x, \sigma_e, D_e, g_m, g_0, g_j, \Delta_0, \Delta_j\Big)^T.
\end{align}
The renormalized couplings $\lambda_a^R(\mu, \epsilon)$ and residues $b_{\lambda_a}(\vec \lambda^R(\mu, \epsilon))$ are analytic in $\epsilon$.
The higher-order poles that generally occur on the right-hand side of Eq.~\eqref{barerenorm} can be set to zero.
The bare couplings $\lambda^B_a$ have engineering dimensions equal to $\Delta_{\lambda_a}(\epsilon)$ while the renormalized couplings are dimensionless. 
Appendix \ref{barerenormalizedappendix} details the calculation of the counterterms $b_{\lambda_a}(\vec \lambda^R)/\epsilon$.

\begin{table}[h!]
\centering
\begin{tabular}{| c | c | c | c |}
\hline
$\lambda_a$ & $\Delta_{\lambda_a}(\epsilon)$ & $\bar \Delta_{\lambda_a}$ & $\rho_{\lambda_a}$ \\ 
 \hline
${g^2 \over N_f}$ & $\epsilon$& $0$ & $1$ \\  
 \hline
$\{v, w_x, \sigma_e \}$ & $z - 1$ & $z - 1$ & $0$ \\
\hline
$m$ & $z$ & $z$ & $0$ \\  
 \hline
$\{\kappa, \Delta_j \}$ & $0$ & $0$ & $0$ \\
\hline
$D_e$ & $z - 2$ & $z - 2$ & $0$ \\
\hline
$\{g_m, g_0, g_j \}$ & $2 z - D$ & $2 z - 2$ & $1$ \\
\hline
$\Delta_0$ & $2 z - 2$ & $2 z - 2$ & $0$ \\
\hline
\end{tabular}
\caption{Engineering dimension $\Delta_{\lambda_a}(\epsilon) = \bar \Delta_{\lambda_a} + \rho_{\lambda_a} \epsilon$ of bare coupling $\lambda^B_a$ ($B$ superscript omitted in the table), where $\bar \Delta_{\lambda_a}$ is independent of $\epsilon = 2 - D$ and $\rho_{\lambda_a}$ is the constant coefficient of $\epsilon$.}
\label{dims}
\end{table}
The engineering dimensions $\Delta_{\lambda_a}(\epsilon)$ of the bare couplings are given in Table \ref{dims}.
These dimensions are determined as follows.
Each term in $S_E^{(1)}$ is dimensionless with the assignments:
\begin{align}
\Delta_\tau = - \Delta_\omega = - z,\quad  \Delta_x = - \Delta_k = - 1,
\end{align} 
\begin{align}
\Delta_{\psi(\tau, x)} = {1 \over 2} - {\epsilon \over 2}, \quad \Delta_{a^E_0(\tau, x)} = z - \eta \epsilon, \quad \Delta_{a^E_j(\tau, x)} = 1 - \eta \epsilon,
\end{align}
and
\begin{align}
\Delta_{g^2/N_f} = 2 \eta \epsilon, \quad \Delta_v = z - 1, \quad \Delta_m = z, \quad \Delta_\kappa = (2 \eta - 1) \epsilon,
\end{align}
where $\eta$ is an arbitrary constant.
We've introduced the dynamical critical exponent $z$ with a value to be determined later; in the absence of $S_E^{(2)}$ and $S_E^{(3)}$, relativistic symmetry requires that $v$ be dimensionless and $z=1$.
In the large $N_f$ expansion, $g$ is fixed and we formally take $\Delta_{N_f}= - 2 \eta \epsilon$.
The effective gauge field propagator is consistent with the engineering dimensions $\Delta_{a^E_0(\omega, k)} = - D - \eta \epsilon$ and $\Delta_{a_j^E(\omega, k)} = (1 - z) - D - \eta \epsilon$ if $\eta=1/2$.
The dimensions of the remaining couplings ensure the terms in $S_E^{(2)}$ and $S_E^{(3)}$ are dimensionless.
 
The beta functions $\beta_{\lambda_a}$ at $\epsilon = 0$ are read off from the residues $b_{\lambda_a}(\vec \lambda^R)$ using
\begin{align}
\label{betafunctions}
\beta_{\lambda_a}(\vec \lambda^R) \equiv - \mu {\partial \lambda^R_a \over \partial \mu} = \bar \Delta_{\lambda_a} \lambda^R_a + \rho_{\lambda_a} b_{\lambda_a}(\vec \lambda^R) - \sum_c \rho_{\lambda_c} \lambda^R_c {\partial b_{\lambda_a}(\vec \lambda^R) \over \partial \lambda^R_c}.
\end{align}
There is no sum over $a$ in Eq.~\eqref{betafunctions}.
The minus sign in front of $\mu {\partial \lambda^R_a \over \partial \mu}$ means that a relevant/irrelevant coupling has a positive/negative beta function.
Notice that only $g^2/N_f$ and the variances $g_m, g_0, g_j$ can contribute to the derivative term on the right-hand side of Eq.~\eqref{betafunctions}. 

We characterize any fixed points $\beta_{\lambda_a}(\vec \lambda^R) = 0$ by the dynamical critical exponent $z$ and correlation length exponent $\nu$, evaluated at the fixed point.
The dynamical critical exponent enters the beta functions \eqref{betafunctions} via $\bar \Delta_{\lambda_a}$ (see Table \ref{dims}) and we determine its value by the condition of vanishing velocity beta function $\beta_v(\vec \lambda^R) =0$\footnote{Nonzero $\beta_v$ implies a quantum correction to the tree-level dynamical exponent, i.e., the engineering dimension $-\Delta_\tau$. 
This follows from the fermion dispersion relation $|\omega| = v |k|$ (see, e.g., \cite{PhysRevB.75.235423}).
We've chosen to introduce an arbitrary $z$ in Table 3 with a value to be determined by vanishing renormalization of the velocity. 
An equivalent choice is to take engineering dimensions consistent with conformal invariance and infer any correction to the tree-level dynamical scaling from a nonzero velocity beta function.}:
\begin{align}
\label{zdef}
z = 1 + {1 \over v^R} \sum_c \rho_{\lambda_c} \lambda^R_c {\partial b_{v}(\vec \lambda^R) \over \partial \lambda^R_c}.
\end{align}
Since the transitions we consider in this paper are tuned by the Dirac mass, we define the correlation length $\xi$ as the inverse momentum scale $\mu_0^{-1}$ at which $v^R(\mu_0)/m^R(\mu_0)$ = 1.\footnote{The factor of $v^R(\mu_0)$ accounts for possible running of the velocity in the equivalent approach where $z=1$ is chosen in Table \ref{dims} and the nonzero velocity beta function determines the correction $z-1$ to dynamical scaling.}
We write the mass beta function as
\begin{align}
\label{massbeta}
\beta_{m}(\vec \lambda^R) = \Big(z - \gamma_{\bar \psi \psi}(\vec \lambda^R) \Big) m^R,
\end{align}
where the anomalous dimension $\gamma_{\bar \psi \psi}$ controls the asymptotic scaling of the correlation function $\langle \bar \psi \psi(\tau, x) \bar \psi \psi(0) \rangle \sim |v^2 \tau^2 + x^2|^{- (D + \gamma_{\bar \psi \psi})}$.
Using Eqs.~\eqref{betafunctions} - \eqref{massbeta}, we find
\begin{align}
\xi = \Lambda^{-1} \Big({m_\Lambda/v_\Lambda \over \Lambda} \Big)^{- \nu}
\end{align}
where $\Lambda$ is an arbitrary momentum cutoff defining the ``initial conditions," $m^R(\Lambda) = m_\Lambda/\Lambda^z$ and $v^R(\Lambda) = v_\Lambda/\Lambda^{z-1}$, and the inverse correlation length exponent
\begin{align}
\nu^{-1} = z - \gamma_{\bar \psi \psi}.
\end{align}
Note that $m^R$ does not enter the residues $b_{\lambda_a}(\vec \lambda^R)$ with $\lambda^R_a \neq m^R$ and only appears linearly in $b_{m}(\vec \lambda^R)$.

In the remainder of the main text, we drop the $B$ and $R$ superscripts for notional clarity.

\subsection{General Analysis}

We now present the results of our renormalization group calculation, which is valid to order $1/N_f$ in the large $N_f$ expansion.
See Appendix \ref{barerenormalizedappendix} for details.

Vanishing velocity beta function determines the dynamical critical exponent to be
\begin{align}
z = 1 + \gmb + \gob + 2\gjb  
- F_w( \wb,\kappa,\seb),
\label{z-general}
\end{align}
where $g_1 = g^2/16$, the rescaled couplings are
\begin{align}
&\overline m = {m \over v}, 
 \wb = \frac{w_x}{v},
\seb = \frac{\sigma_e}{v},
 \gmb= \frac{g_m}{2 \pi v^2},
\gob = \frac{g_0}{2 \pi v^2},
\gjb = \frac{g_j}{2 \pi v^2}, 
\Dob =   \Delta_0, 
\Djb = \Delta_j \, v^2,      
\end{align}
and
\begin{eqnarray}
F_w(\wb, \kappa, \seb) = {1 \over 4 \pi^2 N_f}
\int^{\infty}_{-\infty} dy \;
\frac{ 
 g_1(-1+2y^2)(\sigma_e +|y|) +  \wb |y| \sqrt{1+y^2} }
{(1+y^2)^{2}\, \Big( \longDeno \Big)} .
\label{Coulomb-F-fun}
\end{eqnarray}
The beta functions $\beta_{\lambda_a} = - \mu {\partial \lambda_a \over \partial \mu}$ for the remaining couplings take the form:
\begin{align}
\label{coulombrunning}
\beta_{\wb} & = \wb \big( z -1  \big), \\
\label{conductivityrunning}
\beta_{\seb} & = \seb \big( z - 1 \big), \\
\beta_{D_e} & = D_e \big( z - 2 \big), \\
\label{massbetadetailed}
\beta_{\overline m}  
 & =
\overline m\big(z - \gamma_{\bar \psi \psi} \big)  ,\\
\label{randommassbeta}
\beta_{\gmb} 
& = 2 \gmb \big(z - 1 + {2 \gob \, \gjb \over \gmb} - \gamma_{\bar \psi \psi}  \big), \\
\beta_{\gob} & = 2 \gob \big(z - 1 + {2 \gjb \, \gmb \over \gob} \big), \\
\beta_{\gjb} & = 2 \gjb \big(z - 1 + {\gmb \, \gob \over \gjb} - \gmb - \gob - 2 \gjb + F_w(\wb,\kappa,\seb) \big), \\
\beta_{\Dob} & = \Dob (2 z - 2) + {\gmb \big( \Djb (g_1 + \wb)^2 + \Dob \kappa^2 \big) \over 64 \big( g_1 (g_1 + \wb) \big)^2}  + \frac{\gob \, \gmb \, N_f \, \pi \, v^2 }{32}, \\
\beta_{\Djb} & = \; 
\frac{\gmb \; (g_1^2 \, \Dob + \kappa^2 \, \Djb )}
{ 128 \big(  g_1 (g_1 + \wb)+ \kappa^2 \big)^2 } + \frac{ \gjb \, \gmb N_f \pi v^4}{64  },
\end{align}
where $z$ is given in Eq.~\eqref{z-general}, the mass anomalous dimension
\begin{align}
\label{anomdimexplicit}
\gamma_{\bar \psi \psi} = 2 \gmb + 2 \gob  - {\gob g_1 - \gjb (g_1 + \wb) \over g_1^2 + g_1 \wb + \kappa^2}  
+ F_m(\wb,\kappa,\seb) -  F_w(\wb,\kappa,\seb),
\end{align}
and
\begin{align}
\label{Fmass}
F_m(\wb, \kappa, \seb) & = {1 \over 4 \pi^2 N_f} \int^{\infty}_{-\infty} dy \Big[
\frac{ g_1 (\seb + |y|  ) (-2y^2-3) - \wb \sqrt{1+y^2} |y|}
{(1+y^2)^2 \, \Big[ \longDeno \Big]} \cr
& +
\frac{ (\seb + |y|) 
\bigg(  \sqrt{1+y^2}(g_1^2 -\kappa^2)  ( \seb  + |y|  ) 
+ g_1 \wb |y|  
\bigg)
}
{2 (1+y^2)
\Big[ \sqrt{ 1+y^2 } (g_1^2 +\kappa^2)(\seb+|y|) + g_1  \wb  |y|   \Big]^2
 } \Big].
\end{align}
To simplify the above expressions, we have ignored terms that arise from the screening of the $g_0, g_j$ disorders (see Appendix \ref{disorderscreening}); our detailed analysis below includes such effects whenever relevant.
The gauge coupling $g/\sqrt{N_f}$ is marginal once the large $N_f$ effective gauge field propagator in Fig.~\ref{RPA_series} is adopted and so its beta function is not included.

Let's make a few additional comments about these expressions.
\begin{enumerate}
\item In general, the above beta functions don't have an IR stable solution at $\overline m = 0$, even when disorder screening is included. 
In the remaining sections, we analyze cases for which we have found fixed points when a symmetry is present.

\item We've taken the variances to scale as $1/N_f$ for $N_f \rightarrow \infty$. 
The beta functions have terms that scale as $1/N_f$ and $1/N_f^2$.
The ``classical" contributions to the beta functions arising from the engineering dimensions of couplings scale as $1/N_f$; the ``quantum" corrections generally scale as $1/N_f^2$. 
The exception to the latter appears in the third term in $\beta_{\Dob}$ and the second term in $\beta_{\Djb}$.

\item  The first three beta functions $\beta_{\wb}, \beta_{\seb}, \beta_{D_e}$ characterize the dissipative Coulomb interaction.
In our analysis, we consider $z < 2$ and so the diffusion constant $D_e$ is an irrelevant parameter that will be set to zero.
A nonzero Coulomb interaction allows for two classes of fixed points: 
(1) a {\it finite Coulomb interaction} either with $\wb,\seb \neq 0$ and $z=1$ or with $\wb = \seb = 0$ and $z$ determined by Eq.~\eqref{z-general}; 
(2) an {\it infinite Coulomb interaction} with $\wb \rightarrow \infty$, $\seb \rightarrow \infty$, and $1 < z < 2$ that is controlled by the dissipation parameter $\seb/\wb$.

\item Whenever two of the three disorder variances $g_m, g_0, g_j$ are considered, the third variance is radiatively generated.
When all three variances $g_m, g_0, g_j$ are present, both types of topological disorder $\Delta_0, \Delta_j$ are generated. 
This is consistent with the symmetry assignments in Table \ref{symmetrytable}.\end{enumerate}

\subsection{Finite Coulomb Interaction}

\subsubsection*{No Disorder}

\begin{figure}[h!]
  \centering
\includegraphics[width=0.4\linewidth]{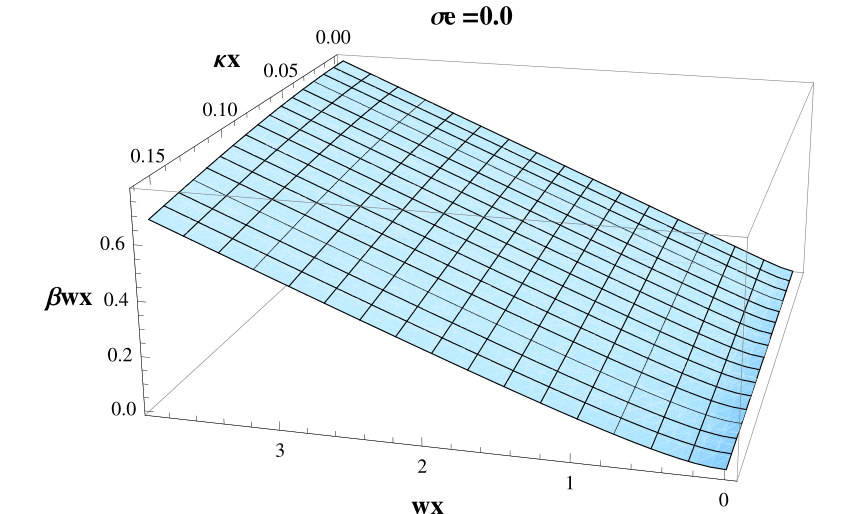} \qquad 
\includegraphics[width=0.4\linewidth]{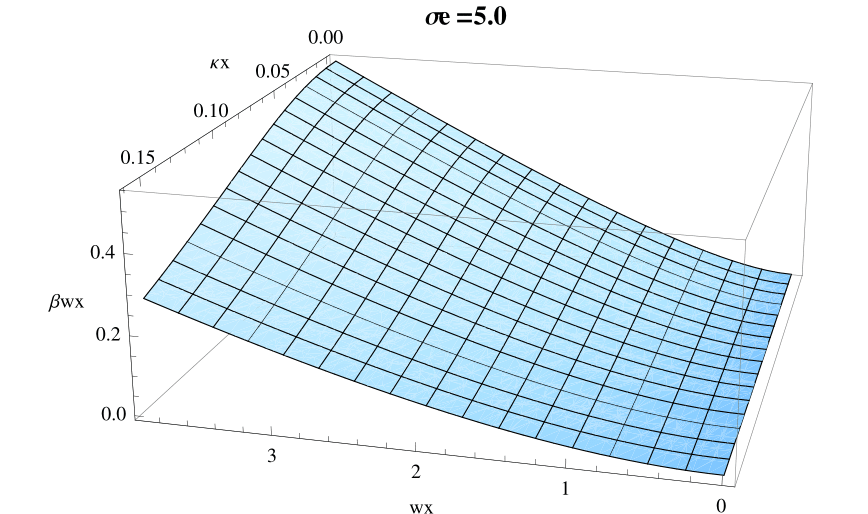}\\
\caption{$\beta_{\wb}$ as a function of $\wb$ and $\kappa$}
\label{fignodisorder}
\end{figure}

In the absence of any disorder, the only nontrivial beta functions are associated to the Coulomb interaction,
\begin{align}
\beta_{\wb} & = - \wb  \;  F_w( \wb,\kappa,\seb).
\end{align}
Since ${1 \over \seb} \beta_{\seb} = {1 \over \wb} \beta_{\wb}$, it's sufficient to consider the behavior of $\wb$ when studying a finite Coulomb interaction.
The integral that defines $F_w$ in Eq.~\eqref{Coulomb-F-fun} can only be evaluated numerically for general $\seb$. 
We've found that $F_w$ is positive for any $\kappa$ when $\wb \neq 0$.
Consequently, the clean fixed point with $\wb = \seb =0$ is perturbatively stable to the addition of a Coulomb interaction and $z = 1$.
Two examples for the behavior of $\beta_{\wb}$ are displayed in Fig.~\ref{fignodisorder}.

In particular, in the limit of $\seb=0$ the beta function for $\wb$ is always negative for any Chern-Simons coupling $\kappa$.
This result should be contrasted with earlier work \cite{PhysRevLett.80.5409} where a critical value of $|\kappa|$ was reported above which the Coulomb interaction was found to be relevant.\footnote{The discrepancy seems to arise from assigning the Chern-Simons coupling $\kappa$ an engineering dimension proportional to $\epsilon = 2 - D$.
This choice, which appears to be inconsistent with the scaling of the effective gauge field propagator in the large $N_f$ expansion, results in additional derivatives with respect to $\kappa$ in the beta function in \eqref{betafunctions}.}
As a check on our calculation, we find the mass anomalous dimension is given by
\begin{align}
\label{nuclean}
\gamma_{\bar \psi \psi}
= {128 \over 3 \pi^2 N_f} {1 - 512 \kappa^2 \over(1 + 256 \kappa^2)^2}
\end{align}
in agreement with \cite{ChenFisherWu1993, PhysRevB.66.144501, 2016JHEP...08..069C, PhysRevB.97.085112} for general $\kappa$. 
For the IQHT ($\kappa = 0$), $\nu^{-1} \approx 1 - 4.3/N_f$; for the SIT ($\kappa = 1/4\pi$), $\nu^{-1} \approx 1 + 1.4/N_f$.

\subsubsection*{${\cal C}$ Symmetry}

\begin{figure}[h!]
  \centering
\includegraphics[width=0.3\linewidth]{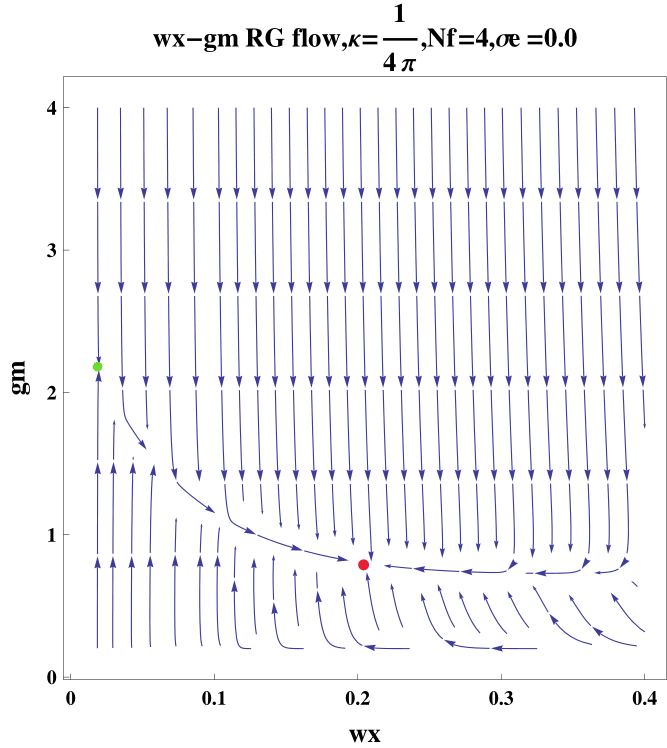} \;\; 
\includegraphics[width=0.3\linewidth]{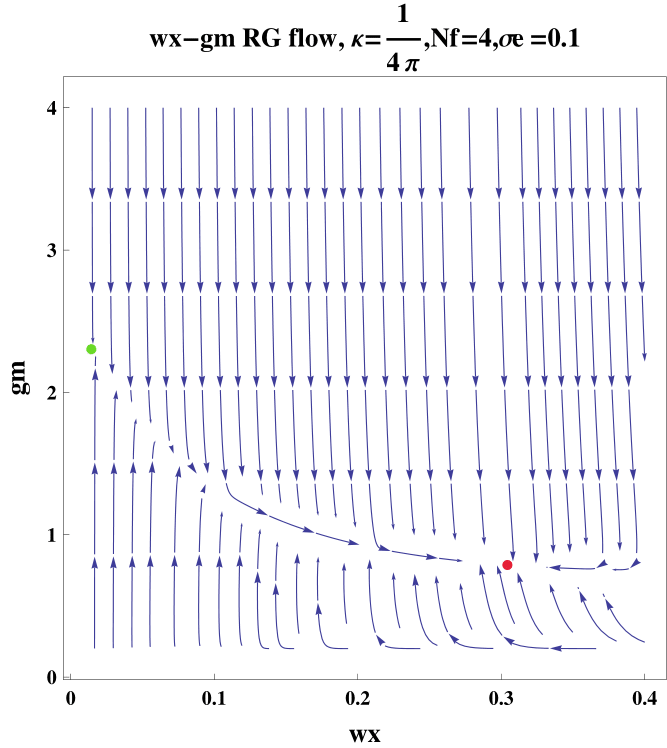}
\includegraphics[width=0.3\linewidth]{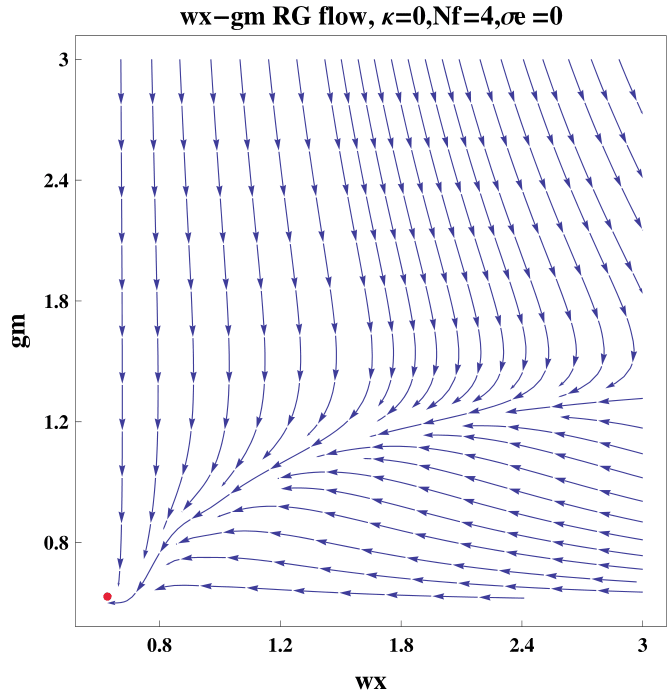}
\\
\caption{RG flow of $w_x$ vs. $g_m $.}
\label{figchargesymmetry}
\end{figure}

According to Table \ref{symmetrytable}, the Coulomb couplings and random mass disorder ($g_m$) are allowed when there is charge-conjugation symmetry. 
The beta functions are
\begin{align}
\beta_{\wb} & = \wb \bigg(
  \gmb    \; 
 -   
 F_w( \wb,\kappa,\seb)
\bigg), \\
 \beta_{\gmb} & = - 2 \gmb \bigg(\gmb + F_m( \wb,\kappa,\seb) \bigg),
 \end{align}
where $F_w$ and $F_{m}$ are defined in Eqs.~\eqref{Coulomb-F-fun} and \eqref{Fmass}.
The flow of the random mass is controlled by the mass anomalous dimension $\gamma_{\bar \psi \psi} = 2 \gmb - F_w( \wb,\kappa,\seb) + F_m( \wb,\kappa,\seb)$.
Within the large $N_f$ approximation, random mass disorder is only a relevant perturbation to the clean fixed point of the previous section ($\wb = \seb = 0$) when $\gamma_{\bar \psi \psi} < 0$, i.e., when $1 < 512 \kappa^2$ (see Eq.~\eqref{nuclean}), in agreement with \cite{PhysRevB.95.235146}.
The presence of a Coulomb interaction does not appear to alter this conclusion within our analysis.
For $\kappa = 1/4\pi$ (or any $\kappa^2 > 1/512$), there exists a line of fixed points with finite disorder and Coulomb interaction parameterized by $\seb$.
Fig.~\ref{figchargesymmetry} shows a few examples of this behavior.
Since $\beta_{\wb} \propto z - 1$ and $\beta_{\gmb} \propto - \gamma_{\bar \psi \psi}$ when $z = 1$, any fixed point with finite disorder and Coulomb interaction has
\begin{align}
\nu^{-1} = z = 1.
\end{align}
At the (generally unstable) fixed point with nonzero random mass disorder and vanishing Coulomb interaction,
\begin{align}
\nu^{-1} = 1 \quad \text{and} \quad z = 1 - {128 \over 3 \pi^2 N_f} {1 - 512 \kappa^2 \over(1 +  256 \kappa^2)^2},
\end{align}
where $\kappa^2 > 1/512$.
For $\kappa = 1/4\pi$, $z \approx 1 + 1.4/N_f$.

It's interesting to compare our results for the critical exponents with recent analytic and numerical studies of the dirty XY model. 
In a large $N_f$ expansion \cite{2019arXiv190909167G} report $\nu = 1$ and $z = 1 + .5/N_f$; numerics \cite{PhysRevB.94.134501} directly probes $N_f = 1$ with the result $\nu = 1.16(5)$ and $z = 1.52(3)$.
In \cite{PhysRevB.95.235145}, a finite disorder fixed point of quantum electrodynamics without Chern-Simons term was found using an $\epsilon$ expansion about $(3+1)$d.
Since our approximation schemes are different, there is no contradiction with our conclusion that random mass disorder is irrelevant when $\kappa = 0$. 
Nevertheless, it would be interesting to consider this issue further.

\subsubsection*{${\cal CT}$ Symmetry}

According to Table \ref{symmetrytable}, the Coulomb coupling, random scalar potential $g_0$, and topological disorder $\Delta_j$ are allowed by ${\cal CT}$ symmetry.
Because a nonzero Chern-Simons term is odd under time-reversal symmetry, we only consider $\kappa = 0$ in the next two subsections that study ${\cal CT}$ and ${\cal T}$ preserving disorder.
The beta functions are
\begin{align}
\beta_{\wb}  & = \wb \, \bigg(
 \gob (1 - \phi_1) 
    - \; F_w( \wb,\kappa=0,\seb) \bigg), \\
\beta_{\gob} &  = {2 \gob \over \wb} \beta_{\wb}, \\
\beta_{\Djb} & = 0,
\end{align}
where $\phi_1$ isolates any terms that arise from the screening of the disorder: $\phi_1 = 0$ means disorder screening is ignored; $\phi_1 = 1$ means that disorder screening is included.
While we're unaware of a general reason to exclude disorder screening, we'll discuss the behavior of the above beta functions both with and without screening to illustrate its effect.

\begin{figure}[h!]
  \centering
\includegraphics[width=0.3\linewidth]{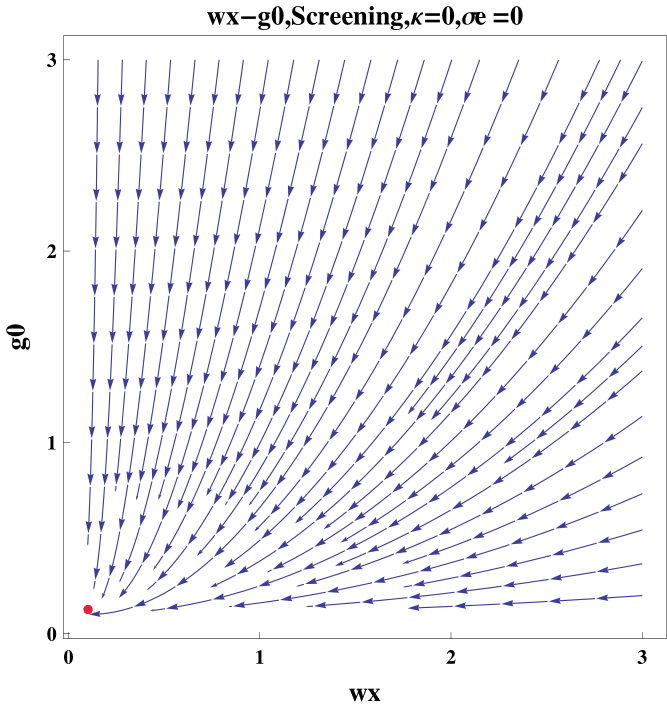} \;\; \;
\includegraphics[width=0.3\linewidth]{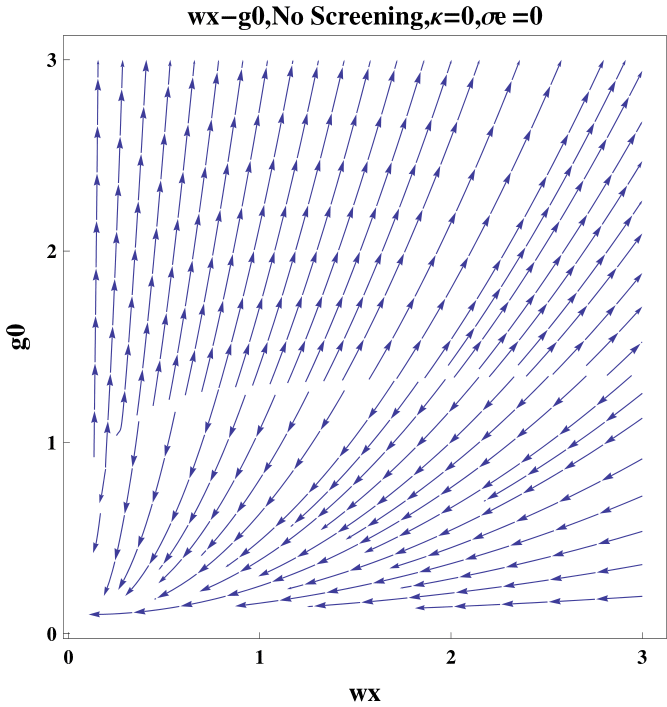} 
\; 
\caption{RG flow of $\wb$ vs. $\gob$.}
\label{CTsymmetryfig}
\end{figure}

As mentioned previously, $F_w$ in Eq.~\eqref{Coulomb-F-fun} is positive for any $\kappa$ when $\wb$ is nonzero; in particular when $\seb = 0$, $F_w(\wb, 0, 0)$ is a monotonically increasing function that approaches ${8 \over N_f \pi^2}$ for $\wb \rightarrow \infty$. 
When disorder screening is ignored ($\phi_1 = 0$), there is a fixed surface defined by $\gob = F_w$, which is parameterized by $(\wb, \gob, \Djb)$, in agreement with \cite{PhysRevB.95.235146}.
This fixed surface is unstable, e.g., consider perturbation to $\gob$ at fixed $\wb$ and $\Djb$.
When disorder screening is included ($\phi_1 = 1$), there is a line of {\it stable} fixed points parameterized by $(\wb, \gob, \Djb) = (0, 0, \Djb)$.
This result is consistent with \cite{PhysRevB.95.235145}.
This behavior is illustrated in Fig.~\ref{CTsymmetryfig}.
The corresponding critical exponents at the stable fixed point ($\phi_1 = 1$) reduce to those of the clean theory without Coulomb interactions.
Recall that $g_0$ and $\Delta_j$ disorders are generated by random electrical vector potential $A_j({\bf x})$ in the free Dirac fermion dual at $N_f =1$ for which a line of diffusive fixed points was found in \cite{PhysRevB.50.7526}.
It's unclear to what extent the line of fixed points parameterized by $\Delta_j$ is related.

If ${\cal CT}$ and ${\cal T}$ are emergent symmetries of the SIT theory we consider, then the beta functions should only have fixed point solutions respecting these symmetries at $\kappa = 1/4\pi$.
Unfortunately, the leading terms in the large $N_f$ beta functions don't produce any such nontrivial fixed points.
Even if $\gmb$ is initially tuned to zero, the random mass beta function receives a positive correction from disorder screening equal to 
\begin{align}
\delta \beta_{\gmb} = {8 \gob^2 g_1^2 \kappa^4 \over (g_1^2 + g_1 \wb + \kappa^2)^4}.
\end{align}
Nonzero $\gob$ and $\gmb$ then results in the generation of all couplings, for which we find runaway flow.

\subsubsection*{${\cal T}$ Symmetry}

According to Table \ref{symmetrytable}, the Coulomb coupling, random vector potential $g_j$, and topological disorder $\Delta_0$ are allowed by ${\cal T}$ symmetry.
The beta functions are
\begin{align}
\beta_{\wb} & = \wb \;   \big(  2 \gjb \;
 \big(1 - \phi_1
  \frac{  g_1 \, (g_1 + 2 \wb)  }{(g_1 + \wb )^2}  \big)
  -  
  F_w( \wb,\kappa =0 ,\seb)
\big), \\
\beta_{\gjb} & = 0, \\ 
\beta_{\Dob} & = {2 \Dob \over \wb} \beta_{\wb},
\end{align}
where we continue to use $\phi_1$ to isolate terms that arise from the screening of the disorder.

\begin{figure}[h!]
  \centering
\includegraphics[width=0.3\linewidth]{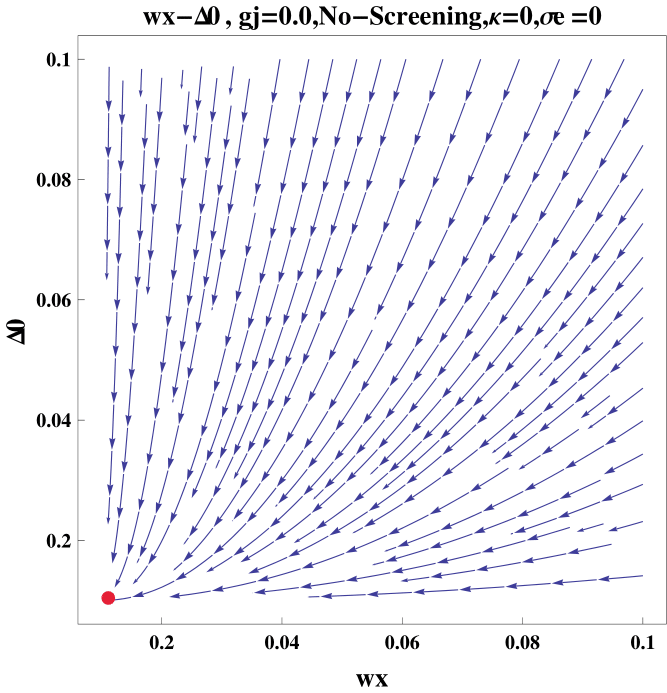} \;\; \;
\includegraphics[width=0.3\linewidth]{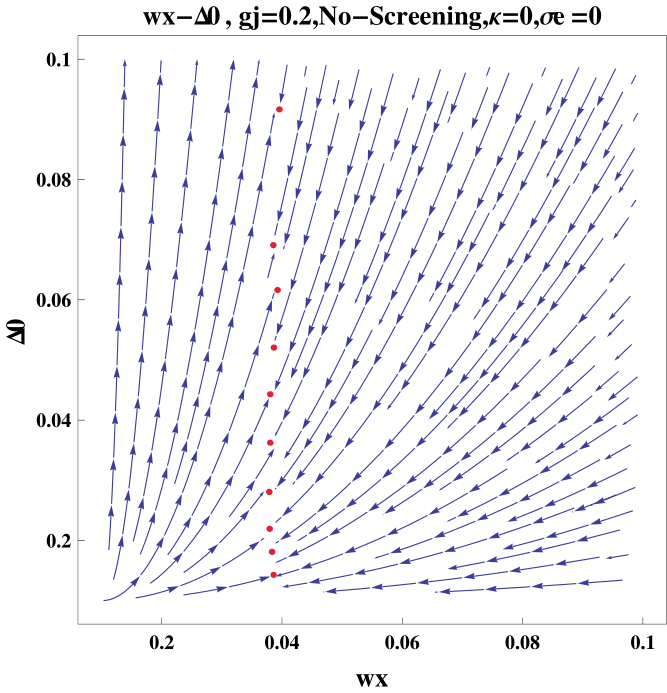} \;\; \; 
\includegraphics[width=0.3\linewidth]{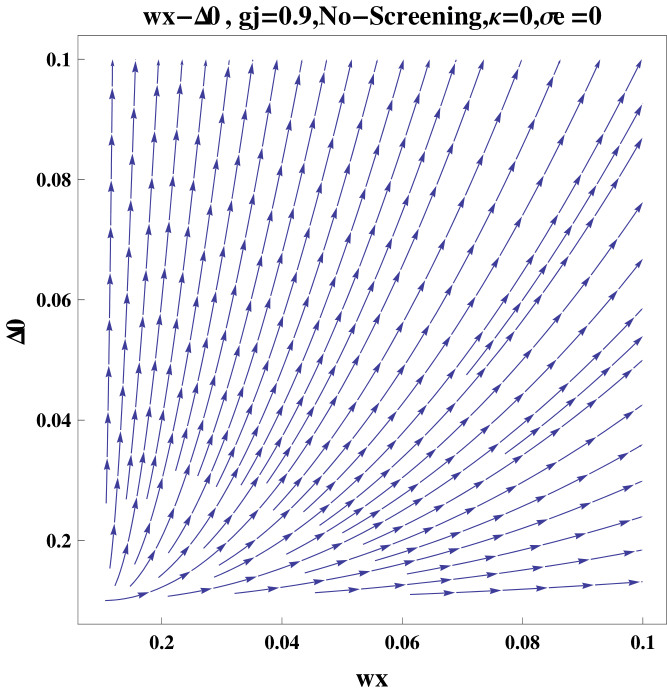} \; 
\caption{RG flow of $\wb$ vs. $\Dob$.}
\label{D0-gj}
\end{figure}

If screening is ignored ($\phi_1 = 0$), there is a surface of fixed points defined by $2 \gjb = F_w(\wb, 0, \seb)$ and parameterized by $(\wb, \gjb, \Dob)$ with $\gjb < 4/N_f \pi^2$ and non-negative $\Dob$.
On this surface $z=1$ and $\nu^{-1} = 1 + 2 \gjb - {\gjb \over g_1} - F_m(\wb, 0, \seb)$; $F_m$ is a monotonically decreasing function of $\wb$ when $\kappa = \seb = 0$: ${128 \over 3 \pi^2 N_f} \geq F_m(\wb) \geq - {8 \over \pi^2 N_f}$.
For fixed $\gjb$, this surface is stable to small deformation by $\wb$ since $F_w$ is an increasing function of $\wb$.
For $\gjb > 4/N_f \pi^2$, we find runaway flows.
This behavior is shown in Fig.~\ref{D0-gj}.
If screening is included ($\phi_1 = 1$), the fixed points are determined by the equation,
\begin{eqnarray}
\gjb^* = \frac{F_w(\wb)}{2[1- \phi_1 \frac{g_1(g_1+2 \wb)}{(g_1+\wb)^2} ]} 
\equiv f_s(\wb)
\end{eqnarray}
where $f_s(\wb)$ monotonically decreases from $f_s(\wb = 0) = \infty$ to $f_s(\wb \to \infty) = \frac{4}{N_f \pi^2}$.
If $\gjb$ is chosen to be smaller than $\frac{4}{N_f \pi^2}$, then the RG flows to
$(\wb^*, \Dob^*) =(0,0)$ because $\beta_{\wb} $ is nonnegative and only vanishes when $\wb  =0$. 
If $\gjb > \frac{4}{N_f \pi^2} $, then there exists a finite value of $\wb$ for which the beta function vanishes, however, the resulting fixed point is IR unstable.

\subsection{Infinite Coulomb Interaction}

Dissipation has played only a minor role in the above analysis.
We'll now discuss how dissipation allows for fixed points with $z \neq 1$ in the presence of a nonzero Coulomb interaction \cite{PhysRevB.75.235423}.

The runnings of the Coulomb interaction parameters $\wb$ and $\seb$ are determined by their engineering dimensions, which are both equal to $z-1$ (see Eqs.~\eqref{coulombrunning} and \eqref{conductivityrunning}), in the large $N_f$ expansion. 
Any situation with nonzero Coulomb interaction and $z > 1$ necessarily requires $\wb$ and $\seb$ individually flowing to strong coupling.
Note, however, that it is their dimensionless ratio $\seb/\wb$ that appears in the action $S_E$ in the limit $\wb, \seb \rightarrow \infty$.
Consequently, we can parameterize this infinite Coulomb interaction limit with the marginal parameter $\alpha = \seb/\wb$.
We refer to $\alpha$ as the dissipation strength. 
$z > 1$ is required for any fixed point with infinite Coulomb interaction to be IR attractive; treating $\alpha$ as a tuning parameter, we'll view any infinite Coulomb interaction fixed point with $0 < z < 1$ as an IR unstable fixed point.

For $\wb \rightarrow \infty$, $F_w$ in (\ref{Coulomb-F-fun}) reduces to 
\begin{eqnarray}
F_\infty(\kappa,\alpha) \equiv 
\int^{\infty}_{-\infty} dy
\frac{g_1 \sqrt{1+y^2} (-1+2y^2) \, \alpha + (1+y^2) \, |y| }
{4\pi^2 N_f \,(1+y^2)^{\frac{5}{2}}  
\, \Big( \sqrt{1+y^2} (g_1^2+\kappa^2)\, \alpha + g_1 \, |y|  \Big)  }  
\end{eqnarray}
For any $\kappa$ and $\alpha \geq 0$, $0 \leq F_\infty(\kappa,\alpha) \leq \frac{8}{N_f \,\pi^2} $.
In this limit, the dynamical critical exponent at infinite Coulomb coupling is 
\begin{eqnarray}
z_\infty =1+ \gmb + \gob + 2 \gjb -\phi_1 \, \gob 
- F_\infty(\kappa,\alpha),
\label{z-infinity}
\end{eqnarray} 
where we've explicitly indicated how screening appears in $z_\infty$.
 
\subsubsection*{${\cal C}$ Symmetry}

\begin{figure}[h!]
  \centering
    \includegraphics[width=0.4\linewidth]{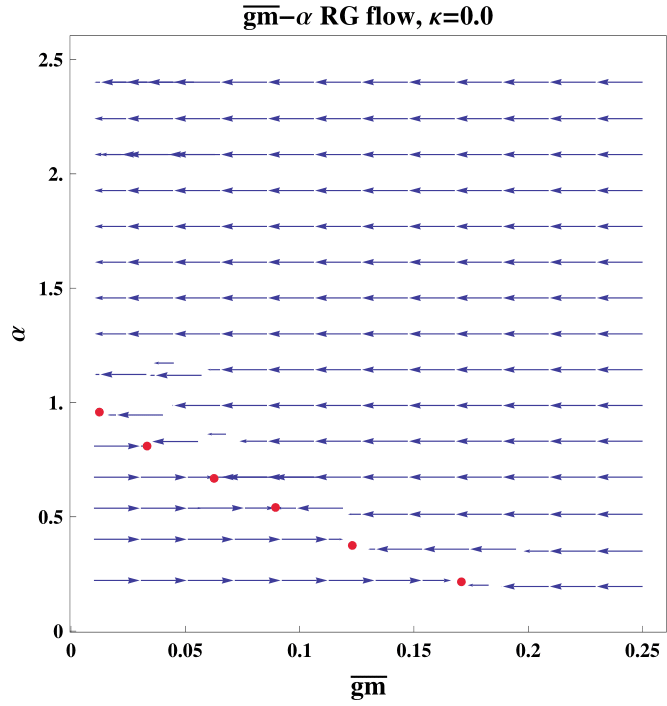} \;\; \; \;\;
\includegraphics[width=0.4\linewidth]{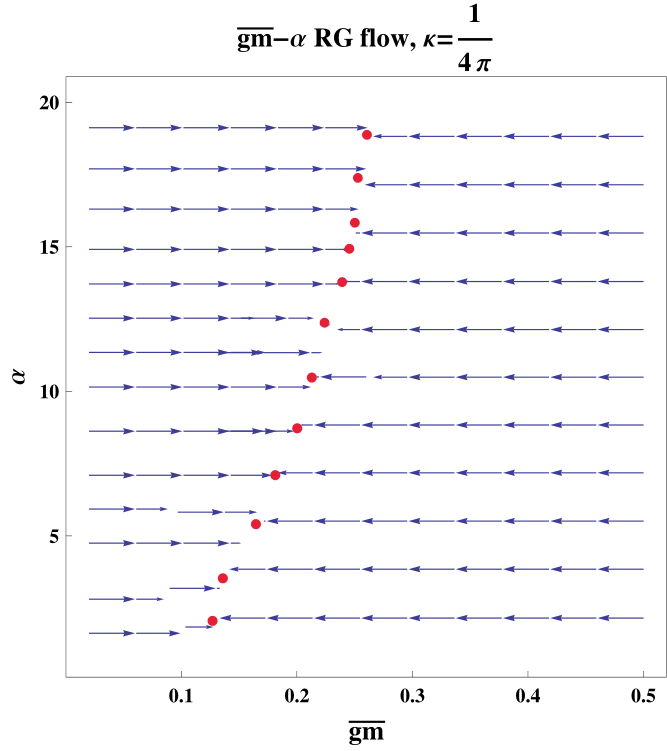}
\caption{RG flow of $\alpha$ vs. $\gmb$. The dissipation parameter $\alpha$
is exactly marginal, so it's a free parameter that can be tuned.
}
\label{Csymmetryinfinity}
\end{figure}
At infinite Coulomb coupling and in the presence of charge-conjugation symmetry, there exist nontrivial fixed points for any $\kappa$.
These occur at small values of $\gmb$ and are found by solving $\beta_{\gmb}=0$ from \eqref{randommassbeta} using Eq.~\eqref{z-infinity}:
\begin{align}
\gmb^*(\alpha)  & = F_m(\alpha, \kappa) \cr
& \equiv 
\int^{\infty}_{-\infty} dy
\frac{ \sqrt{1+y^2} \, |y| + g_1 \, \alpha \, (3+2y^2)}
{4 N_f \pi^2(1+y^2)^2  \big[ g_1 |y| + \alpha \, \sqrt{1+y^2} (g_1^2+\kappa^2) \big]} 
\nn \\
& -
\frac{\alpha  
\big[ 
g_1 \sqrt{1+y^2} \, |y| + (1+y^2)(g_1^2-\kappa^2) \, \alpha
\big]}
{8 N_f \pi^2 (1+y^2)^{\frac{3}{2}}
\big[g_1 |y| + \alpha \, \sqrt{1+y^2} (g_1^2+\kappa^2)   \big]^2
}.
\label{gmstr-alpha}
\end{align}
Fig.~\eqref{Csymmetryinfinity} shows the behavior of the renormalization group flows for $\kappa = 0$ and $\kappa = 1/4\pi$.

\begin{figure}[h!]
  \centering
  \includegraphics[width=0.5\linewidth]{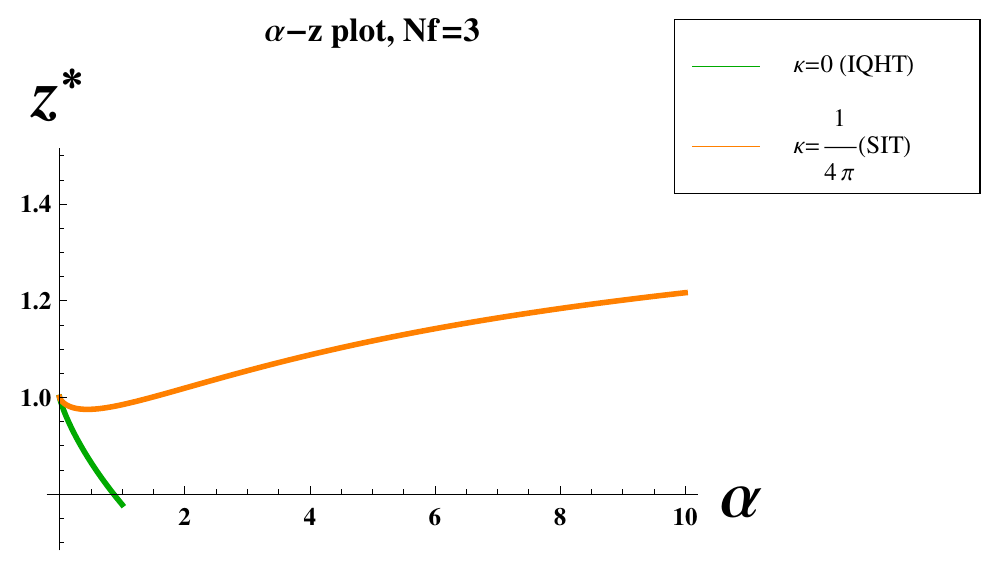} \;\;  
\caption{Dynamical critical exponent $z_\infty$ versus effective dissipation strength $\alpha$ evaluated on the fixed points when $N_f = 3$. The green line corresponds to a $z_\infty < 1$, which is an unstable $\wb \rightarrow \infty$ fixed point.
When $\kappa=1/4\pi$, $1< z_\infty < 2$ for $\alpha \geq 1.47$.
}
\label{zinfinitycsymmetry}
\end{figure}

Note that the beta functions for $\wb \to \infty$ are different from the case of finite $\wb$, even when $\alpha =0$: nontrivial fixed points exist for any value of $\kappa$.
The correlation length exponent $\nu^{-1} = 1$ because the fixed points are solved from $\beta_{\gmb}=0$ (see Eq.~\eqref{anomdimexplicit}). 
The dynamical critical expoennt $z_\infty$ is found using (\ref{z-infinity}): 
\begin{eqnarray}
z_\infty(\alpha) = \gmb^*(\alpha) -F_\infty(\kappa,\alpha).
\end{eqnarray}
To guarantee the irrelevancy of the diffusion constant $D_e$ of the 2DEG bath, $z_\infty <2$ is required. 
Fortunately, $z_\infty$ does not exceed two for the values of $\alpha$ we've considered---see Fig.~\eqref{zinfinitycsymmetry}.

\subsubsection*{${\cal CT}$ Symmetry}

As with finite Coulomb coupling, we focus on $\kappa = 0$ in this and the next subsection because the Chern-Simons term is odd under ${\cal CT}$ and ${\cal T}$.

\begin{figure}[h!]
  \centering
    \includegraphics[width=0.4\linewidth]{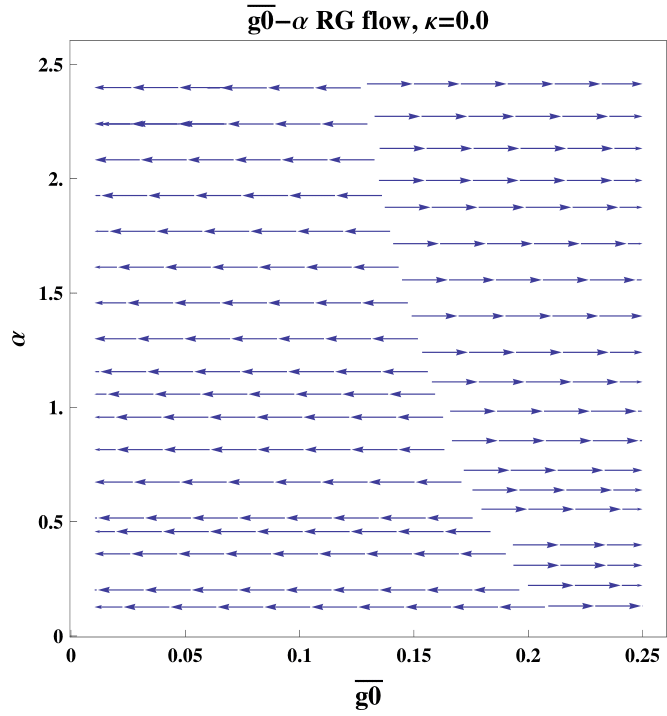} \;
\caption{The RG flow of $\gob$ with respect to $\alpha$. 
}
\label{CTcleanfig}
\end{figure}

In the $\wb \to \infty$ limit, only $\beta_{\gob} = 0$ is nontrivial:
\begin{eqnarray}
2 \gob^2(-1+ \phi_1)
+ 2  \, \gob \,  F_\infty(\kappa=0,\alpha) =0
\end{eqnarray} 
Including disorder screening ($\phi_1=1$), $\gob$ is marginally irrelevant.
If disorder screening is ignored, an unstable fixed point lies at $\gob^* = F_\infty(\kappa=0,\alpha)$.
Perturbation by $\gob$ about this fixed point results either in flow (along the $\gob$ direction) towards strong coupling or towards the infinite Coulomb interaction clean fixed point with critical exponents:
\begin{eqnarray}
&&  z_\infty(\alpha) = 1- F_\infty(\kappa=0,\alpha)<1  ,
\label{clean-z-inf}
\\
&& 
\nu^{-1} = 1 - F_m(\alpha, \kappa=0).
\label{nu-clean}
\end{eqnarray}
Since $z_\infty < 1$, these infinite Coulomb coupling fixed points are IR unstable.
This renormalization group flow is shown in Fig.~\ref{CTcleanfig}.

\subsubsection*{${\cal T}$ Symmetry}

\begin{figure}[h!]
  \centering
  \includegraphics[width=0.35\linewidth]{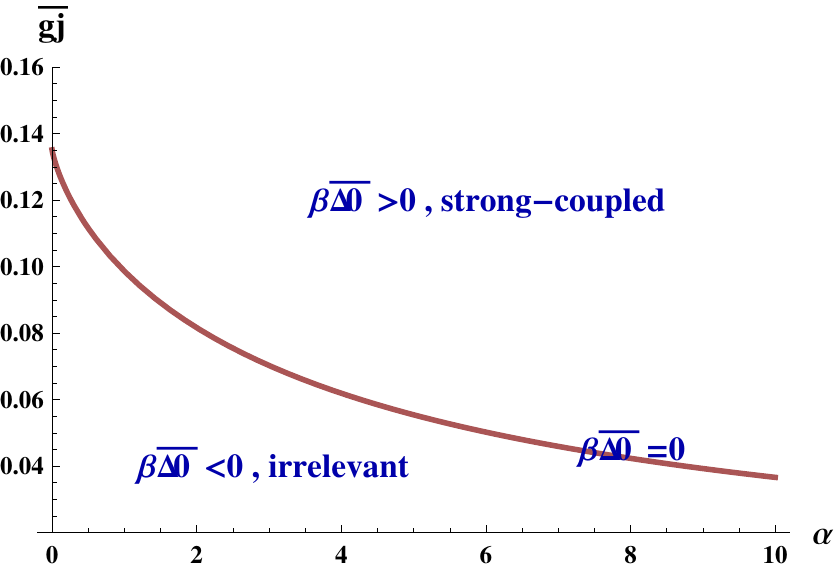} 
\caption{In the case of $\wb \to \infty$, the fixed point solution for $\Dob$ is obtained by tuning $\gjb$ and $\alpha$ so as to sit on the curve above. 
}
\label{Tsymmetryinfinitecoulombfig}
\end{figure}

When time-reversal symmetry is preserved and $\kappa = 0$, the only nontrivial beta function is $\beta_{\Dob}$. 
In the limit of strong Coulomb coupling, the disorder screening terms vanish. 
Solving $\beta_{\Dob}=0$ gives the condition
\begin{eqnarray}
\gjb^* = \frac{1}{2} F_\infty(\alpha,\kappa=0)
\label{gjstr}
\end{eqnarray}
on the marginal couplings $\gjb$ and $\alpha$.
The resulting fixed point is IR unstable along the $\Dob$ direction and, depending on the values of $\gjb$ and $\alpha$, flows either to strong coupling or to zero when perturbed about this fixed point.
This is shown in Fig.~\ref{Tsymmetryinfinitecoulombfig}.
Using Eq.~\eqref{z-infinity}, we see that $z_\infty = 1$ at the fixed point defined by $\beta_{\Dob} = 0$.
The correlation length exponent is given by $\nu^{-1} = 1 + 2 \gjb - F_m(\alpha, \kappa = 0) - {\gjb \over g_1} = 1 - 7 F_\infty(\alpha, \kappa = 0) - F_m(\alpha, \kappa = 0)$, where the second equality is obtained after evaluating on the fixed point defined by Eq.~(\ref{gjstr}).

\section{Discussion}
\label{discussion}

In this paper, we studied the influence of quenched disorder and a dissipative Coulomb interaction on two different quantum phase transitions: an integer quantum Hall transition (IQHT) and a superconductor-insulator transition (SIT).
We considered both transitions using effective theories that consist of a Dirac fermion coupled to a $U(1)$ Chern-Simons gauge field at level $(\theta - 1/2)$: $\theta = 1/2$ corresponds to the IQHT, while $\theta = 1$ corresponds to the SIT.
We performed a renormalization group analysis using a large $N_f$ expansion in which the number fermion flavors $N_f \rightarrow \infty$ to study the critical properties of these theories.
We found both theories to be stable to the addition of a Coulomb interaction.
The IQHT was stable to ${\cal C}$ preserving disorder and exhibited a line of diffusive fixed points with ${\cal CT}$ disorder.
(${\cal C}$ is charge-conjugation symmetry and ${\cal T}$ is time-reversal symmetry.) 
The SIT exhibited a line of fixed points parameterized by the Coulomb coupling when ${\cal C}$ is preserved.
Other cases resulted in runaway flow.

Without disorder, the free Dirac fermion in \eqref{dirac} has a correlation length exponent $\nu^{-1}_{\rm Dirac} = 1$, while the 3d XY model in \eqref{bose} has a correlation length exponent $\nu^{-1}_{\rm Bose} \approx 3/2$.
In the large $N_f$ expansion, we find $\nu^{-1}(\theta = 1/2) \approx 1 - 4.3/N_f$ and $\nu^{-1}(\theta = 1) \approx 1 + 1.4/N_f$, in agreement with \cite{ChenFisherWu1993, PhysRevB.66.144501, 2016JHEP...08..069C, PhysRevB.97.085112}.
Evidently the leading order term in the large $N_f$ expansion provides a poor approximation to the critical exponents of the clean fixed points \cite{ChenFisherWu1993}.
Comparing our results ($\nu^{-1} = 1, z = 1 + 1.4/N_f$) for the correlation length and dynamical critical exponents with recent numerical ($\nu \approx 1.2, z \approx 1.5$) \cite{PhysRevLett.92.015703, PhysRevLett.114.255701, PhysRevLett.108.055701, PhysRevB.94.134501} and analytic ($\nu = 1, z \approx 1 + .5/N_f$) \cite{2019arXiv190909167G} studies (without a Coulomb interaction) suggests this may also be the case for the dirty 3d XY model.
Higher-order ${\cal O}(1/N_f^2)$ terms may improve the comparison.
Interestingly, the free Dirac fermion and 3d XY model admit duals involving a Dirac fermion coupled to a {\it non-Abelian} $U(N)$ Chern-Simons gauge field for any $N > 1$ \cite{Hsin:2016blu, PhysRevB.99.125135}.
Such formulations suggest alternative approximation schemes.
For instance, without disorder, these theories have a correlation length exponent equal to unity at 2-loop order in the planar limit \cite{GMPTWY2012}.
Might the planar limit furnish better approximations to such theories, compared with the large $N_f$ expansion?

There are a variety of other observables and generalizations to consider.
For example, scaling dimensions of the lowest dimension monopole operators in the Chern-Simons theories we studied should correspond to the $\eta$ exponents of the Dirac and XY models.
In $(2+1)$d, the dc $T\rightarrow 0$ conductivity tensor can be universal \cite{Fisher1990, damlesachdev97}; it would be interesting to calculate and compare across the duality \cite{2016JHEP...07..090G}.
Perhaps considering the effects of finite density is most pressing, given that the electronic systems inspiring this work have a finite density of states.

One of the motivations of the current work was to better understand the emergent symmetries that are found at IQHTs and SITs via electrical transport experiments.
For concreteness, consider the magnetic field-tuned SIT at which a ``self-duality"\footnote{``Self-duality" requires that the electrical conductivity tensor of the 3d XY model satisfy: $\sigma_{xx}^2 + \sigma_{xy}^2 = (e_\ast^2/h)^2$, where $e_\ast$ is the electromagnetic charge of the bosons \cite{Fisher1990a}.} with dc $\sigma_{xx} \approx (2e)^2/h$ and $\sigma_{xy} \approx 0$ is found at low temperatures \cite{Breznay2016}.
It was argued in \cite{PhysRevB.95.045118} that PV symmetry (see \S\ref{symmetry}) of the ``fermionic dual" to the XY model in \ref{unite} results in self-dual transport.
How this symmetry might be preserved quantum mechanically is unclear \cite{2018arXiv180906886H}.
This question is related to the emergent time-reversal symmetry of this ``fermionic dual" at zero Dirac composite fermion density.
Perhaps unsurprisingly, the leading order large $N_f$ beta functions that we studied do not appear to respect the emergent time-reversal symmetry; at least, we haven't found nontrivial solutions with an emergent time-reversal invariance at $\kappa = 1/4\pi$.
It would be interesting to further understand this apparent shortcoming.


\section*{Acknowledgments}

We thank Hart Goldman, Sri Raghu, and Alex Thomson for useful conversations and correspondence.
M.M. is supported by the Department of Energy Office of Basic Energy Sciences contract DE-SC0020007.
This research was supported in part by the National Science Foundation under Grant No.~NSF PHY-1748958.

\appendix

\section{Calculation Overview}
\label{barerenormalizedappendix}

In this appendix we derive the residues $b_{\lambda_a}(\vec \lambda^R(\mu, \epsilon))$ in Eq.~\eqref{barerenorm},
\begin{align}
\label{barerenormappendix}
\lambda^B_a \mu^{- \Delta_a(\epsilon)} = \lambda^R_a(\mu, \epsilon) + {b_{\lambda_a}(\vec \lambda^R(\mu, \epsilon)) \over \epsilon},
\end{align}
that determine the beta functions at $\epsilon = 0$ via Eq.~\eqref{betafunctions}:
\begin{align}
\label{betafunctionsappendix}
\beta_{\lambda_a}(\vec \lambda^R) \equiv - \mu {\partial \lambda^R_a \over \partial \mu} = \bar \Delta_{\lambda_a} \lambda^R_a + \rho_{\lambda_a} b_{\lambda_a}(\vec \lambda^R) - \sum_c  \rho_{\lambda_c} \lambda^R_c {\partial b_{\lambda_a}(\vec \lambda^R) \over \partial \lambda^R_c}.
\end{align}
After establishing notation, we'll list the main results used in the main text. 
Later sections provide algebraic details.

\subsection*{Setup}

Identify $S_E$ in Eqs.~\eqref{action1} - \eqref{action3} with the bare action $S_B$ by endowing all fields and couplings with bare ($B$) subscripts/superscripts.
To simplify notation, we'll leave replica and flavor indices implicit.
Define renormalized ($R$) fields and couplings,
\begin{align}
\label{bareandrenorm}
\psi_B = Z_f^{1/2} \psi_R, \quad a_{B}^0 = Z_{a,0}^{1/2} a_R^0, \quad a_B^j = Z^{1/2}_{a,j} a_R^j, \quad \lambda^B_c = Z^{-1/2}_c \lambda^R_c
\end{align}
where the vector of couplings (either $B$ or $R$)
\begin{align}
\vec \lambda = \Big({g^2 \over N_f}, v, m, \kappa, w_x, \sigma_e, D_e, g_m, g_0, g_j, \Delta_0, \Delta_j\Big)^T.
\end{align}
Separate $S_B$ into physical and counterterm actions:
\begin{align}
\label{bareisphysplusct}
S_B = S_{\rm phys}^{(1)} + S_{\rm phys}^{(2)} + S_{\rm phys}^{(3)} + S_{\rm CT}^{(1)} + S_{\rm CT}^{(2)} + S_{\rm CT}^{(3)}
\end{align}
with
\begin{align}
\label{physaction1}
S^{(1)}_{\rm phys} & = \int d\tau d^D x \Big[ \bar{\psi}_R\Big( \gamma_\tau  (\partial^\tau + i {g_R \mu^{\epsilon/2} \over \sqrt{N_f}} a_R^{\tau})
+ v_R \mu^{z - 1} \gamma_j  (\partial^j+ i {g_R \mu^{\epsilon/2} \over \sqrt{N_f}} a_R^j)
 \Big) \psi_R \cr
 & + \mu^z m_R \bar \psi_R \psi_R + {i \kappa_R \over 2} a_R d a_R \Big], \\
 \label{physaction2}
S^{(2)}_{\rm phys} & = \int d\omega d^Dk\ {w_R \mu^{z-1} \over 2} a_R^T(- \omega, - k) {k^2 \over |k| + f_R(\omega, k)} a_R^T(\omega, k), \\
\label{physaction3}
S^{(3)}_{\rm phys}  & = - {1 \over 2} \int d\tau d\tau' d^Dx \Big[(g_m)_R \mu^{2 z - D} (\bar \psi_R \psi_R) (\bar \psi_R \psi_R) + (g_0)_R \mu^{2z - D} (\bar \psi_R \gamma^0 \psi_R) (\bar \psi_R \gamma^0 \psi_R) \cr 
& + (g_j)_R \mu^{2 z - D} (\bar \psi_R \gamma^j \psi_R)  (\bar \psi_R \gamma^j \psi_R) + (\Delta_0)_R \mu^{2 z - 2} b_R b_R + (\Delta_j)_R {\bf e}_R \cdot {\bf e}_R \Big],
\end{align}
\begin{align}
\label{CTaction1}
S^{(1)}_{\rm CT} & = \int d\tau d^D x \Big[ \bar{\psi}_R\Big( \gamma_\tau  (\partial^\tau + i {g_R \mu^{\epsilon/2} \over \sqrt{N_f}} a_R^{\tau}) \delta_1
+ v_R \mu^{z - 1} \gamma_j  (\partial^j+ i {g_R \mu^{\epsilon/2} \over \sqrt{N_f}} a_R^j) \delta_2
 \Big) \psi_R \cr
 & + \mu^z m_R \bar \psi_R \psi_R \delta_m + {i \kappa_R \over 2} a_R d a_R \delta_\kappa \Big], \\
 \label{CTaction2}
S^{(2)}_{\rm CT} & = \int d\omega d^Dk\ {w_R \mu^{z-1} \delta_w \over 2} a_R^T(- \omega, - k) {k^2 \over |k| + f_R(\omega, k)} a_R^T(\omega, k), \\
\label{CTaction3}
S^{(3)}_{\rm CT}  & = - {1 \over 2} \int d\tau d\tau' d^Dx \Big[(g_m)_R \mu^{2 z - D} \delta_{g_m} (\bar \psi_R \psi_R) (\bar \psi_R \psi_R) + (g_0)_R \mu^{2z - D} \delta_{g_0} (\bar \psi_R \gamma^0 \psi_R) (\bar \psi_R \gamma^0 \psi_R) \cr 
& + (g_j)_R \mu^{2 z - D} \delta_{g_j} (\bar \psi_R \gamma^j \psi_R)  (\bar \psi_R \gamma^j \psi_R) + (\Delta_0)_R \mu^{2 z - 2} \delta_{\Delta_0} b_R b_R + (\Delta_j)_R \delta_{\Delta_j} {\bf e}_R \cdot {\bf e}_R \Big],
\end{align}
\begin{align}
f_R = {e_\ast^2 (\sigma_e)_R \mu^{z-1} \over |\omega| + (D_e)_R \mu^{z - 2} k^2},
\end{align}
and the renormalization group scale $\mu$ enters in accord with the engineering dimensions listed in Table \ref{dims}.
The counterterms $\delta_X$ have poles in $\epsilon$ with coefficients determined by the requirement that correlation functions of physical fields have no divergences as $\epsilon \rightarrow 0$.
We focus exclusively on the terms in $\delta_X$ proportional to $1/\epsilon$.
Using Eq.~\eqref{bareandrenorm} to impose Eq.~\eqref{bareisphysplusct}, we relate the bare and renormalized couplings:
\begin{align}
\label{barecouplingtorenormalizedcouplingintermediate}
v^B \mu^{1 - z} & = v^R (1 + \delta_2 - \delta_1), \\
m^B \mu^{-z} & = m^R (1 + \delta_m - \delta_1), \\
\kappa^B & = \kappa^R(1 + \delta_\kappa), \\
w_x^B \mu^{1 - z} & = w_x^R(1 + \delta_w), \\
g_X^B \mu^{D - 2 z} & = g_X^R (1 - 2 \delta_1 + \delta_{g_X}), \quad g_X \in \{g_m, g_0, g_j \}, \\
\Delta_0^B \mu^{2 - 2 z} & = \Delta_0^R (1 + \delta_{\Delta_0}), \\
\Delta_j^B & = \Delta_j^R (1 + \delta_{\Delta_j}), \\
\sigma_e^B \mu^{1 - z} & = \sigma_e^R (1 + \delta_{\sigma_e}), \\
D_e^B & = D_e^R (1 + \delta_D). \\
\end{align}
Thus, we can read off the residues:
\begin{align}
\label{residuesCTs}
b_v & = v^R(\delta_2 - \delta_1) \epsilon \\
b_m & = m^R (\delta_m - \delta_1) \epsilon \\
b_\kappa & = \kappa^R \delta_\kappa \epsilon \\
b_w & = w^R_x \delta_w \epsilon \\
b_{g_X} & = - g_X^R (2 \delta_1 - \delta_{g_X}) \epsilon, \quad g_X \in \{g_m, g_0, g_j \}, \\
b_{\Delta_X} & = \Delta_X^R \delta_{\Delta_X} \epsilon, \quad \Delta_X \in \{\Delta_0, \Delta_j \}, \\
b_{\sigma_e} & = \sigma_e^R \delta_{\sigma} \epsilon, \\
b_{D_e} & = D_e^R \delta_{D} \epsilon.
\end{align}

\section{Counterterms}

As discussed in the main text, we choose the dynamical critical exponent $z$ in such a way that the fermion velocity $v$ does not run, i.e., the velocity beta function is zero.
In the expressions below, it's convenient to redefine couplings to absorb the velocity dependence as follows:
\begin{eqnarray}
&&   
\gmb= \frac{g_m}{2 \pi v^2} \; , \;
\gob = \frac{g_0}{2 \pi v^2} \; , \;
\gjb = \frac{g_j}{2 \pi v^2} \; , \; 
\Dob =   \Delta_0  \; , \;  
\Djb = \Delta_j \, v^2  \; , \;      \nn \\
&& 
\wb = \frac{w_x}{v}  \; , \;
\seb = \frac{\sigma_e}{v}     \; , \; 
\int^{\infty}_{-\infty} dz F(z) =  \int^{\infty}_{-\infty} v \, dy\, F( y=\frac{z}{v}),
\end{eqnarray}
where the function $F(z)$ is introduced below.
We also define $g_1 = g/16$ and $\overline{m} = m/v$.

Let us make a few remarks about the expressions below.
\begin{itemize}
\item We use $\phi_1$ to parameterize the screening of the disorder described in Appendix \ref{disorderscreening}. 
$\phi_1=0$ means the screening is ignored; $\phi_1=1$ means the screening is included.
\item Terms proportional to $\xi$ are divergent. 
However, this is an unphysical divergence due to our gauge choice: This divergence does not appear in physical quantities such as critical exponents.
\item The Ward identity guarantees the gauge field corrections to $\delta_1$ and $\delta g_0$ cancel; the ones in $\delta_2$ and $\delta_{g_j}$ likewise cancel. 
In the absence of the Coulomb interaction, the equality of the gauge corrections in $\delta_1$ and $\delta_2$ is a coincidence, which makes $\beta_{g_0},\beta_{g_j}$ independent of the gauge corrections. 
When the Coulomb interaction is included, $\beta_{g_0}$ receives corrections from the gauge field, while $\beta_{g_j}$ does not.
\end{itemize}

\subsubsection*{$\delta_\kappa$, $\delta_w$, $\delta_{\sigma}$, and $\delta_{D}$ Counterterms}

Quantization of the Chern-Simons level and finiteness of the gauge field self-energy in $3$d implies
\begin{align}
\delta_\kappa = \delta_w = \delta_{\sigma} = \delta_{D} = 0.
\end{align}
Consequently, renormalizations of $\kappa, w_x, \sigma_e$, and $ D_e$ are controlled by their engineering dimensions. 

\subsubsection*{$\delta_1$ Counterterm}

The diagrams that contribute to $\delta_1$ are given by taking the temporal component of Fig.~\ref{F-Self-Energy}:
\begin{align}
\delta_1 \, \epsilon
& = -( \gmb+ \gob + 2 \gjb)
+  \,\phi_1  \; \bigg[
\frac{ g_1^2 \wb \;( g_1 \gjb -g_1 \gob -\gob \;  \wb ) }
   { (g_1^2 + g_1 \wb+ \kappa^2 )^2}
 + \frac{ g_1 \; ( g_1 \gjb + g_1 \gob + 2 \gob \;  \wb ) \;  }
   {g_1^2 + g_1 \wb+ \kappa^2 }  
\bigg]\;  \qquad  \nn \\
& \;\;
+ \int^{\infty}_{-\infty}\,
\frac{1}{ 4 \pi^2  N_f}
\frac{(1-y^2) (g_1 \,\seb \, y^2 + \wb \sqrt{1+y^2} \, |y|\,)
        + g_1\, |y|\, y^2\,(1-y^2 \, \xi  ) }    
{ (1+y^2)^2 \Big[ \sqrt{ 1+y^2 } (g_1^2 +\kappa^2)(\seb+|y|) + g_1 \, \wb \, |y|   \Big] }.
\label{dlt-1} 
\end{align}

\subsubsection*{$\delta_2$ Counterterm}

\begin{figure}[h!]
  \centering
\includegraphics[width=0.8\linewidth]{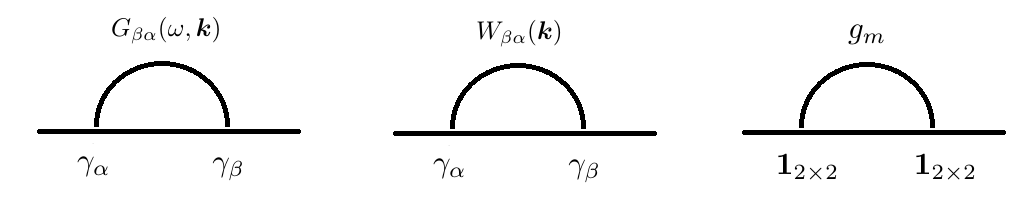} \\
\caption{Diagrams contributing to $\delta_1,\delta_2 $.}
\label{F-Self-Energy}
\end{figure}

The diagrams that contribute to $\delta_2$ are given by taking the spatial component of   Fig.~\ref{F-Self-Energy}: 
\begin{align}
\delta_2 \, \epsilon
& = - \phi_1  \;
\frac{ g_1^2 
\big[ g_1^2 \gjb+ 2 g_1 \gjb\, \wb- (\gob- 2 \gjb) \, \kappa^2   \big]    }
{ (g_1^2 + g_1 \wb+ \kappa^2 )^2 } \,    \nn \\
& \qquad
+
\int^{\infty}_{-\infty} dy \,
\frac{1}{ 4 \pi^2  N_f}
\frac{ g_1 (1-y^2-y^4) \, \seb  - (\wb \, y^2 \sqrt{1+y^2}    )  \,|y|  
+   g_1 (1-y^2-y^4 \xi ) \, |y|    }     
{ (1+y^2)^2 \Big[ \sqrt{ 1+y^2 } (g_1^2 +\kappa^2)(\seb+|y|) + g_1 \, \wb \, |y|
 \Big] }.
\label{dlt-2}
\end{align}

\begin{figure}[h!]
  \centering
\includegraphics[width=0.8\linewidth]{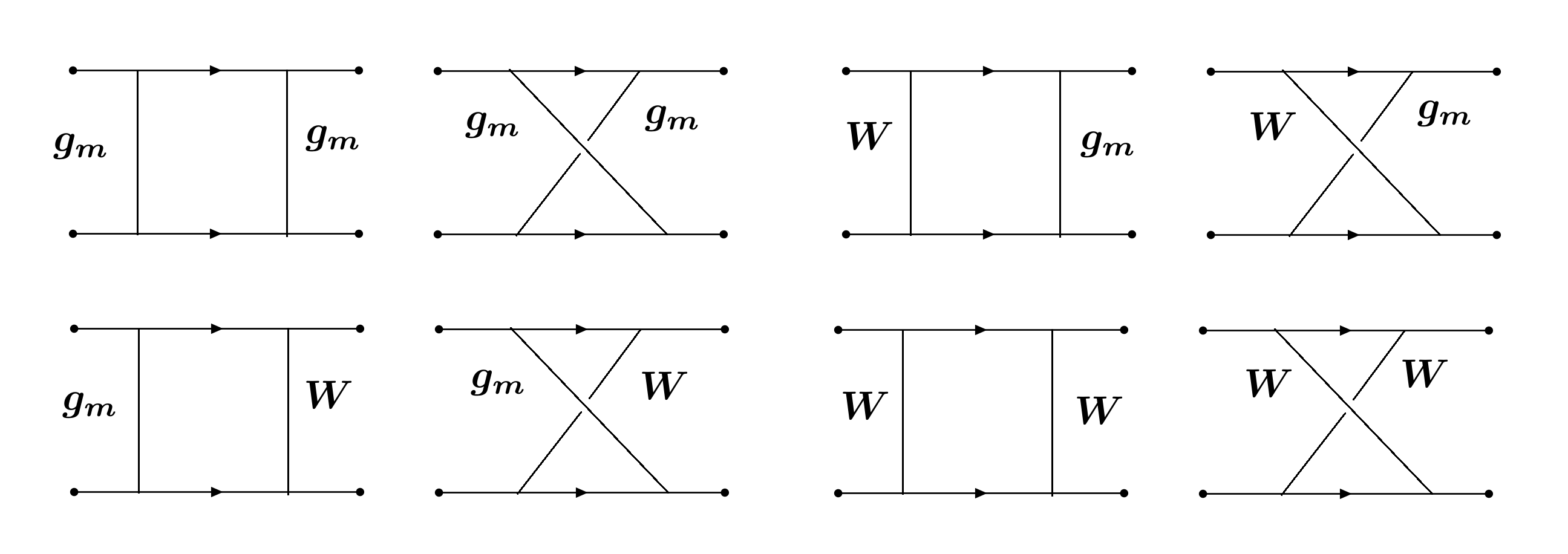} \\
\caption{2-PI diagrams contributing to $\delta_{g_m},\delta_{g_0},\delta_{g_j}$.}
\label{2-PI-Box}
\end{figure}

\subsubsection*{$\delta_{g_m}$ Counterterm}

\begin{figure}[h!]
  \centering
\includegraphics[width=1.0\linewidth]{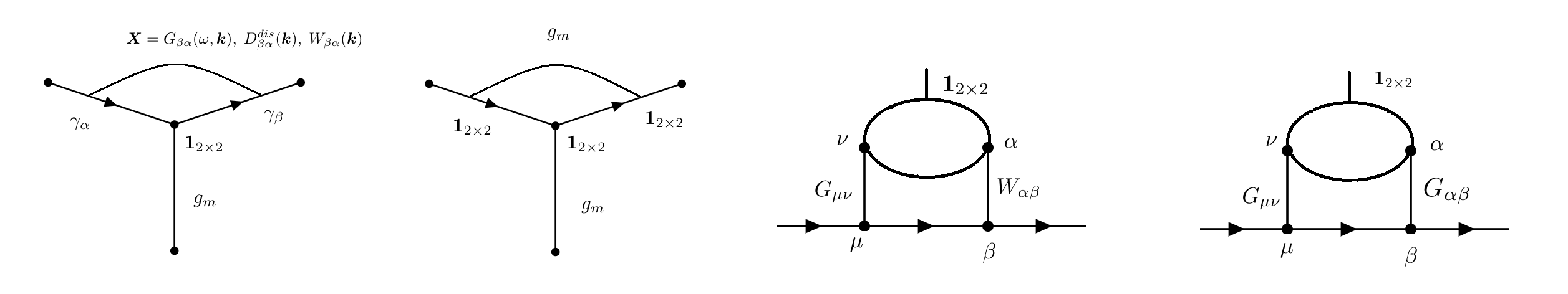}
\caption{The mass vertex contributions to $\delta_{g_m}$ }
\label{gm-vtx}
\end{figure}

$\delta_{g_m}$ is extracted from the diagrams in Figs.~\ref{2-PI-Box} and \ref{gm-vtx}:
\begin{eqnarray}
&&
\delta_{g_m} \epsilon
= \bigg[\frac{2(\gob + \gmb ) (\gmb-2 \gjb ) }{\gmb}
+ 2 \; \frac{g_1 (\gjb-\gob) + \gjb \, \wb }  {g_1^2 + g_1 \, \wb + \kappa^2 }
\bigg]  \,          \nn \\
&&
+ \phi_1 \, \bigg[
4 g_1 \; \,
\frac{\; g_1 \kappa^2 ( -\gob^2-\gjb^2) 
+ \gjb \, \gob \, 
\big[ 2 g_1^3 + 4 g_1^2 \wb \, + 2\wb \kappa^2 + g_1 (\wb^2 + 4 \kappa^2) 
\big]
}
{ \gmb \; [g_1^2 + g_1 \, \wb + \kappa^2]^2}
\bigg]  \,     \nn \\
&&  + \phi_1 \, \bigg[
\frac{ g_1^5 (\gob-\gjb) + g_1^4\, \wb\, (2 \gob  -3 \gjb )
+ g_1^3 \, ( \gob \, \wb^2 -2 \gjb \, \wb^2 + 3 \gob\, \kappa^2 -3 \gjb \, \kappa^2 \,)
}
{ [g_1^2 + g_1 \, \wb + \kappa^2]^3}     \nn \\
&&  \qquad  \qquad  \qquad \qquad \qquad   +  \frac{ 
 g_1^2 \kappa^2 (3 \gob \, \wb )
+ 2 g_1 \kappa^4 (\gjb-\gob)   }
{[g_1^2 + g_1 \, \wb + \kappa^2]^3}
   \bigg]  \,    \nn \\
&&  + \phi_1 \, \bigg[
\frac{ 2 g_1^4( \gjb-\gob) + 4 g_1^3 \, \wb (\gjb-\gob)
+ g_1^2 (-2 \gob \, \wb^2- 6 \gob \kappa^2 + 6 \gjb \, \kappa^2  )
-4 g_1 \, \gob \, \wb \, \kappa^2
}
{ [g_1^2 + g_1 \, \wb + \kappa^2]^2}
\bigg]  \,   \nn \\
&& +
 \phi_1^2  \Bigg[
\frac{4 g_1^2 \gjb^2\kappa^2[g_1^2 (g_1^2 + 2 g_1 \wb-\wb^2 ) 
         +2 g_1(g_1-\wb)\kappa^2 -\kappa^4  \; ]    \;       }
{\gmb [g_1^2 + g_1 \, \wb + \kappa^2]^4 }    \nn \\
&&  \qquad \qquad  +
\frac{ 4 g_1^2 \gob^2 \kappa^2 
[g_1^2(g_1^2 + 2 g_1 \wb + \wb^2)+ 2 g_1(g_1+\wb)\kappa^2 -\kappa^4 ]}
{ \gmb [g_1^2 + g_1 \, \wb + \kappa^2]^4}     \nn \\
&& 
+  
\frac{ 4 \gob \; \gjb g_1^2 \big[ 
-g_1^3 (g_1+ \wb)^2(g_1+ 2 \wb)- 2 g_1^2(g_1 + \wb) (2 g_1+ 3 \wb	) \kappa^2
-g_1(5 g_1 + 2 \wb ) \kappa^4 + 2\kappa^6
\big]}
{\gmb [g_1^2 + g_1 \, \wb + \kappa^2]^4}
\, 
\Bigg]     \nn \\
&&
- \int^{\infty}_{-\infty} dy \,
\frac{ g_1 (y^2+2)(1+y^2) \, \seb
    + \Big( \wb \,(1+y^2) \, \sqrt{1+y^2}  +g_1 (2+3 y^2 +y^4 \xi )  \Big)  |y| \;  }
{2 \pi^2 \, N_f \, (1+y^2)^2 
\Big[ \sqrt{ 1+y^2 } (g_1^2 +\kappa^2)(\seb+|y|) + g_1 \, \wb \, |y|   \Big]
 }   \;  \nn \\
 && 
 +  \int^{\infty}_{-\infty} dy \,
\frac{ (\seb + |y|) \; 
\bigg(  (1+y^2)(g_1^2 -\kappa^2) \, ( \seb  + |y|  \, ) 
+ ( g_1 \, \wb \, \sqrt{1+y^2}    ) \, |y|
\bigg)
}
{4 \pi^2 \, N_f \, (1+y^2)^{\frac{3}{2}}
\Big[ \sqrt{ 1+y^2 } (g_1^2 +\kappa^2)(\seb+|y|) + g_1 \, \wb \, |y|   \Big]^2
 }   \;\;    .
\end{eqnarray}

\subsubsection*{$\delta_m$ Counterterm}

$\delta_{m}$ is extracted from the diagrams in \ref{gm-vtx}:
\begin{align}
m\, \epsilon \,\delta_m   &= {1 \over 2}  g_m \, \epsilon\,
\Big(  \delta_{g_m} - \text{[2-PI boxes]} \Big) \nn \\
&=  \gob -2 \gjb+ \gmb
- \frac{ g_1(\gob-\gjb ) -\gjb\, \wb  }{2 (g_1^2 + g_1 \, \wb + \kappa^2 )}
+  \phi_1 \;
\Big( -
\frac{ g_1(\gob-\gjb)+ 3 g_1(\gob-\gjb)+ 2 \gob \, \wb }{g_1^2 + g_1 \, \wb + \kappa^2 }
  \nn  \\
& - g_1^2 \, \frac{ 
 \gjb [ 4g_1^2+ 4 \wb + g_1( 7+ 2\wb )\;    ]
 -\gob [ 4 g_1^2 + \wb (7+ 2 \wb) + g_1 (7+ 6 \wb)  ]    }
{2 (g_1^2 + g_1 \, \wb + \kappa^2)^2}    \nn \\
& - \frac{  -2 g_1^3 (g_1 + \wb) [-g_1 \gjb + \gob (g_1 + \wb)  \; ] }
  {(g_1^2 + g_1 \, \wb + \kappa^2)^3}   
  \qquad  \Big)  \nn \\
& 
- \int^{\infty}_{-\infty} dy \,
\frac{ g_1 (y^2+2)(1+y^2) \, \seb
    + \Big( \wb \,(1+y^2) \, \sqrt{1+y^2}  +g_1 (2+3 y^2 +y^4 \xi )  \Big)  |y| \;  }
{ 4 \pi^2 \, N_f \, (1+y^2)^2 
\Big[ \sqrt{ 1+y^2 } (g_1^2 +\kappa^2)(\seb+|y|) + g_1 \, \wb \, |y|   \Big]
 }   \;  \nn \\
&
 +  \int^{\infty}_{-\infty} dy \,
\frac{ (\seb + |y|) \; 
\bigg(  (1+y^2)(g_1^2 -\kappa^2) \, ( \seb  + |y|  \, ) 
+ ( g_1 \, \wb \, \sqrt{1+y^2}    ) \, |y|
\bigg)
}
{ 8 \pi^2 \, N_f \, (1+y^2)^{\frac{3}{2}}
\Big[ \sqrt{ 1+y^2 } (g_1^2 +\kappa^2)(\seb+|y|) + g_1 \, \wb \, |y|   \Big]^2
 }  ,
\qquad
\end{align}
where $\text{[2-PI boxes]}$ refers to the contributions to $\delta_{g_m}$ from the diagrams in Fig.~\ref{2-PI-Box}.

\subsubsection*{$\delta_{g_0}$ Counterterm}

\begin{figure}[h!]
  \centering
\includegraphics[width=0.6\linewidth]{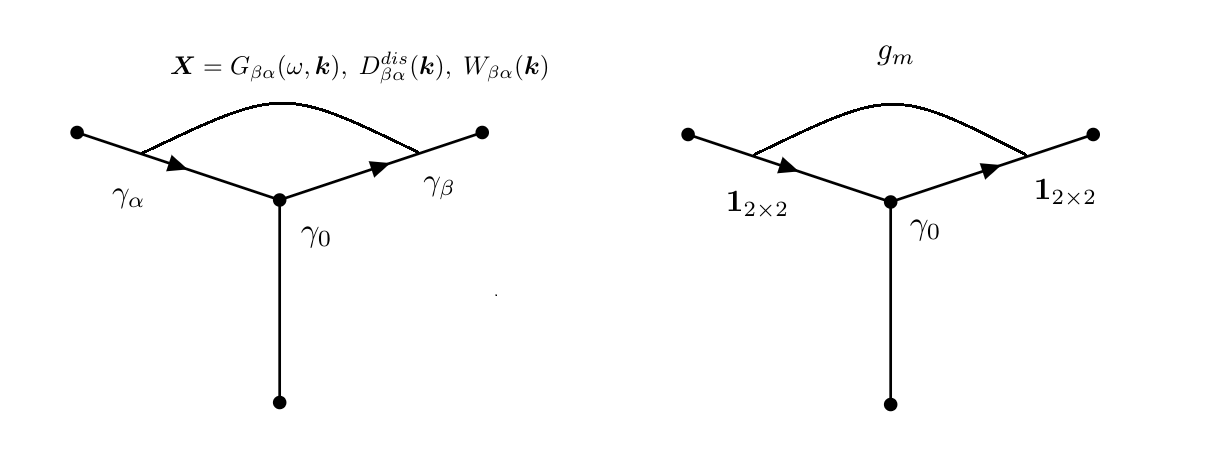} \\
\caption{ $\gamma_{\mu=0}$ vertex component contributions to $\delta_{g_0} $}
\label{mu0-vtx}
\includegraphics[width=0.6\linewidth]{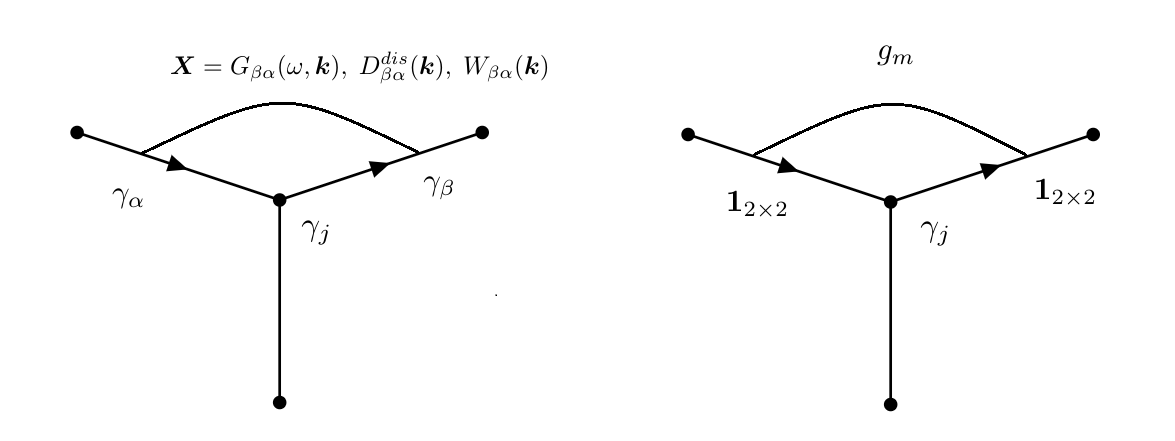} \\
\caption{ $\gamma_{\mu=j} $ vertex component contributions to $\delta_{g_j} $}
\label{muj-vtx}
\end{figure}

$\delta_{g_0}$ is extracted from the diagrams in Figs.~\ref{2-PI-Box} and \ref{mu0-vtx}:
\begin{eqnarray}
&& \delta_{g_0} \epsilon  =
\frac{-2( \gob+ 2 \gjb)\, ( \gob + \gmb)}{\gob} \,      \nn   \\
&&
+ \phi_1 \bigg[
\frac{ 4 g_1^2 \, \gjb \; \gmb \; (g_1^2+2 g_1 \, \wb + 2\kappa^2 )}
    {\gob \; [g_1^2 + g_1 \, \wb + \kappa^2]^2}
 + \frac{ 2 g_1^2 \big( g_1^2 \gjb + 2g_1 \gjb \, \wb+ (\gjb-2\gmb)\,\kappa^2  \big) }
 {[g_1^2 + g_1 \, \wb + \kappa^2]^2}       \nn \\
 &&   \qquad \qquad\qquad\qquad\qquad\qquad\qquad\qquad
+ 
\frac{ 2 \gob g_1 \; \big(
g_1^3 +2 g_1^2 \wb  + 2 \kappa^2 \, \wb + g_1 \wb^2 + g_1\kappa^2  
\big)
}
{[g_1^2 + g_1 \, \wb + \kappa^2]^2}  
\bigg]  \;        \nn \\
&& 
+   \int^{\infty}_{-\infty} dy \;\;  
\frac{
g_1 \,y^2 (1-y^2)\, \seb + g_1 \,y^2 (1-y^2\, \xi ) \, |y|
 + (1-y^2) \, \wb \sqrt{1+y^2} \, |y|    \;     }
{2 \pi^2 \, N_f \, (1+y^2)^2 
\Big[ \sqrt{ 1+y^2 } (g_1^2 +\kappa^2)(\seb+|y|) + g_1 \, \wb \, |y|   \Big]  } .
\end{eqnarray}

\subsubsection*{$\delta_{g_j}$ Counterterm}

$\delta_{g_j}$ is extracted from the diagrams in Figs.~\ref{2-PI-Box} and \ref{muj-vtx}:
\begin{eqnarray}
&& 
\delta_{g_j} \epsilon  =
\frac{-2 \gob \; \gmb}{\gjb} \, 
+ \phi_1  \bigg[
\frac{2 g_1^2 (\gob -\gmb ) \, \kappa^2 }{[g_1^2 + g_1 \, \wb + \kappa^2]^2 }
\bigg] \,  \qquad \qquad\qquad\qquad\qquad\qquad\qquad\qquad     \nn \\
&& + \phi_1  \bigg[
  \frac{2 g_1 \; \gob \; \gmb 
  \Big(  g_1^3 +2 g_1^2\, \wb + 2 \wb\, \kappa^2 +g_1 \wb^2 + 2 g_1 \kappa^2   \Big)  }
{[g_1^2 + g_1 \, \wb + \kappa^2]^2 }
\bigg]  \;   \nn   \\
&& + 
\int^{\infty}_{-\infty} dy \,
\frac{1}{ 2 \pi^2  N_f}
\frac{ g_1 (1-y^2-y^4) \, \seb  - (\wb \, y^2 \sqrt{1+y^2}    )  \,|y|  
+   g_1 (1-y^2-y^4  \xi  ) \, |y|    }
{ (1+y^2)^2 \Big[ \sqrt{ 1+y^2 } (g_1^2 +\kappa^2)(\seb+|y|) + g_1 \, \wb \, |y|
 \Big] } .
\end{eqnarray}

\begin{figure}[h!]
  \centering
\includegraphics[width=0.45\linewidth]{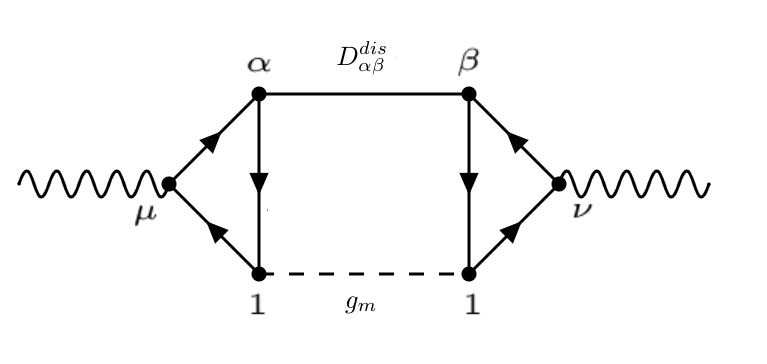} \\
\caption{Diagrams contributing to $\delta_{\Delta_0},\delta_{\Delta_j}$}
\label{Dlt0j}
\end{figure}

\subsubsection*{$\delta_{\Delta_0}$ Counterterm}

$\delta_{\Delta_0}$ is extracted from the diagram in Fig.~\ref{Dlt0j}:
\begin{eqnarray}
&& 
\delta_{\Delta_0} \epsilon =
\frac{- \gmb \, (g_1^2 \,\Djb + 2 g_1 \, \wb\, \Djb + \wb^2\, \Djb+ \Dob \,\kappa^2) }
{64 \Dob \; (g_1^2 + g_1 \, \wb + \kappa^2 )^2 }  \;      \nn \\
&&  
+   \Big[
\frac{\; - \gob\, \gmb N_f \, \pi \, v^2  }{32 \Dob }
+ \phi_1 \; 
\frac{ g_1 \gmb \, N_f \,  \pi \,v^2
[-g_1 \gjb \kappa^2 + \gob(g_1 + \wb ) (g_1^2 + g_1 \, \wb + 2\kappa^2 )  \;] 
  }
{32 \Dob \,(g_1^2 + g_1 \, \wb + \kappa^2 )^2 } 
\Big].
\end{eqnarray}

\subsubsection*{$\delta_{\Delta_j}$ Counterterm}

$\delta_{\Delta_j}$ is extracted from the diagram in Fig.~\ref{Dlt0j}: 
\begin{eqnarray}
\delta_{\Delta_j} \epsilon && =
\frac{ -\gmb (g_1^2 \Dob + \Djb \, \kappa^2 ) }
{128 \Djb \; (g_1^2 + g_1 \, \wb + \kappa^2 )^2} \;   \nn \\
&& +  \Big[
\frac{\; - \gjb\, \gmb \, N_f\, \pi v^2 }{ 64 \, \Djb}
+ \phi_1 \;
\frac{ g_1^2 \,\gmb \, N_f \, \pi  v^2 
[ g_1^2 \, \gjb + 2g_1 \, \gjb \, \wb - (\gob-2\gjb) \kappa^2  \; ]    }
{ 64 \Djb \,(g_1^2 + g_1 \, \wb + \kappa^2 )^2} 
\Big].
\end{eqnarray}

\section{Feynman Rules for Disorder and Screening}
\label{disorderscreening}

\subsection*{Feynman Rules for Disorder Vertices}
From the action (\ref{physaction3}), we can read the Feynman rules for the various types of disorder.
\begin{itemize}
\item[] 4-fermion mass vertex:
\begin{eqnarray}
\frac{1}{2} (\bar{\psi}\psi)(\bar{\psi}\psi) \;\;
 \Rightarrow\;\;
  \boxed{
 + g_m\, 2\pi \delta(\omega=0)
 }.
 \label{f-rule-gm}
\end{eqnarray}
\item[] 4-fermion density vertex:
\begin{eqnarray}
\frac{1}{2} (\bar{\psi}  \gamma^0 \psi)(\bar{\psi}  \gamma^0 \psi) \;\;
 \Rightarrow\;\;   
 \boxed{
 (g_0) (\gamma^0) (\gamma^0) 2 \pi \delta(\omega=0) 
 }.
  \label{f-rule-g0}
\end{eqnarray}
\item[] 4-fermion current vertex along the $k$-direction, $k=x$ or $y$:
\begin{eqnarray}
\frac{1}{2} (\bar{\psi} i \gamma^k \psi)(\bar{\psi} i \gamma^k \psi) \;\;
 \Rightarrow\;\;
 \boxed{
 (+g_j) (i\gamma^k) \, (i\gamma^k) 2 \pi \delta(\omega=0)
 }.
 \label{f-rule-gj}
\end{eqnarray}
\item[]
$b(\tau) b(\tau')$ disordered 2-pt vertex rule: 
\begin{eqnarray}
\frac{1}{2} b_z\, b_z  \;\;   \Rightarrow\;\;
\boxed{
\Delta_0 \, (\delta_{ij} {\bf k}^2 -k_i k_j) 2\pi \delta(\omega)    
 a_i({\bf k}) a_j(-{\bf k}) 
 }.
\label{bsq}
\end{eqnarray}
\item []
$e_j(\tau) e_j(\tau')$ disordered 2-pt vertex: 
\begin{eqnarray}
\frac{1}{2}  (e_x\, e_x +e_y \, e_y)
  \;\; \Rightarrow \;\;
  \boxed{
(-\Delta_j	\, c^2) \; (k_x^2+k_y^2) \,2\pi \delta(\omega)   
\;   a_0( { \bf k }) a_0(- { \bf k } ) 
}.
\label{exsq} 
\end{eqnarray}
\end{itemize}
The factor of $\frac{1}{2}$ factor is canceled by symmetry factor equal to two.
The $2\pi$ factor always cancels with the $1/2\pi$ that accompanies any frequency integral $\int \frac{d\omega}{2\pi}$.

\subsection*{Gauge Propagator}

\subsubsection*{Vacuum Polarization Tensor}
\begin{eqnarray}
&&
\Pi_{\mu \nu}(k_0 , {\bf k} )
=(\frac{-ig}{\sqrt{N_f}})^2 (\frac{v}{c})^{2-\delta_{\mu 0}-\delta_{\nu 0}}
\times N_f \times (-1)
\int \frac{d^2 p_E}{(2\pi)^3}
Tr[\gamma_\mu S_F(k+p) \gamma_\nu S_F(p) ]   \qquad  \qquad \\
&&
=(+g^2)(-i)^2 \;
(\frac{v}{c})^{2-\delta_{\mu 0}-\delta_{\nu 0}}  \;
\frac{1}{v^2}
\int \frac{d^2 \bar{p}_E d p_0}{(2\pi)^3} \;
Tr[\gamma_\mu \gamma_\alpha \gamma_\nu \gamma_\beta]
\frac{(\bar{k}+\bar{p})_\alpha  (\bar{p})_\beta }
       { [x(\bar{k}+\bar{p})^2+(1-x) \bar{p}^2]^2},
\end{eqnarray}

\begin{eqnarray}
\Pi_{\mu \nu}(k_0 , {\bf k} )
= \frac{ -1  }{16} \frac{g^2}{v^2}
(\frac{v}{c})^{2-\delta_{\mu 0}-\delta_{\nu 0}}
\frac{1}{|\bar{k}|}
[ \delta_{\mu \nu} \bar{k}^2 -\bar{k}_\mu \bar{k}_\nu ]
\;\;\;,\;\;\;
\bar{k} = (\omega, v \, {\bf k}) \;\;,\;\;
|\bar{k}| = \sqrt{ \omega^2+ v^2 {\bf k}^2 }.\qquad 
\end{eqnarray}
The minus sign comes from the fermion loop. The ratio $v/c$ can be set to $v$ in future equations.

\subsubsection*{``1-$\mu$ "Vacuum Polarization Vector}
\begin{eqnarray}
&&
\Pi_{\mu}(k_0 , {\bf k} )
= \sqrt{g_m} (\frac{-ig}{\sqrt{N_f}})  (\frac{v}{c})^{1-\delta_{\mu 0} }
\times N_f \times (-1)
\int \frac{d^2 p_E}{(2\pi)^3}
Tr[\gamma_\mu S_F(k+p) \; { \bf 1} \; S_F(p) ]  \\
&&
= \sqrt{g_m} \;(+ i g)(-i)^2 \;
(\frac{v}{c})^{1-\delta_{\mu 0} }  \;
\frac{1}{v^2}
\int \frac{d^2 \bar{p}_E d p_0}{(2\pi)^3} \;
Tr[\gamma_\mu \gamma_\alpha \;\; \gamma_\beta]
\frac{(\bar{k}+\bar{p})_\alpha  (\bar{p})_\beta }
       { [x(\bar{k}+\bar{p})^2+(1-x) \bar{p}^2]^2}   = 0.
       \qquad  \qquad  
\end{eqnarray}
The momentum part is proportional to $\delta_{\alpha \beta}\; p^2$, while the trace is proportional to the $\epsilon_{\mu \alpha \beta}$ tensor, and so it vanishes.

\subsubsection*{``1-1"Vacuum polarization scalar}
\begin{eqnarray}
&&
\Pi_{m}(k_0 , {\bf k} )
=  ( \sqrt{g_m})^2 N_f \times (-1)
\int \frac{d^2 p_E}{(2\pi)^3}
Tr[\; {\bf 1} \; S_F(k+p) \; {\bf 1} \; S_F(p) ]  \\
&&
= 
g_m
\frac{1}{v^2}
\int \frac{d^2 \bar{p}_E d p_0}{(2\pi)^3} \;
Tr[ \; \gamma_\alpha \; \gamma_\beta]
\frac{(\bar{k}+\bar{p})_\alpha  (\bar{p})_\beta }
       { [x(\bar{k}+\bar{p})^2+(1-x) \bar{p}^2]^2}
    = \frac{- |k| \,g_m}{8v^2}.
\end{eqnarray}
Although nonzero, when connecting external fermion lines, the resulting diagram would be proportional to the number of replicas $n_R$ and vanish in the $n_r \rightarrow 0$ limit.

\subsubsection*{Effective Gauge Propagator}

In Coulomb gauge (longitundal component $a_L = 0$), the kinetic term for the gauge field is
\begin{eqnarray}
&&  
S_{gauge} =\frac{1}{2} \int dk^2 d \omega \;
\begin{pmatrix}
a_0  &  a_T
\end{pmatrix}
\begin{pmatrix} {g^2 \over 16} {k^2 \over \sqrt{\omega^2 + v ^2 k^2}} & i \kappa |k| \cr i \kappa |k| & w_x {k^2 \over |k| + f(\omega, k)} + {g^2 \over 16} \sqrt{\omega^2 + v^2 k^2} \end{pmatrix}
\begin{pmatrix}
a_0 \\
a_T
\end{pmatrix},
\qquad \qquad 
\end{eqnarray}
where $(k^2 \equiv | {\bf k}|^2 )$.
Recall that $g_1 \equiv  \frac{g^2}{16} = \frac{1}{16}   $,
the effective coulomb coupling $w_x \equiv   \frac{+e^2}{4 \pi^2} $, and
$ f(k,\omega) \equiv \frac{ \sigma_e \, {\bf k}^2    }{ |\omega| +D_e {\bf k}^2}$.
The transverse component of the gauge field is
$a_T(k,\omega) \equiv  i \hat{k}_x  a_y(k,\omega) - i \hat{k}_y a_x(k,\omega)$, where $\hat k_j = k_j/|k|$.

When dealing with the gamma matrix contraction in Feynman diagram calculations, we have to write the effective gauge propagator obtained from $S_{gauge}$ in the $a_0,a_x,a_y$ basis ($i, j = x,y$):
\begin{eqnarray}
&&
G_{00} =  \frac{1}{ {\bf k}^2 } \; 
\frac{\;   g_1 \;\sqrt{\omega^2+ v^2 {\bf k}^2} +   \,  F(k,\omega)  }
{g_1^2 +   \kappa^2  + \frac{g_1    \, F(k,\omega)}{ \sqrt{\omega^2+ v^2 {\bf k}^2}  }
} \label{gauge-prop-00}  \\ 
&&
G_{0 i } =\frac{ \kappa\,    }{ {\bf k}^2 }
\frac{- \epsilon_{ij} k_j}
{
g_1^2 +  \kappa^2  + \frac{g_1    \, F(k,\omega)}{ \sqrt{\omega^2+ v^2 {\bf k}^2}  }
 } 
 \qquad  , \qquad  \qquad 
G_{i0} = -G_{0i} 
= 
\frac{ \kappa\,    }{ {\bf k}^2 }
\frac{+ \epsilon_{ij} k_j}
{
g_1^2 +   \kappa^2  + \frac{g_1    \, F(k,\omega)}{ \sqrt{\omega^2+ v^2 {\bf k}^2}  }
 }   \\
&&
G_{ij} = 
(  \delta_{ij} - \frac{ k_i k_j  }{ {\bf k}^2 }) \;
\frac{ \frac{ g_1 \,  }{ \sqrt{\omega^2+ v^2 {\bf k}^2} }       }
{ \bigg[ 
  g_1^2 +   \kappa^2  + \frac{g_1    \, F(k,\omega)}{ \sqrt{\omega^2+ v^2 {\bf k}^2}  }
    \bigg]  }  
    \label{gauge-prop-ij}  
    \\
&& \text{where  } \;\;
F(k,\omega) \equiv 
 \frac{+e^2}{4 \pi^2} \frac{|{\bf k}|^2}{ | {\bf k} |+ f(k,\omega) }   
 =
  \frac{+e^2}{4 \pi^2} \frac{| {\bf k} |^2}
  { | {\bf k} |+  \frac{ \sigma_e \, {\bf k}^2  }{ |\omega| +D_e {\bf k}^2} } 
 \xrightarrow[]{\text{critical limit}, D_e=0 }
  \frac{+e^2}{4 \pi^2} \frac{| {\bf k} |^2}
  { | {\bf k} |+  \frac{ \sigma_e \, {\bf k}^2  }{|\omega|}       }     
  \qquad \qquad 
\end{eqnarray}

\subsection*{Screened Disorder $W_{\mu \nu}$}

The disorders $g_0, g_j$ are screened by the fermion polarization.
The Feynman rules in (\ref{f-rule-g0})-(\ref{f-rule-gj}) have to be adjusted to account for this screening:
\begin{eqnarray}
W_{\mu \nu} = W^{(0)}_{\mu \nu} + \phi_1 W^{(sc)}_{\mu \nu},
\end{eqnarray}
where $W^{(0)}_{\mu \nu} = \text{Diag} \;  (g_0, i^2g_j, i^2 g_j )$ is the bare part in
(\ref{f-rule-g0}),(\ref{f-rule-gj}) and $W^{(sc)}_{\mu \nu}$ is the screening part from the summation of fermion bubbles.

\noindent \begin{tabular}{lll}
 & \includegraphics[scale=0.48]{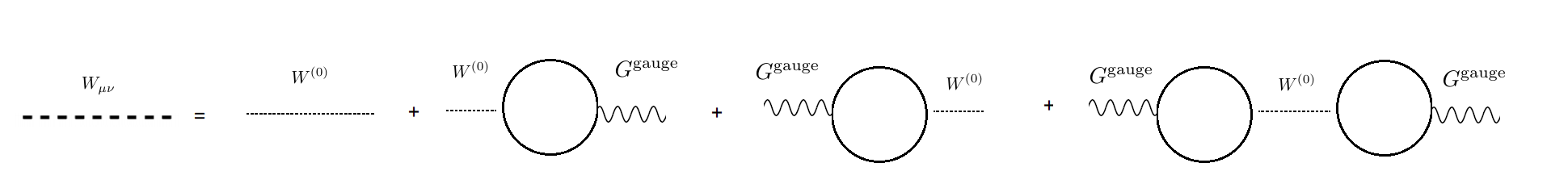}   &   
\end{tabular}
The prefactor $\phi_1$ isolates the screened and un-screened contributions: $\phi_1 = 0$ means that disorder screening is ignored; $\phi_1 = 1$ means that disorder screening is included.
When disorder connects with the gauge propagator, we should set $\sigma_e = 0$ before setting $\omega = 0$ (due to the presence of the $\delta(\omega)$) factor.
Otherwise, there is no disorder screening.
Note that the vertex factors are included in $W_{\mu \nu}$, so when applying the Feynman rules, we only need to multiply by $\gamma_\mu$ without any constant or velocity factor.

We separate the screening part into symmetric and antisymmetric components:
\begin{eqnarray}
&&  W^{(sc)} = W^{\text{sym}} + W^{\text{as}},  \\
&& W_{00}^{\text{sym}} =  
g_1 \; \frac{ -g_1 g_j \kappa^2 + g_0(g_1+ \wb)\; (g_1^2+ g_1 \wb + 2 \kappa^2) }
{(g_1^2+g_1 \wb + \kappa^2)^2},    \\
&& W_{ij}^{\text{sym}} =
\frac{g_1^2 (g_1^2 g_j + 2 g_1 g_j \wb - (g_0 - 2 g_j) \kappa^2)}
{ (g_1^2+g_1 \wb + \kappa^2)^2  }
\; \frac{1}{k^2} (k^2 \delta_{ij} - k_i k_j ),   \\
&& W_{0i}^{\text{as}} 
= \frac{g_1 \kappa  \; (g_1 g_j \wb + (-g_0 + g_j)  \kappa^2)}{ [g_1^2+g_1 \wb + \kappa^2]^2 }
\; \frac{- \epsilon_{ij} k_j  }{k}   .   
\end{eqnarray}
Other components of $W^{(sc)}$ not included above vanish.





\subsection*{Effective Gauge Disorder $D^{dis}_{\mu \nu}$ }
The expressions in (\ref{bsq}) and (\ref{exsq}) 2-point vertex rules: each side of the vertex connects with dressed propagator found in 
(\ref{gauge-prop-00})-(\ref{gauge-prop-ij}).
The effective gauge disorder is defined by
\begin{equation}
D^{dis}_{\mu \nu} = G_{\mu \alpha}\,  D^{0,dis}_{\alpha \beta} \,  G_{\beta \nu},
\end{equation}
where $D^{0,dis}_{00}= -\Delta_j \, \bm{k}^2$ defined in (\ref{exsq}), 
$D^{0,dis}_{ij} =  \Delta_0 \, (\delta_{ij} \bm{k}^2 -k_i k_j) $  defined in (\ref{bsq}), and $D^{0,dis}_{0i} =D^{0,dis}_{i0} =0 $. 
We decompose $D^{dis}_{\mu \nu}$ into symmetric and antisymmetric components:
\begin{eqnarray}
&& D^{dis}_{\mu \nu}= D^{\text{S}}_{\mu \nu} + D^{\text{AS}}_{\mu \nu}, \\
&& D^{\text{S}}_{00} =
 -\frac{ g_1^2 v^2 \Delta_j + 2g_1 v^2 \wb \Delta_j + v^2 \wb^2\Delta_j 
 + \Delta_0 \kappa^2} { (g_1^2+g_1 \wb + \kappa^2)^2  },  \\
&& D^{\text{S}}_{ij} =
\frac{ (g1^2 \Delta_0 + v^2 \Delta_j \kappa^2 )  } 
{ v^2   (g_1^2+g_1 \wb + \kappa^2)^2 }  \frac{ k^2\delta_{ij} - k_i k_j   }{k^2},  \\
&&
D^{\text{AS}}_{0i} = \frac{ \kappa\ ( -g_1 \Delta_0 + g_1 v^2 \Delta_j + v^2 \wb \Delta_j )  }{ v \; (g_1^2+g_1 \wb + \kappa^2)^2}
\frac{\epsilon_{ij} k_j  }{k}.
\end{eqnarray} 
Components of $D^{dis }$ not listed above are zero.

Since $G_{\mu \nu}$ is constructed by the RPA sum of fermion loops, $G_{\mu \nu}$ can no longer connect any more fermion loop. Consequently, $D^{dis}_{\mu \nu}$
does not include any fermion loops.
Note that $D^{dis}_{\mu \nu}$ generates
$\frac{\Delta_{0,j}}{N_f}$.
This disorder renormalizes $\Delta_0,\Delta_j$ at 3-loop order.


\section{Fermion Self-energy}
\label{Self-energy}

\subsection*{Self-energy---Screened Disorder Correction $W_{\mu \nu }$}

\begin{eqnarray}
&&
\Sigma_d(p_0,\bm{p})=
\int \frac{d^2 \bm{k}}{(2\pi)^2} 
\; (\bm{1}_{2\times 2}) \; S_F(p_0 -0,\bm{p}-\bm{k}) \;(\bm{1}_{2\times 2})
g_m
+
\int \frac{d^2 \bm{k}}{(2\pi)^2} 
\gamma_\nu   S_F(p_0 -0,\bm{p}-\bm{k}) \gamma_\mu  W_{\mu \nu }(\bm{k})\nn \\
&&
=
\int \frac{d^2 \bm{k}}{(2\pi)^2} 
\; (\bm{1}_{2\times 2}) \; S_F(p_0 -0,\bm{p}-\bm{k}) \;(\bm{1}_{2\times 2})
g_m
+ 
   \int \frac{d^2 \bm{k}}{(2\pi)^2} 
\gamma_\nu
\bigg[
(+i)
\frac{\gamma_0 p_0 + v(\bm{p}-\bm{k})_i \gamma_i}{p_0^2 +v^2 (\bm{p}-\bm{k})^2}
\bigg]
 \gamma_\mu 
 \;  W_{\mu \nu }(\bm{k})     \nn    \\ 
&&
=
 \frac{+i}{v^2} \frac{p_0 \gamma_0}{2\pi \epsilon} \;
[g_m ] 
+
   \int \frac{d^2 \bm{k}}{(2\pi)^2} 
\gamma_\nu
\bigg[
(+i)
\frac{\gamma_0 p_0 + v(\bm{p}-\bm{k})_i \gamma_i}{p_0^2 +v^2 (\bm{p}-\bm{k})^2}
\bigg]
 \gamma_\mu 
 \; W_{\mu \nu }(\bm{k})  \\
 &&
 =
\frac{+ i p_0 \gamma_0 }{2 \pi \epsilon v^2} \, g_m 
+ \frac{+ i p_0 \gamma_0}{2 \pi \epsilon v^2} \,(g_0 + 2 g_j)
\,   \nn \\
&& \; 
 +   \phi_1  \bigg[
\frac{ g_1^2 \wb \;( g_1 \gjb -g_1 \gob -\gob \;  \wb ) }
   { (g_1^2 + g_1 \wb+ \kappa^2 )^2}
 + \frac{ g_1 \; ( g_1 \gjb + g_1 \gob + 2 \gob \;  \wb ) \;  }
   {g_1^2 + g_1 \wb+ \kappa^2 }  
\bigg]\; \frac{-i}{\epsilon} p_0\, \gamma_0   \nn \\
&& \;
+   \phi_1  \;
\frac{ (-1)g_1^2 
\big[ g_1^2 \gjb+ 2 g_1 \gjb\, \wb- (\gob- 2 \gjb) \, \kappa^2   \big]    }
{ (g_1^2 + g_1 \wb+ \kappa^2 )^2 } \, 
\frac{-i}{\epsilon} \, v  p_j\, \gamma_j .
\end{eqnarray}

\subsection*{Self-energy---Gauge Correction}
Only the symmetric part of the gauge propagator produces a divergence at ${\cal O}(\frac{1}{N_f})$.
\begin{eqnarray}
&&
\Sigma_{g}(p_0 ,\bm{p})
=(\frac{-i g}{\sqrt{N_f}})^2  
\big( \frac{v}{c} \big)^{2-\delta_{\mu 0}-\delta_{\nu 0}}
\int \frac{d^3k}{(2\pi)^3} \; \gamma_\nu S_F(p-k) \gamma_\mu \; G_{\mu \nu}(k_0,\bm{k})
\\
&&
= \frac{-g^2}{N_f} (+i)
\big( \frac{v}{c} \big)^{2-\delta_{\mu 0}-\delta_{\nu 0}}
\; \frac{1}{v^2} \int 
\frac{d^2 \bar{k} \;dk_0 }{(2\pi)^3} 
\bigg[
\gamma_\nu 
\frac{(p_0-k_0)\gamma_0 + (\bar{p}-\bar{k})_a \gamma_a  } 
{(p_0-k_0)^2+(\bar{p}-\bar{k})^2}
\; \gamma_\mu  
\bigg]
G_{\mu \nu}(k_0,\bar{k}).
\end{eqnarray}
Carrying out the momentum integral and setting $c=1$:
\begin{flalign}
& - \Sigma_g(p_0, \bm{p}) & \nn \\
& = \, \int_{-\infty}^{\infty}   dz  
\frac{
ig^2 
\bigg(
(v^2-z^2) 
g_1 \;z^2 \sigma_e 
+ \big[  g_1 z^2 v^2 -  g_1 z^4 \times  \xi  
+ w_x \sqrt{v^2 + z^2}  (v^2-z^2) 
   \big]  \big{|} z \big{|}   
\bigg)                ( p_0 \gamma_0 ) 
}
{\longNum}     \qquad      & \nn  \\
&
+
 \int_{-\infty}^{\infty}   dz 
\frac{    i g^2 
\bigg(
g_1  (v^4-v^2 z^2 -z^4)  \sigma_e 
- 
\big[ 
w_x  z^2 \sqrt{v^2 + z^2} 
+ \, g_1 (-v^4 + v^2 z^2 + z^4 \times \xi   )
\big]   \big{|} z \big{|}  
\bigg)      (v p_j \gamma_j  )
}
{\longNum}  
\label{Self-energy-gauge}
&  
\end{flalign}
To obtain the above expression, we first perform a gradient expansion of 
$\Sigma(p_0, \bm{p})$ around $p_0=\bm{p}=0$. 
Next, focus on the linear term in $p_0,\bm{p}$ and replace the frequency integral $k_0 \to z\, |\bm{k}|$. 
When the 2d spatial momentum integral is done, the result is the expression shown above. 
The above expression is integrable only at $\sigma_e=0$. 
The $\xi$ term is a divergent integral that arises from the choice of Coulomb gauge. Physical observables are free from any $\xi$ dependence.

\subsection*{Self-Energy---Effective Gauge Disorder $D^{dis}_{\mu \nu}$, $\mathcal{O}(\frac{\Delta_{X}}{N_f})$}
\begin{eqnarray}
&&
\Sigma_b(p_0 ,\bm{p}) =
(\frac{-i  \, g \,    }{\sqrt{N_f}})^2  
\big( \frac{v}{c} \big)^{2-\delta_{\mu 0}-\delta_{\nu 0}}
\int \frac{d^2 \bm{k}}{(2\pi)^2} \;
\gamma_\nu S_F(p_0-0,\bm{p}-\bm{k}) \gamma_\mu  \, D^{dis}_{\mu \nu}(\bm{k} )   \\
&& 
= \frac{+i   (\Delta_0 + v^2 \Delta_j)}
{2 N_f \pi v^2 \epsilon (g_1^2 +  \kappa^2)} \; \gamma_0 p_0
+ 
\frac{ -i   
        \,(g_1^2 \Delta_0 +  \kappa^2 v^2 \Delta_j)	}
 {2 N_f \pi v^2 \epsilon (g_1^2 +   \kappa^2)^2 }
  \;   (v\, \gamma_j p_j ).
\end{eqnarray}

\section{3-point Vertex $\bar{u}(q)\;  \delta \Gamma^\mu \;  u(p)$}
\label{mu-vertex}

\begin{tabular}{lll}
 & \includegraphics[scale=0.5]{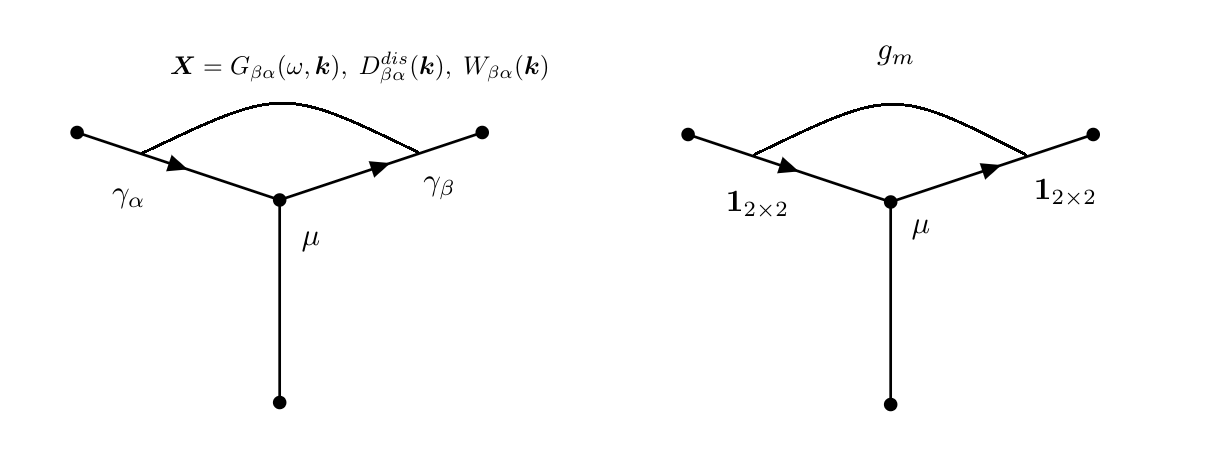}  &
\end{tabular}
~\\


\subsection*{$\Gamma^\mu$---Gauge Correction}

\begin{eqnarray}
&& 
\Gamma_1^\mu=
\big( \frac{-i g}{\sqrt{N_f}}  \big)^{3}
\big(\frac{v}{c} \big)^{3-\delta_{\alpha 0}-\delta_{\beta 0}-\delta_{\mu 0}}
\int \frac{d^2 \bm{k} d\omega}{(2\pi)^3}
\; \gamma_\alpha S_F(q-k) \gamma_\mu S_F(p-k) \gamma_\beta G_{ \beta \alpha}(k,\omega)
\\
&&
=
\big( \frac{-i g}{\sqrt{N_f}}  \big)^{3}
\big(\frac{v}{c} \big)^{3-\delta_{\alpha 0}-\delta_{\beta 0}-\delta_{\mu 0}}
\frac{1}{v^2}\; \nn\\
&& \times
\int \frac{d^2 \bm{\bar{k} } d\omega}{(2\pi)^3}
\; \gamma_\alpha 
(+i)\frac{\gamma_0(q_0-k_0)+ \gamma_c (\bar{q}-\bar{k})_c }{(q_0-k_0)^2+(\bar{q}-\bar{k})^2} 
\gamma_\mu 
(+i)\frac{\gamma_0(p_0-k_0)+ \gamma_d (\bar{p}-\bar{k})_d }{(p_0-k_0)^2+(\bar{p}-\bar{k})^2}
\gamma_\beta  G_{ \beta \alpha}(k,\omega)    \qquad \qquad    
\end{eqnarray}
To isolate the divergent part, one can set the external momentum $p=q=0$.
Following the same steps we used in the self energy diagram evaluation, we obtain
\begin{eqnarray}
&& 
\Gamma_1^\mu
=\frac{- i g}{\sqrt{N_f}}
\frac{
(-g^2 )  
\bigg(
(v^2-z^2)
g_1 \;z^2 \sigma_e 
+ \big[  g_1 z^2 v^2 -  g_1 z^4 \times  \xi 
+ w_x \sqrt{v^2 + z^2} (v^2-z^2) 
   \big] \big{|} z \big{|}  
\bigg)  
}
{\longNum}  ( \gamma_0 )      \nn  \\
&&
+
\frac{- i g}{\sqrt{N_f}}  \frac{v}{1} 
\frac{
(- g^2) \; 
\bigg(
g_1  (v^4-v^2 z^2 -z^4)  \sigma_e 
- 
\big[ 
w_x z^2 \sqrt{v^2 + z^2} +  g_1 (-v^4 + v^2 z^2 + z^4 \times \xi  )
\big]   \big{|} z \big{|}  
\bigg)      
}
{\longNum}   ( \gamma_j ).          \nn \\
&&         ~~ 
\end{eqnarray}
As before, $\xi$ labels the divergent part.
Gauge invariance is easy to check by comparing with Eq.~(\ref{Self-energy-gauge}):
$\Gamma^t_1= \frac{-g}{\sqrt{N_f}} \frac{\partial \Sigma_g }{\partial p_{0}} \;,\;\;
\Gamma^j_1= \frac{-g}{\sqrt{N_f}} \frac{\partial \Sigma_g }{\partial   p_{j}}
$.
%
%
 
\subsection*{$\Gamma^\mu$---Effective Gauge Disorder Correction}
\begin{eqnarray}
&&
\Gamma_2^\mu=
\big( \frac{-i g}{\sqrt{N_f}}  \big)^{3}
\big(\frac{v}{c} \big)^{3-\delta_{\alpha 0}-\delta_{\beta 0}-\delta_{\mu 0}}
\int \frac{d^2 \bm{k} }{(2\pi)^2}
\; \gamma_\alpha S_F(q-k) \gamma_\mu S_F(p-k) \gamma_\beta  D^{dis}_{\beta \alpha }(\bm{k})
\\
&&
=
\frac{ -i     (\Delta_0+ v^2 \Delta_j)}
 {2 N_f^{3/2} \pi v^2 \epsilon (g_1^2 + \kappa^2) }  \; \gamma^0 \; 
+ 
\frac{ i   (g_1^2 \Delta_0 +   v^2 \kappa^2 \Delta_j )}
{2 N_f^{3/2} \pi v \epsilon (g_1^2 +  \kappa^2)^2}
\; \gamma^j .
\end{eqnarray}

\subsection*{$\Gamma^\mu$---Screened Disorder Correction $W_{\mu \nu }$}
\begin{eqnarray}
&&   
\Gamma_3^\mu =
(\frac{-i g}{\sqrt{N_f}})  
\big( \frac{v}{c} \big)^{1-\delta_{\mu 0}} \times \int \frac{d^2 \bm{k}}{(2\pi)^2}
\gamma_\alpha S_F(q-k)  \gamma_\mu S_F(p-k) \gamma_\beta W_{ \beta  \alpha}  \\
&&
=
\frac{ -i g }{ \sqrt{N_f} }
\frac{1}{ \epsilon}   
\bigg(
\frac{g_0+ 2 g_j  }{2  \pi v^2}
+ \phi_1  
\big[
\frac{ g_1^2 \wb \;(- g_1 \gjb +g_1 \gob +\gob \;  \wb ) }
   { (g_1^2 + g_1 \wb+ \kappa^2 )^2}
 - \frac{ g_1 \; ( g_1 \gjb + g_1 \gob + 2 \gob \;  \wb ) \;  }
   {g_1^2 + g_1 \wb+ \kappa^2 }  
\big]
\bigg)    \; \gamma^0    \nn \\
&&
+
\frac{ -i g }{ \sqrt{N_f} }
\; \phi_1  \; 
\frac{1}{\epsilon} \; 
\Big(
\frac{ g_1^2 
\big[ g_1^2 \gjb+ 2 g_1 \gjb\, \wb- (\gob- 2 \gjb) \, \kappa^2   \big]    }
{ (g_1^2 + g_1 \wb+ \kappa^2 )^2 }
\Big) 
 \; \gamma^j.
\end{eqnarray}

\subsection*{$\Gamma^\mu$-Random Mass Correction $g_m$}
\begin{eqnarray}
&&
\Gamma_4^\mu =
(\frac{-i g}{\sqrt{N_f}})  
\big( \frac{v}{c} \big)^{1-\delta_{\mu 0}} 
 \times \int \frac{d^2 \bm{k}}{(2\pi)^2}
\bm{1} \; S_F(q-k) \gamma_\mu S_F(p-k) \; \bm{1}  \, g_m
=
\frac{ -i g  }{ \sqrt{N_f} }
\frac{  g_m }{2  \pi v^2 \epsilon }  \gamma^0 + 0\; \gamma^j.
\qquad  \qquad 
\end{eqnarray}

\section{3-point Vertex $\bar{u}(q)\;  1_{2\times 2} \;  u(p)$}
\label{1-vertex}

\begin{tabular}{lll}
 & \includegraphics[scale=0.5]{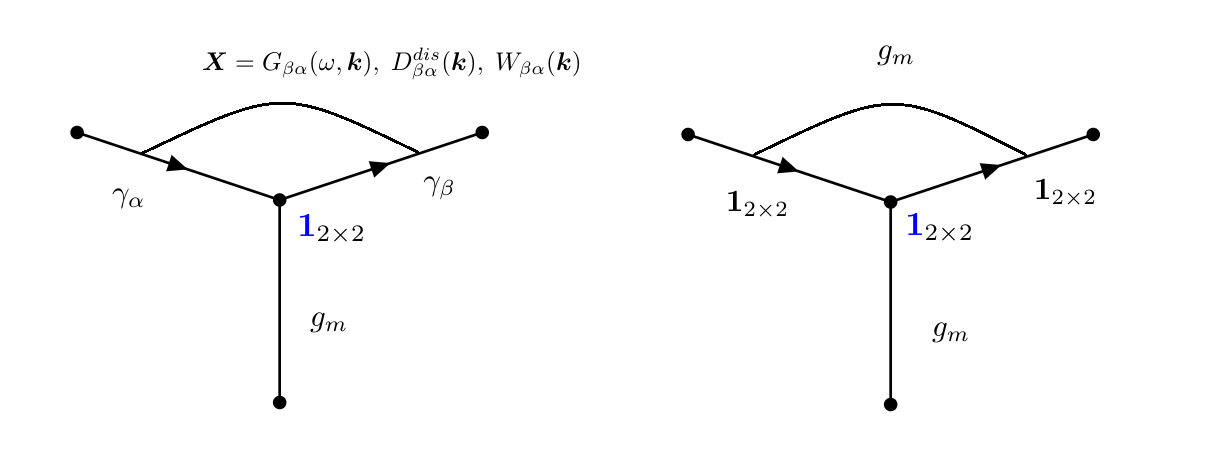}  &
\end{tabular}
 
\subsection*{$1_{2\times 2}$---Gauge Correction}
 
\begin{eqnarray}
&& 
 \Gamma^m_1
=  (\frac{-i g}{\sqrt{N_f}})^2 
(\frac{v}{c})^{2-\delta_{\alpha 0}-\delta_{\beta 0}}
\; \frac{1}{v^2}  (+i)^2     \nn  \\
&&  \qquad \qquad \times  
\int \frac{d^2 \bar{k} d\omega}{(2\pi)^3}
\gamma_\alpha 
\frac{ \gamma_0 ( q_0- \omega)+ \gamma_c(\bar{q}-\bar{k})_c }
   {(q_0-\omega)^2+(\bar{q}-\bar{k})^2} 
   \;
  \bm{1}  
  \;
\frac{\gamma_0 (p_0 -\omega )+ \gamma_d (\bar{p}-\bar{k})_d }
{(p_0 - \omega)^2+ (\bar{p}-\bar{k})^2}
\, \gamma_\beta \; G_{\alpha \beta}.
\qquad  
\end{eqnarray}
\begin{eqnarray}
&& 
\Gamma_1^m 
=
\frac{   
(v^2+z^2) \;\; g^2 
\bigg[
g_1 (2v^2 +z^2) \sigma_e    +
w_x \sqrt{v^2 +z^2}  + g_1 (2v^2 +z^2)\; |z| \;\; 
\bigg]
}
{\longNum}  \;  \; \bm{1}_{2 \times 2}    \qquad \;  \nn \\
&&
=
\frac{   
\; g^2 
\bigg[
g_1 (2v^2 +z^2) (v^2+z^2)  \;  \sigma_e    +
w_x \, (v^2+z^2)  \, \sqrt{v^2 +z^2}  + g_1 
 (2v^4 + 3 z^2 v^2 + \xi\;  z^4  )   \; |z| \;\; 
\bigg]
}
{\longNum}  \;  \; \bm{1}_{2 \times 2}.     
\label{Fm}
\end{eqnarray}

\subsection*{$1_{2\times 2}$---Effective Gauge Disorder Correction}

\begin{eqnarray}
&&
\Gamma_2^m=
\big( \frac{-i g}{\sqrt{N_f}}  \big)^{2}
\big(\frac{v}{c} \big)^{2-\delta_{\alpha 0}-\delta_{\beta 0} }
\int \frac{d^2 \bm{k} }{(2\pi)^2}
\; \gamma_\alpha S_F(q-k) 
\; \bm{1} \;
S_F(p-k) \gamma_\beta  D^{dis}_{ \beta \alpha}(\bm{k})   \qquad \qquad \nn \\
&&
=
\frac{  (\Delta_0 -v^2\Delta_j) (g_1^2-c^4 \kappa^2)}
{2\pi N_fv^2 \epsilon (g_1^2 + c^4 \kappa^2)^2}
\;  \; \bm{1}_{2 \times 2}.
\end{eqnarray}

\subsection*{$1_{2\times 2}$---Screened Disorder Correction $W_{\mu \nu }$}

\begin{eqnarray}
&&
\Gamma_3^m =
 \int \frac{d^2 \bm{k}}{(2\pi)^2}
\gamma_\alpha S_F(q-k) 
\bm{1} 
  S_F(p-k) \gamma_\beta W_{\beta \alpha }  \nn \\
 &&
=   \bigg[
    \frac{-g_0 + 2 g_j }{2 \pi v^2 \epsilon}
+ \phi_1   \frac{1}{\epsilon}
\Big(
- \frac{ g_1^2( 2 g_1+ \wb ) (-g_1 g_j+ \gob \, g_1 + \gob \,\wb  }
{ (g_1^2 + g_1 \wb+ \kappa^2  )^2}
+ \frac{g_1 (3 \gob \, g_1 -3 g_1 \gjb + 2 \gob \, \wb)}{ g_1^2 + g_1 \wb+ \kappa^2 }
\Big)
\bigg]    \bm{1}_{2 \times 2} .   \nn \\
&&  
\end{eqnarray}

\subsection*{$1_{2\times 2}$---Random Mass Correction $g_m$}
\begin{eqnarray}
\Gamma_4^m =
(\frac{-i g}{\sqrt{N_f}})  
 \times \int \frac{d^2 \bm{k}}{(2\pi)^2}
\bm{1} \; S_F(q-k) 
\; \bm{1} \;
S_F(p-k) \; \bm{1}  \, g_m
=
 \frac{-g_m}{2 \pi v^2 \epsilon}  \;  \; \bm{1}_{2 \times 2}.
\end{eqnarray}

\section{4-point Fermion-Fermion Interaction}
 
Define
\begin{equation}
H_{\mu \nu}  \equiv D_{\mu \nu}^{dis} (\frac{-i g}{\sqrt{N_f}})^2
(\frac{v }{c})^{2-\delta_{\mu 0}-\delta_{\nu 0}}.
\end{equation}
Note that $W_{\mu \nu} \sim \mathcal{O}(\frac{1}{N_f^0})$ and
$H_{\mu \nu} \sim \mathcal{O}(\frac{1}{N_f})$.
Take the external three momenta to be $p_1,p_2, p_3,p_4$, where
$p_i = (\omega_i, \bm{p_i})$.  
Schematically, the interaction has the form,
$[\overline{\psi}(p_3)...\psi(p_1)] \; [\overline{\psi}(p_4)..\psi(p_2)]$.
Define:
\begin{eqnarray}
&&\Gamma_A \equiv (\gamma_7, \gamma_0, + \gamma_x , + \gamma_y) \;\;, 
\gamma_7 \equiv \bm{1}_{2\times 2}  
\,\;,\; 
T^{A_1} = (\bm{1}\, g_m, W_{\mu \nu}, H_{\mu \nu}) 
\,\;,\; 
\tilde{T}^{A_1} = (\bm{1}\, g_m, W^{(0)}_{\mu \nu} , H_{\mu \nu})  \qquad\;\,  \\
&& 
W^{(0)}_{\mu \nu} 
= Diag(g_0, i^2 g_j, i^2 g_j)
\end{eqnarray}
We use $A,B,C,D = \{ 1,2,3,4 \} $ indices to label $\bm{1},\gamma_0, \gamma_x , \gamma_y$ and number subscripts, e.g., $A_1, A_2$, to label which interaction we choose:
$A_1 =1$ for the $ g_m$ interaction; $A_1 =2$ for the $W_{\mu \nu}$ interaction;
$A_1= 3 $ for the $H_{\mu \nu}$ interaction.


The diagrams below correspond to the following expressions:
\begin{flalign}
&
B_1= \int \frac{d^2 k}{(2\pi)^2}
\psb(p_3) \Gamma_B S_F(p_1+k) \Gamma_A \psi(p_1) \;\;
\psb(p_4) \Gamma_D S_F(p_2-k) \Gamma_C \psi(p_2) \; \nn \\
& \qquad \qquad \times 
T^{A_1}_{CA} (\bm{k},\omega=0) \; T^{A_2}_{BD} (p_1+k-p_3, \omega=0),
& \\
&
B_2=  \int \frac{d^2 k}{(2\pi)^2}
\psb(p_3) \Gamma_B S_F(p_1+k) \Gamma_A \psi(p_1) \;\;
\psb(p_4) \Gamma_C S_F(p_4+k) \Gamma_D \psi(p_2) \; \nn \\
& \qquad \qquad \times 
T^{A_1}_{CA}(\bm{k},\omega=0) \; 
T^{A_2}_{BD}(p_1-p_3+k, \omega=0),   &\\
&
B_3 = \int \frac{d^2 k}{(2\pi)^2}
\psb(p_3) \Gamma_D S_F(p_3-k) \Gamma_A S_F(p_1-k) \Gamma_B \psi(p_1) \; \nn \\
& \qquad \qquad \times 
T^{A_2}_{BD}(\bm{k},\omega=0),\;\;\;
\psb(p_4) \Gamma_C \psi(p_2) \; \tilde{T}^{A_1}_{AC}(p_3-p_1 ,\omega=0),
&  \\
& 
B_4=
\int \frac{d^2 k}{(2\pi)^2}
\psb(p_3) \Gamma_A \psi(p_1) \; \tilde{T}^{A_1}_{AC}(p_3-p_1,\omega=0)
\nn \\
& \qquad \qquad \times
\psb(p_4) \Gamma_D S_F(p_4-k) \; \Gamma_C S_F(p_2-k) \Gamma_B \psi(p_2) \;
T_{BD}^{A_2} (k,\omega=0).
&
\end{flalign}
\begin{tabular}{lll}
 & \includegraphics[scale=0.34]{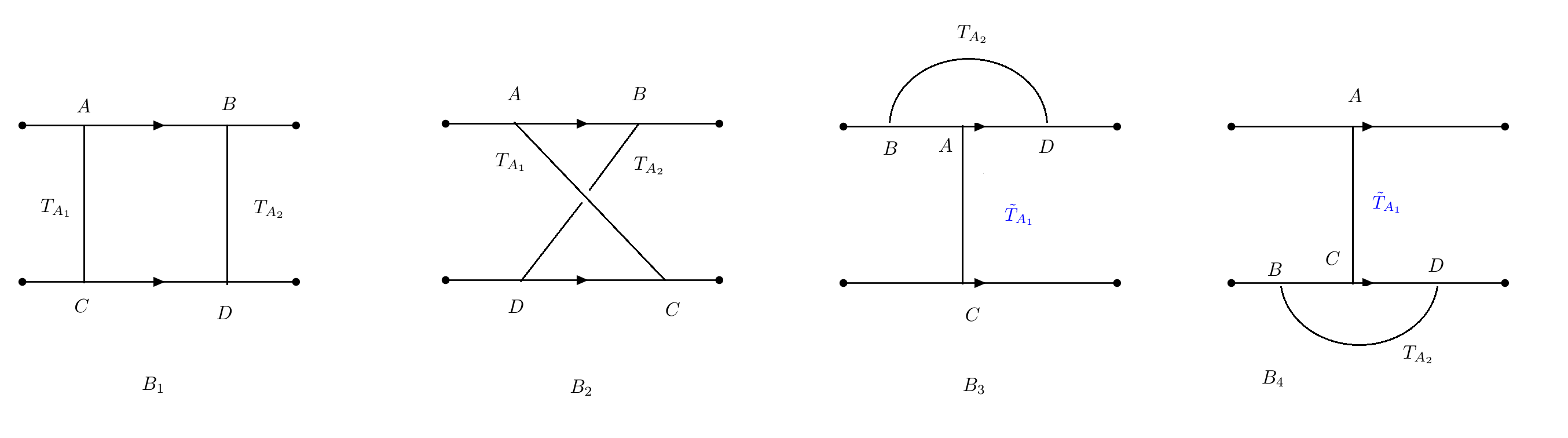}  &
\end{tabular}
For diagrams $B_3, B_4$, the $T_{A1}$ vertex is un-dressed, i.e., $W^{0}_{\mu \nu}$, which is directly related to the random coupling being renormalized. 

\subsection*{4-point Interaction---Boxes $B_3, B_4$}
Diagrams of type $B_3,B_4$ can be directly obtained from the 3-point vertex corrections in Appendices (\ref{mu-vertex}) and (\ref{1-vertex}) with symmetry factor $2$ (counting upper or lower vertices), so we don't have to recompute them at here.
The terms in $\Gamma^\mu$ renormalize $g_0,g_j$ and the terms in 
$\Gamma^m$ renormalize $g_m$.

\subsection*{4-point Fermion Interaction---Boxes $B_1, B_2$}
 
Diagrams for boxes $B_1,B_2$ are presented below.

\begin{tabular}{lll}
 & \includegraphics[scale=0.31]{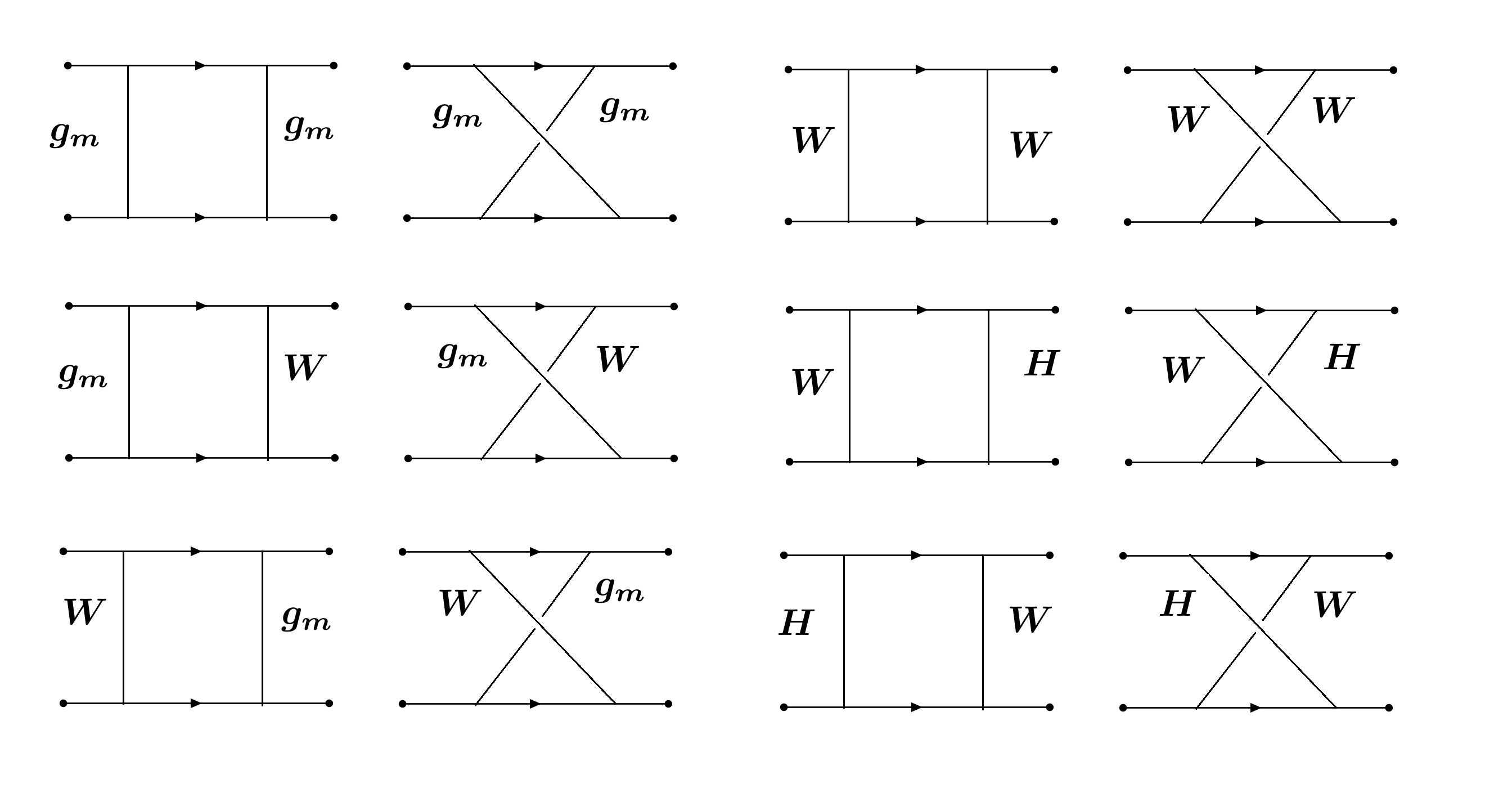}  &
\end{tabular}
The $W$-$H$ diagrams are $\mathcal{O}(\frac{\Delta_X \, g_X}{N_f} )$.
The $H$-$H$ diagrams are $\mathcal{O}(\frac{\Delta_X^2}{N_f^2})$.

For each interaction,
$(\bar{\psi} \psi) (\bar{\psi} \psi), \; 
(\bar{\psi} \gamma_0 \psi) (\bar{\psi} \gamma_0 \psi) , \;
(\bar{\psi} i \gamma_j \psi) (\bar{\psi} i \gamma_j \psi),
$
we sum all these diagrams with the help of computer software to $\mathcal{O}(g_X^2, \frac{g_X}{N_f})$.
The contribution from diagrams $B_1,B_2$ are the following.
\begin{eqnarray}
&&
\text{BOX}_{11} = 
\Bigg[
 \frac{2 g_0 g_j }{\pi v^2 \, \epsilon\; g_m}   
+ \phi_1 
(-4 g_1) \frac{ 
g_1  \, \gob \; \gjb ( 2g_1^2 + 4 g_1 \wb + \wb^2)
- \kappa^2 [ \gob^2 g_1 + g_1 \gjb^2 - 2\gob \; \gjb (2g_1 + \wb) ]
} 
 {\epsilon \, (g_1^2 + g_1 \wb+ \kappa^2 )^2}    \nn \\
&&
- \frac{\phi_1^2}{\epsilon}  \Big( 
\frac{4 g_1^2 \gjb^2\kappa^2[g_1^2 (g_1^2 + 2 g_1 \wb-\wb^2 ) 
         +2 g_1(g_1-\wb)\kappa^2 -\kappa^4  \;  ]  }
{\gmb [g_1^2 + g_1 \, \wb + \kappa^2]^4   }    \nn \\
&& \qquad \qquad
+
\frac{ 4 g_1^2 \gob^2 \kappa^2 
[g_1^2(g_1^2 + 2 g_1 \wb + \wb^2)+ 2 g_1(g_1+\wb)\kappa^2 -\kappa^4 ]}
{ \gmb [g_1^2 + g_1 \, \wb + \kappa^2]^4}     \nn \\
&& 
+  
\frac{ 4 \gob \; \gjb g_1^2 \big[ 
-g_1^3 (g_1+ \wb)^2(g_1+ 2 \wb)- 2 g_1^2(g_1 + \wb) (2 g_1+ 3 \wb	) \kappa^2
-g_1(5 g_1 + 2 \wb ) \kappa^4 + 2\kappa^6
\big]}
{\gmb [g_1^2 + g_1 \, \wb + \kappa^2]^4}
\, 
\Big)  \,
\Bigg ] \;\nn \\
&& \qquad \qquad \qquad
\times  (\bar{\psi} \bm{1}  \psi)(\bar{\psi}  \bm{1}  \psi) .
\end{eqnarray}
\begin{eqnarray}
&&
 \text{BOX}_{\gamma_0 \gamma_0} = 
\Bigg[ 
 \frac{2 g_j g_m  }{\pi v^2 \, \epsilon \, g_0} 
- \phi_1  \; 
\frac{ 4 g_1^2 \, \gmb[ g_1^2 \gjb + 2 g_1\gjb\; \wb -(\gob-2 \gjb) \kappa^2  ]}
{\gob \, \epsilon [g_1^2 + g_1 \, \wb + \kappa^2]^2} 
\Bigg] \; (\bar{\psi}  \gamma_0   \psi)(\bar{\psi}   \gamma_0  \psi).
\qquad \qquad \qquad 
 \end{eqnarray}
\begin{eqnarray}
 &&
 \text{BOX}_{\gamma_j \gamma_j} =  
 \Bigg[
 \frac{2 g_0 g_m  }{\pi v^2  \epsilon g_j}    
 -
 \phi_1
\frac{ g_1 \gmb [-g_1 \gjb \kappa^2
+ \gob( g_1^3+ 2g_1^2 \wb + 2 \wb \kappa^2) + g_1(\wb^2+ 2\kappa^2)\;  ]}
{ \gjb \epsilon [g_1^2 + g_1 \wb + \kappa^2]^2}    
\Bigg]
(\bar{\psi}    i\gamma_j   \psi)(\bar{\psi}    i\gamma_j   \psi).
    \nn \\
&&  
\end{eqnarray}
As mentioned before, the index $j=x$ or $y$; there is no index sum here. 
And we assume the random current disorder variance $g_x =g_y \equiv g_j$(isotropic).

 
\section{2-loop Vertex Corrections}
At leading order, the generic two-loop diagram has the form pictured below.

\noindent 
\begin{tabular}{lll}
 & \includegraphics[scale=0.4]{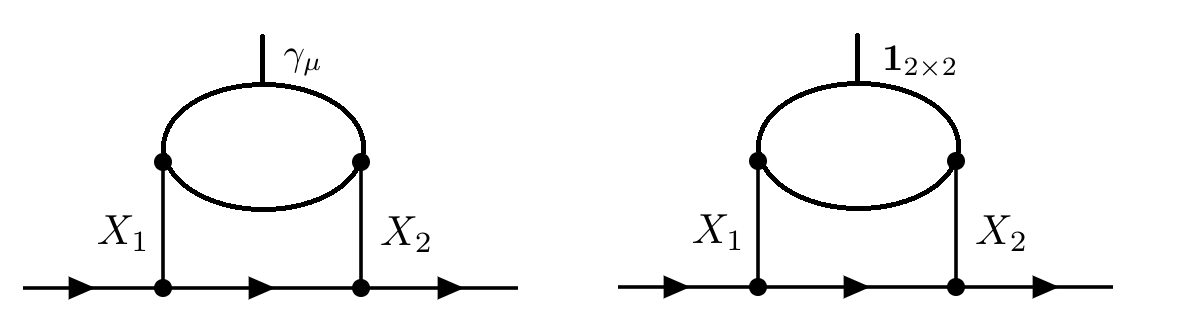}  &
\end{tabular}\\
The interaction legs $X_1$ and $X_2$ can be chosen to be the gauge propagator $G_{\mu \nu}$
or disorder $E_{\mu \nu} \in \{    W_{\mu \nu}, \; g_m   \} $. 
In principal there are four possibible choice: 
$(X_1,X_2)= (G,G), (E,G) , (G,E)$, or $(E,E)$. 
In the replica limit $n_r \rightarrow 0$, the $(E,E)$ diagram vanishes because the fermion bubble is proportional to $n_R$.
Also, $(E,G)$ and $(G,E)$ are the same diagrams so we only need to compute one of them.
The top vertex can be either $\gamma^\mu$ or $\bm{1}_{2\times2}$. 
However, we'll see below that diagrams using the $\gamma^\mu$ vertex are zero.

\subsection*{Mass Vertex $\overline{u}\,  1_{2\times 2} \,u$---one leg gauge, one leg disorder}

\begin{tabular}{lll}
 & \includegraphics[scale=0.4]{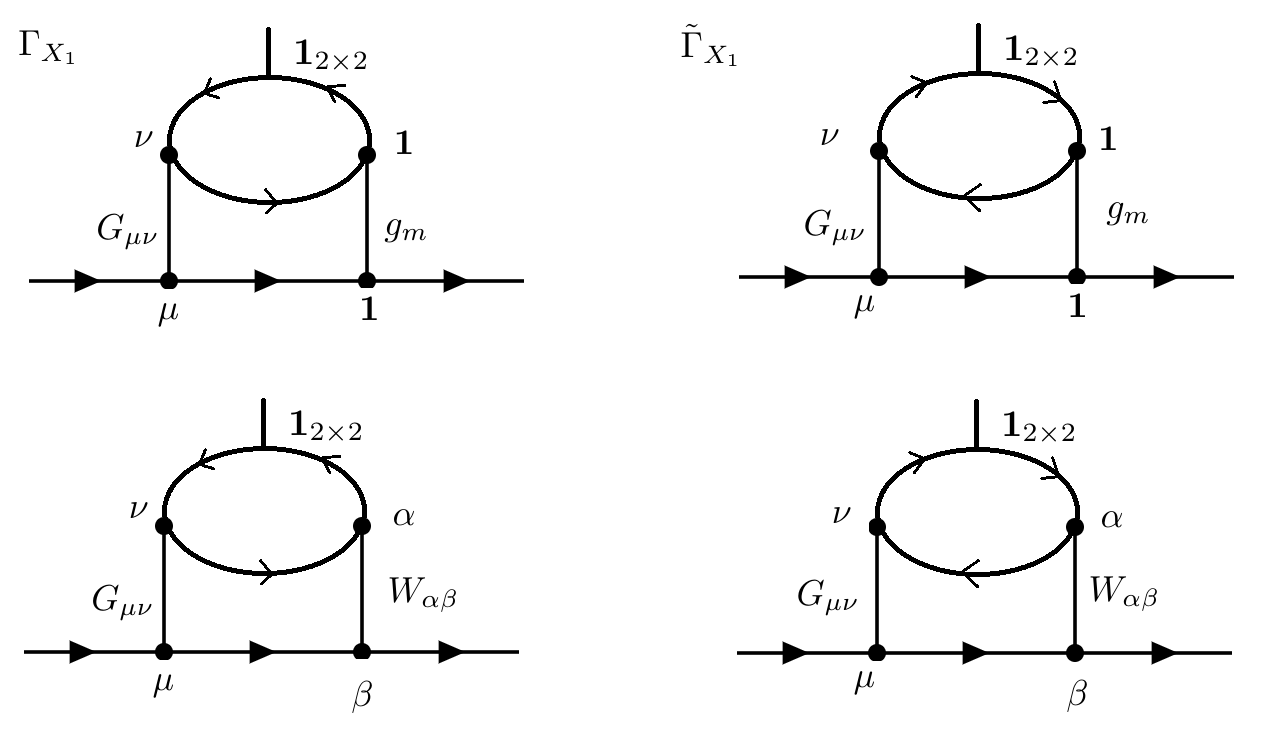}  &
\end{tabular}

\begin{flalign}
&
\Gamma_{X_1}
=
\frac{1}{v^2}\int \frac{d^2 \bar{\bm{q}} dq_0}{(2\pi)^3}
\frac{1}{v^2}\int \frac{d^2 \bar{\bm{k}} }{(2\pi)^2}  \;\;
\overline{u}(p_3)\,
\begin{pmatrix}   \bm{1} \\ \gamma_\beta  \end{pmatrix}
S_F(p_1-k)[  \frac{-ig}{\sqrt{N_f}}(\frac{v}{c})^{1-\delta_{\mu 0}} \gamma_\mu  ]\, u(p_1)
\;
G_{\mu \nu} (k) \;    (-1) \times N_f   & \nn\\
&    \times
Tr
\bigg[
[  \frac{-ig}{\sqrt{N_f}}(\frac{v}{c})^{1-\delta_{\nu 0}} \gamma_\nu  ]
S_F(q-p_1+p_3) \; \bm{1}   \;
S_F (q)   \begin{pmatrix}   \bm{1} \\ \gamma_\alpha  \end{pmatrix}
S_F(k+q-p_1+p_3)  
\bigg]
\begin{pmatrix}   g_m \\ W_{\alpha \beta}(k-p_1+p_3)  \end{pmatrix} .
  &
\end{flalign}
\begin{flalign}
&
\tilde{\Gamma}_{X_1}
=
\frac{1}{v^2}\int \frac{d^2 \bar{\bm{q}} dq_0}{(2\pi)^3}
\frac{1}{v^2}\int \frac{d^2 \bar{\bm{k}} }{(2\pi)^2}  \;\;
\overline{u}(p_3)\,
\begin{pmatrix}   \bm{1} \\ \gamma_\beta  \end{pmatrix}
S_F(p_1-k)[  \frac{-ig}{\sqrt{N_f}}(\frac{v}{c})^{1-\delta_{\mu 0}} \gamma_\mu  ]\, u(p_1)
\;
G_{\mu \nu} (k) \;  (-1) \times N_f    & \nn\\
&   \times
Tr
\bigg[
[  \frac{-ig}{\sqrt{N_f}}(\frac{v}{c})^{1-\delta_{\nu 0}} \gamma_\nu  ]
S_F(q-k)                   
 \begin{pmatrix}   \bm{1} \\ \gamma_\alpha  \end{pmatrix}
S_F (q-p_1+p_3)  
 \; \bm{1}   \;
S_F (q)  
\bigg]
\begin{pmatrix}   g_m \\ W_{\alpha \beta}(k-p_1+p_3)  \end{pmatrix} .
  &
\end{flalign}
The direction of the fermionic loop momenta is different in $\Gamma$ and $\tilde \Gamma$.
We use the upper/lower components to distinguish the diagrams that arise from either $g_m$/$W_{\mu \nu}$.

To extract the UV divergence, we can set $p_1 =p_3 =0$. 
For $g_m$, the divergences in $\Gamma_{X_1}$ and $\tilde{\Gamma}_{X_1}$ cancel (upon changing variables $q \to -q$ in $\tilde{\Gamma}_{X_1}$ and using basic properties of the trace).
For $W_{\alpha \beta}$, 
$\Gamma_{X_1}$ and $\tilde{\Gamma}_{X_1}$ have identical divergences.
\begin{eqnarray}
&&\qquad \qquad\qquad\qquad\qquad\qquad\qquad\qquad\qquad\qquad\qquad\qquad
\qquad\qquad\qquad\qquad  \qquad \qquad \qquad \qquad \qquad \qquad \qquad\qquad  
   \nn \\
&&
\Gamma_{X_1}
= \frac{- g^2 }{1} \; (\frac{v}{c})^{2-\delta_{\mu 0} -\delta_{\nu 0}}
\int \frac{d^3k }{(2\pi)^3 } 
\bar{u}(p_3) \; \bigg[ \gamma_\beta \; 
 \frac{k_0 \gamma_0 + v \, k_c \gamma_c}{k_0^2 + v^2 k^2 } \; \gamma_\mu \; \bigg] u(p_1)
 ~\qquad \qquad \qquad   \nn \\
&&  \times \;\; 
 G_{\mu \nu} (k)  \;\times  W_{\alpha \beta} (k) 
   \times  
\int  \frac{  d^3 q}{ (2\pi)^3}  \;\;
Tr \bigg[
\gamma_\nu  \;  \frac{q_0 \gamma_0 + v \,  q_d \gamma_d}{q_0^2 + v^2 q^2 } \; 
\bm{1} 
\;  \frac{q_0 \gamma_0 + v \, q_e \gamma_e}{q_0^2 + v^2 q^2 } \; 
\; \gamma_\alpha \; 
 \frac{ (k_0+q_0) \gamma_0 +v \, (k_f+ q_f) \gamma_f}{(k_0+q_0)^2 + v^2 (k+q)^2 } \; 
\bigg].
\end{eqnarray}

Refer to the calculations in (\ref{Gamma-z1-reduction}) to compute
\begin{eqnarray}
\int \frac{  d^3 q}{(2\pi)^3 } 
Tr \bigg[
\gamma_\nu  \;  \frac{q_0 \gamma_0 + v \,  q_d \gamma_d}{q_0^2 + v^2 q^2 } \; 
\;  \frac{q_0 \gamma_0 + v \, q_e \gamma_e}{q_0^2 + v^2 q^2 } \; 
\; \gamma_\alpha \; 
 \frac{ (k_0+q_0) \gamma_0 +v \, (k_f+ q_f) \gamma_f}{(k_0+q_0)^2 + v^2 (k+q)^2 } \; 
\bigg]  
=  \frac{ i \; \epsilon_{\nu \alpha \sigma} \; (k_0, v\, \bm{k})_\sigma }
    {8 v^2 \sqrt{ k_0^2 + v^2 \bm{k}^2}}.    \qquad  \;
\end{eqnarray}
After setting $g = c = 1$,
\begin{eqnarray}
&&
\Gamma_{X_1}
= \frac{- g^2 }{1} \; (\frac{v}{c})^{2-\delta_{\mu} -\delta_{\nu 0}}
\int \frac{d^3k }{(2\pi)^3 } 
\bar{u}(p_3) \; \bigg[ \gamma_\beta \; 
 \frac{k_0 \gamma_0 + v \, k_c \gamma_c}{k_0^2 + v^2 k^2 } \; \gamma_\mu \; \bigg] u(p_1)
 \times 
G_{\mu \nu} (k)   
 \times  
  W_{\alpha \beta} (k) 
   \times  
 \frac{ i \; \epsilon_{\nu \alpha \sigma} \; (k_0, v\, \bm{k})_\sigma }
    {8 v^2 \sqrt{ k_0^2 + v^2 \bm{k}^2}}    \nn  \\
&& = 
\frac{  (g_0 \, g_1 - g_j g_1 - g_j \wb )}
{4 \pi v^2 \epsilon (g_1^2 + g_1 \wb+ \kappa^2)}    \nn \\
&&
+ \phi_1
\Bigg[
\frac{ 2 g_1^3(g_1+ \wb) ( -g_1 \, \gjb + g_1 \gob + \gob \, \wb ) }
{ \epsilon \; (g_1^2 + g_1 \wb+ \kappa^2)^3}
-
\frac{ g_1^2  [7 \gob (g_1 + \wb ) -\gjb (7g_1+ 4 \wb)]  }
{2  \epsilon \; (g_1^2 + g_1 \wb+ \kappa^2)^2}
+
\frac{g_1 (\gob-\gjb)}
{\epsilon \; (g_1^2 + g_1 \wb+ \kappa^2)}
\Bigg] . \nn \\
&&
\end{eqnarray}
In total, we need to multiply by a factor of $4$ to count the clockwise/counterclockwise fermion loops and the exchange of $W \leftrightarrow G$ in the diagrams. 
\begin{eqnarray}
\Gamma_{X_1} + \tilde{\Gamma}_{X_1}
+ \Big(  \Gamma_{X_1} + \tilde{\Gamma}_{X_1}  \Big)_{W \leftrightarrow G}
= 4 \Gamma_{X_1}
\end{eqnarray}



\subsection*{Vector Vertex $\overline{u}\, \gamma^\rho \,u$---one leg gauge, one leg disorder}

Replace the mass-vertex expressions $\bm{1}$ by $\gamma_\rho$

\begin{tabular}{lll}
 & \includegraphics[scale=0.4]{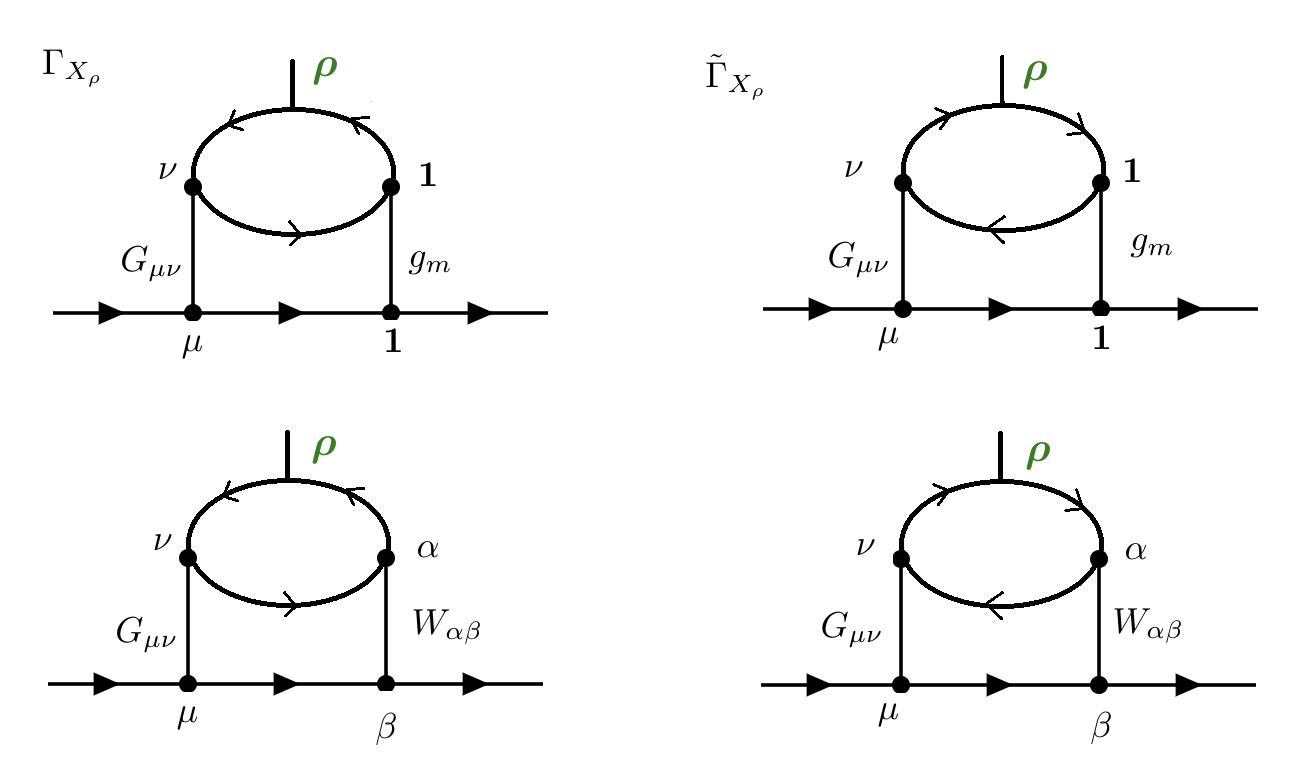}  &
\end{tabular}
%
%
\begin{flalign}
&
\Gamma_{X_\rho}
=
\frac{1}{v^2}\int \frac{d^2 \bar{\bm{q}} dq_0}{(2\pi)^3}
\frac{1}{v^2}\int \frac{d^2 \bar{\bm{k}} }{(2\pi)^2}  \;\;
\overline{u}(p_3)\,
\begin{pmatrix}   \bm{1} \\ \gamma_\beta  \end{pmatrix}
S_F(p_1-k)[  \frac{-ig}{\sqrt{N_f}}(\frac{v}{c})^{1-\delta_{\mu 0}} \gamma_\mu  ]\, u(p_1)
\;
G_{\mu \nu} (k) \;     (-1) \times N_f   & \nn\\
&   \qquad \times
Tr
\bigg[
[  \frac{-ig}{\sqrt{N_f}}(\frac{v}{c})^{1-\delta_{\nu 0}} \gamma_\nu  ]
S_F(q-p_1+p_3) \; 
\gamma_\rho   
 \;
S_F (q)   \begin{pmatrix}   \bm{1} \\ \gamma_\alpha  \end{pmatrix}
S_F(k+q-p_1+p_3)  
\bigg]
\begin{pmatrix}   g_m \\ W_{\alpha \beta}(k-p_1+p_3)   \end{pmatrix} .
  &
\end{flalign}
\begin{flalign}
&
\tilde{\Gamma}_{X_\rho}
=
\frac{1}{v^2}\int \frac{d^2 \bar{\bm{q}} dq_0}{(2\pi)^3}
\frac{1}{v^2}\int \frac{d^2 \bar{\bm{k}} }{(2\pi)^2}  \;\;
\overline{u}(p_3)\,
\begin{pmatrix}   \bm{1} \\ \gamma_\beta  \end{pmatrix}
S_F(p_1-k)[  \frac{-ig}{\sqrt{N_f}}(\frac{v}{c})^{1-\delta_{\mu 0}} \gamma_\mu  ]\, u(p_1)
\;
G_{\mu \nu} (k) \;   (-1) \times N_f   & \nn\\
&   \qquad \times
Tr
\bigg[
[  \frac{-ig}{\sqrt{N_f}}(\frac{v}{c})^{1-\delta_{\nu 0}} \gamma_\nu  ]
S_F(q-k)                   
 \begin{pmatrix}   \bm{1} \\ \gamma_\alpha  \end{pmatrix}
S_F (q-p_1+p_3)  
 \; \gamma_\rho   
S_F (q)  
\bigg]
\begin{pmatrix}   g_m \\ W_{\alpha \beta}(k-p_1+p_3)  \end{pmatrix} .
  &
\end{flalign}
By similar argument, the term with an even number of $\gamma$'s in the trace would cancel between
$\Gamma$ and $\tilde{\Gamma}$, so in this case we only need to 
compute upper component $(g_m)$.
Set $p_1=p_3=0$, straightforward calculation gives 
\begin{flalign}
&
\Gamma _{X_\rho}
=(-N_f) 
(-i)^4 (\frac{-ig}{\sqrt{N_f}})^2 
(\frac{v}{c})^{2-\delta_{\mu 0}-\delta_{\nu 0}}
\frac{1}{v^2}\int \frac{d^3 q}{(2\pi)^3}
\frac{1}{v^2}\int \frac{d^2 k}{(2\pi)^2}   & \nn \\
& \; \; \times 
\big[ 
\bar{u}(p_3)\;  \frac{ (-k)^c }{k^2} \gamma_c  \gamma_\mu \; u(p_1)    \big] \;
Tr[\gamma_\nu \slashed{q} \gamma_\rho \slashed{q} (\slashed{q}+\slashed{k})]\;
\frac{1}{q^2 \, q^2 (k+q)^2}
G_{\mu \nu} (\bm{k}, \omega=0) \; g_m     \nn  & \\
& 
 =0. &  
\end{flalign}
So there is no contribution from $\Gamma_{X_\rho}, \tilde{\Gamma}_{X_\rho}$

\subsection*{Mass Vertex $\overline{u}\,  1_{2\times 2} \,u$---both legs are gauge propagators}

\begin{tabular}{lll}
 & \includegraphics[scale=0.45]{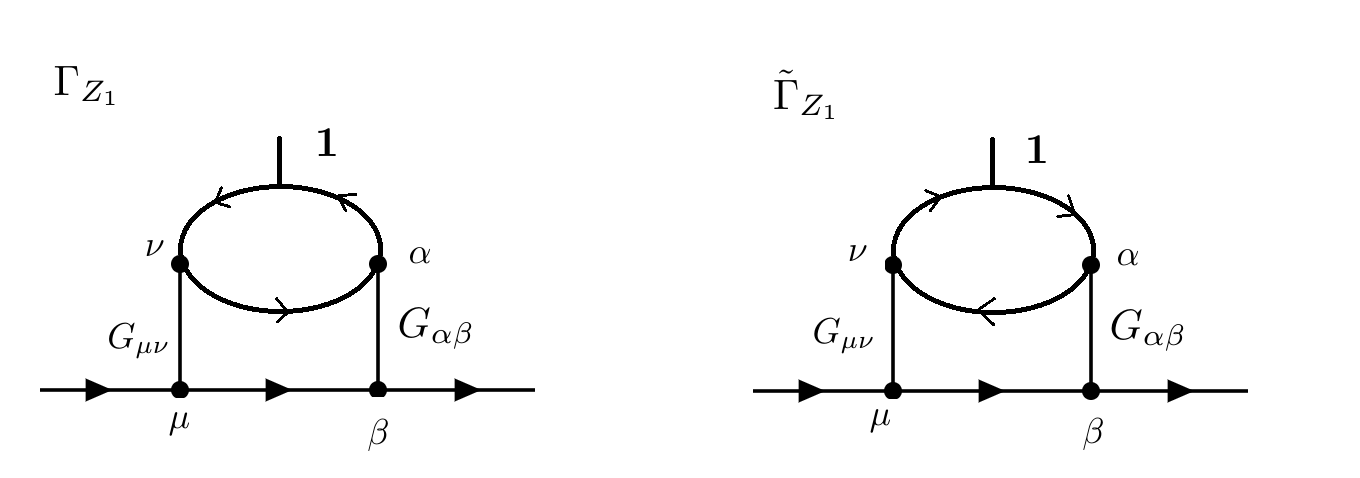}  &
\end{tabular}
%
\begin{flalign}
&
\Gamma_{Z_1}
=(\frac{-ig}{\sqrt{N_f}})^{4} \,
(\frac{v}{c} )^{4-\delta_{\mu 0}-\delta_{\nu 0}-\delta_{\alpha 0}-\delta_{\beta 0} }
 \times (-1) \times (N_f) \;
\int \frac{d^3k d^3 q}{(2\pi)^3(2\pi)^3}
\overline{u}(p_3) \; \gamma_\beta \; S_F(p_1-k) \; \gamma_\mu \; u(p_1)
\;   & \nn \\
& \qquad \times
G_{\mu \nu} (k)
Tr \bigg[
\gamma_\nu S_F(q-p_1+p_3) \; 
\bm{1}  
\; S_F(q) \; \gamma_\alpha \; 
S_F(k+q-p_1+p_3)  
\bigg]  \;\;  G_{\alpha \beta}(k-p_1+p_3).
\end{flalign}
%
\begin{flalign}
&
\tilde{\Gamma}_{Z_1}
=(\frac{-ig}{\sqrt{N_f}})^{4} \,
(\frac{v}{c} )^{4-\delta_{\mu 0}-\delta_{\nu 0}-\delta_{\alpha 0}-\delta_{\beta 0} }
 \times (-1) \times (N_f) \;
\int \frac{d^3k d^3 q}{(2\pi)^3(2\pi)^3}
\overline{u}(p_3) \; \gamma_\beta \; S_F(p_1-k) \; \gamma_\mu \; u(p_1)
\;   & \nn \\
& \qquad \times
G_{\mu \nu} (k)
Tr \bigg[
\gamma_\nu S_F(-k+ q+p_1-p_3) \;
\gamma_\alpha \; 
S_F(q) \; 
\bm{1} 
\; 
S_F(q+p_1-p_3)  
\bigg]  \;\;  G_{\alpha \beta}(k-p_1+p_3).
\end{flalign}
Upon taking the external momenta to zero,
\begin{eqnarray}
&&
\Gamma_{Z_1} = \frac{g^4}{N_f}
(\frac{v}{c} )^{4-\delta_{\mu 0}-\delta_{\nu 0}-\delta_{\alpha 0}-\delta_{\beta 0} }
\int \frac{d^3k }{(2\pi)^3 } 
\bar{u}(p_3) \; \bigg[ \gamma_\beta \; 
 \frac{k_0 \gamma_0 + v \, k_c \gamma_c}{k_0^2 + v^2 k^2 } \; \gamma_\mu \; \bigg] u(p_1)
 ~\qquad \qquad \qquad   \nn \\
&& 
\times   G_{\mu \nu} (k)  \;G_{\alpha \beta} (k) 
  \times  
\int  \frac{  d^3 q}{ (2\pi)^3}  \;\;
Tr \bigg[
\gamma_\nu  \;  \frac{q_0 \gamma_0 + v \,  q_d \gamma_d}{q_0^2 + v^2 q^2 } \; 
\bm{1}
\;  \frac{q_0 \gamma_0 + v \, q_e \gamma_e}{q_0^2 + v^2 q^2 } \; 
\; \gamma_\alpha \; 
 \frac{ (k_0+q_0) \gamma_0 +v \, (k_f+ q_f) \gamma_f}{(k_0+q_0)^2 + v^2 (k+q)^2 } \; 
\bigg] . \nn \\
&&
\end{eqnarray}
Perform the $q$ integral first, 
\begin{eqnarray}
&& 
 F_{\Gamma_z}(k) \equiv 
\int \frac{  d^3 q}{(2\pi)^3 } 
Tr \bigg[
\gamma_\nu  \;  \frac{q_0 \gamma_0 + v \,  q_d \gamma_d}{q_0^2 + v^2 q^2 } \; 
\;  \frac{q_0 \gamma_0 + v \, q_e \gamma_e}{q_0^2 + v^2 q^2 } \; 
\; \gamma_\alpha \; 
 \frac{ (k_0+q_0) \gamma_0 +v \, (k_f+ q_f) \gamma_f}{(k_0+q_0)^2 + v^2 (k+q)^2 } \; 
\bigg]     \qquad \qquad  \\
&& =
\frac{1}{v^2}
\int \frac{  d^3Q }{(2\pi)^3 } 
Tr \bigg[
\gamma_\nu  \;  \frac{ Q_\lambda \gamma_\lambda }{Q^2} \; 
\;  \frac{ Q_\rho \gamma_\rho }{ Q^2 } \; 
\; \gamma_\alpha \; 
 \frac{ (K+Q)_\sigma \; \gamma_\sigma  }{ (K+Q)^2} \; 
\bigg]. \\
&& \text{Here we define~~~}  Q \equiv (q_0, v\, \bm{q}) \;, \;
d^3 Q  \equiv d q_0 d^2 (v \bm{q}) \;\; , \;\; 
 K \equiv (k_0, v \, \bm{k}) \;\;    \nn  
\end{eqnarray}
Standard Feynman tricks give
\begin{eqnarray}
 F_{\Gamma_z}(k)  = 
 \frac{ i \; \epsilon_{\nu \alpha \sigma} \; (k_0, v\, \bm{k})_\sigma }
    {8 v^2 \sqrt{ k_0^2 + v^2 \bm{k}^2}}.
 \label{Gamma-z1-reduction}   
\end{eqnarray}

So we have $(k_0 = \omega)$
\begin{eqnarray}
&&
\Gamma_{Z_1} = \frac{g^4}{N_f}
(\frac{v}{c} )^{4-\delta_{\mu 0}-\delta_{\nu 0}-\delta_{\alpha 0}-\delta_{\beta 0} }
\int \frac{d^3k  }{(2\pi)^3 } 
\bar{u}(p_3) \; \bigg[ \gamma_\beta \; 
 \frac{k_0 \gamma_0 + v \, k_c \gamma_c}{k_0^2 + v^2 k^2 } \; \gamma_\mu \; \bigg] u(p_1)
\,   \nn \\
&& \qquad \qquad 
\times  G_{\mu \nu} (k )  \;G_{\alpha \beta} (k) \;
\bigg(
 \frac{ i \; \epsilon_{\nu \alpha \sigma} \; (k_0, v\, \bm{k})_\sigma }
    {8 v^2 \sqrt{ k_0^2 + v^2 \bm{k}^2}}
\bigg)   \;\;    \nn  \\
&&
= \bar{u}(p_3) \; \bm{1}_{2\times 2} \;  u(p_1) \;    \nn \\
&&    \times
\int^{\infty}_{-\infty}\, dz \,
\frac
{\;\;   -g^4 v^2 (\sigma_e + |z| )
\bigg[  (v^2+z^2) (g_1^2 -  \kappa^2) \, \sigma_e 
+    |z| \; 
     \Big(   g_1 w_x \sqrt{ v^2+z^2 }  
     + (g_1^2 -    \kappa^2 ) \; (v^2+z^2)\;  
     \Big) \; 
\bigg]
}
{16 \, \epsilon \, N_f \, \pi^2 (v^2+z^2)^{\frac{3}{2}}
\bigg[
\sqrt{v^2 +z^2} (g_1^2 +   \kappa^2) \, \sigma_e  +
|z| \,  \Big(
g_1^2 \sqrt{v^2+ z^2} 
+     ( g_1 \, w_x  + \sqrt{v^2 +z^2} \, \kappa^2 )
\Big)
\bigg]^2
} .  \nn \\
&& 
\end{eqnarray}
The same manipulations are used in the computations of $\delta_1,\delta_2$.
Note that unlike the case of $\delta_1,\delta_2$, this term renormalizes 
$g_m$ without any divergent integration, labeled by $\xi$.


Taking the limit $\sigma_e =w_x =0$, the expression reduces to 
\begin{eqnarray}
&&
 \Gamma_{Z_1}\Big{|}_{\sigma_e =w_x =0}
 = \bar{u}(p_3) \; \bm{1}_{2\times 2} \;  u(p_1) \; \times
 \frac{ \; -  g^4 \; (g_1^2-   \kappa^2)  }
 { 8 \, \epsilon \, N_f  \, \pi^2 (g_1^2 + \kappa^2 )^2},
\end{eqnarray}
which agrees with the result in \cite{ChenFisherWu1993}
Unlike the case of $\Gamma_{X_1}$, the two legs are identical so the symmetry factor is $2$:
\begin{eqnarray}
\Gamma_{Z_1} + \tilde{\Gamma}_{Z_1}  = 2 \Gamma_{Z_1}.
\end{eqnarray}

\subsection*{Vector Vertex $\overline{u}\, \gamma^\rho \,u$---
both legs are gauge propagators}

Replace $\bm{1} $ by $\gamma_\rho$ 
to obtain the vector counterparts

\begin{tabular}{lll}
 & \includegraphics[scale=0.45]{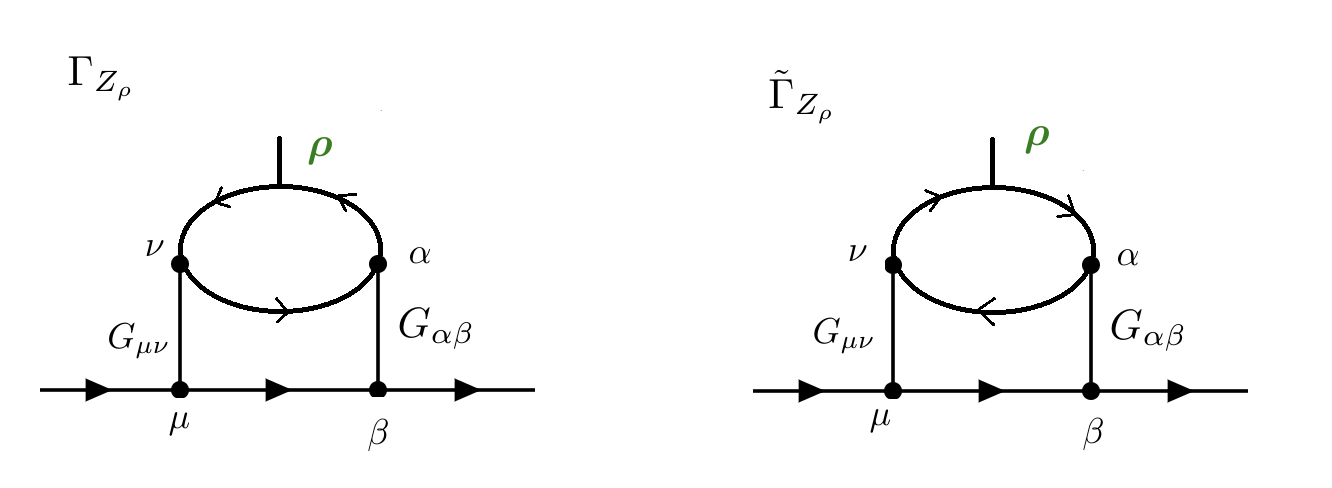}  &
\end{tabular}


\begin{flalign}
&
\Gamma_{Z_\rho}
=(\frac{-ig}{\sqrt{N_f}})^{4} \,
(\frac{v}{c} )^{4-\delta_{\mu 0}-\delta_{\nu 0}-\delta_{\alpha 0}-\delta_{\beta 0} }
\, \times (-1) \times (N_f) \;
\int \frac{d^3k d^3 q}{(2\pi)^3(2\pi)^3}
\overline{u}(p_3) \; \gamma_\beta \; S_F(p_1-k) \; \gamma_\mu \; u(p_1)
\;   & \nn \\
& \qquad \times
G_{\mu \nu} (k)
Tr \bigg[
\gamma_\nu S_F(q-p_1+p_3) \; 
\gamma_\rho 
\; S_F(q) \; \gamma_\alpha \; 
S_F(k+q-p_1+p_3)  
\bigg]  \;\;  G_{\alpha \beta}(k-p_1+p_3).
\end{flalign}
\begin{flalign}
&
\tilde{\Gamma}_{Z_\rho}
=
(\frac{-ig}{\sqrt{N_f}})^{4} \,
(\frac{v}{c} )^{4-\delta_{\mu 0}-\delta_{\nu 0}-\delta_{\alpha 0}-\delta_{\beta 0} } 
\, \times (-1) \times (N_f) \;
\int \frac{d^3k d^3 q}{(2\pi)^3(2\pi)^3}
\overline{u}(p_3) \; \gamma_\beta \; S_F(p_1-k) \; \gamma_\mu \; u(p_1)
\;   & \nn \\
& \qquad \times
G_{\mu \nu} (k)
Tr \bigg[
\gamma_\nu S_F(-k+ q+p_1-p_3) \;
\gamma_\alpha \; 
S_F(q) \; 
\gamma_\rho 
\; 
S_F(q+p_1-p_3)  
\bigg]  \;\;  G_{\alpha \beta}(k-p_1+p_3).
\end{flalign}

By the same argument as before, there are six $\gamma$'s in the trace, so $\Gamma_{Z_\rho}$ and $\tilde{\Gamma}_{Z_\rho}$ cancel one another:
\begin{eqnarray}
\Gamma_{Z_\rho} +  \tilde{\Gamma}_{Z_\rho} =0  .
\end{eqnarray}

\section{3-loop Corrections of Disorders $\Delta_0,\Delta_j$}
\label{Delta0j-disorder}

\begin{tabular}{lll}
 & \includegraphics[scale=0.55]{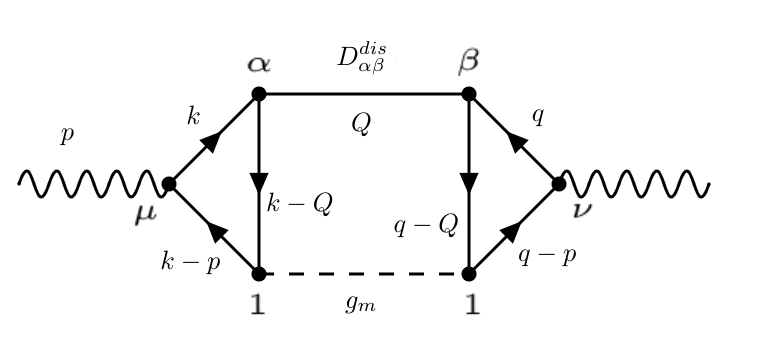}  &
\end{tabular}

\subsection*{With $D_{\alpha \beta}^{dis}$ Propagator}

\begin{flalign}
&
\pi_1^{\mu \nu}
=\int \frac{d^3 k}{(2\pi)^3}
\int \frac{d^3 q}{(2\pi)^3}
\int \frac{d^2 Q dQ_0}{(2\pi)^2}
\big( \frac{-i g}{\sqrt{N_f}} \big)^{4}
\big(\frac{v}{c} \big)^{4-\delta_{\mu 0}-\delta_{\nu 0}-\delta_{\alpha 0}-\delta_{\beta 0}}
\;(-1)^2 (N_f)^2 \, \nn  &  \\
&  \times 
Tr \bigg[
\gamma_\mu \, S_F(k-p) \,\bm{1}\, S_F(k-Q) \gamma_\alpha \,S_F(k)
\bigg]  
&  \nn \\
& \times \, g_m \, \delta(p_0-Q_0)\delta(Q_0) \, D_{\alpha \beta}^{dis}(\bm{Q},Q_0=0)
\; \times 
 Tr \bigg[
\gamma_\nu S_F(q) \gamma_\beta S_F(q-Q) \, \bm{1}\, S_F(q-p)
\bigg]
&  \\
&\text{(flipping the signs for $k$ and $q$ variables )}& \nn \\
&
=
\int \frac{d^2 Q }{(2\pi)^2} \; \;
\bigg[
\int \frac{d^3 k}{(2\pi)^3} 
\big(\frac{v}{c} \big)^{2-\delta_{\mu 0}-\delta_{\alpha 0} }
\text{Tr}\big[
\gamma_\mu \, S_F(k+p) \,\bm{1}\, S_F(k+Q) \gamma_\alpha \,S_F(k)  \big] 
\bigg]
 & \nn \\
 &  \times 
\bigg[
\int \frac{d^3 q}{(2\pi)^3}
\big(\frac{v}{c} \big)^{2-\delta_{\nu 0}-\delta_{\beta 0}}
 \text{Tr} \big[
\gamma_\nu S_F(q) \gamma_\beta S_F(q+Q) \, \bm{1}\, S_F(q+p) \big]
\bigg]   \nn & \\
& 
\times 
\bigg[ g_m \, \big( \frac{-i g}{\sqrt{N_f}} \big)^{4}   \, N_f^2  \,
D_{\alpha \beta}^{dis}(\bm{Q},Q_0=0)\, \delta(p_0=0)
 \bigg].
&
\label{pi-munu}
\end{flalign}
Naively evaluating this diagram is problematic because the Feynman parameter integrals are not doable. 
To extract the divergence, we Taylor expand the expression to second order in $p$.
First, we define 
\begin{eqnarray}
T^{\mu \alpha}(Q,p)=
\bigg[
\int \frac{d^3 k}{(2\pi)^3} 
\big(\frac{v}{c} \big)^{2-\delta_{\mu 0}-\delta_{\alpha 0} }
\text{Tr}\big[
\gamma_\mu \, S_F(k+p) \,\bm{1}\, S_F(k+Q) \gamma_\alpha \,S_F(k)  \big] 
\bigg].
\end{eqnarray}
By reversing the trace order , we have
\begin{eqnarray}
&&
\bigg[
\int \frac{d^3 q}{(2\pi)^3}
\big(\frac{v}{c} \big)^{2-\delta_{\nu 0}-\delta_{\beta 0}}
 \text{Tr} \big[
\gamma_\nu S_F(q) \gamma_\beta S_F(q+Q) \, \bm{1}\, S_F(q+p) \big]
\bigg]  \\
&& = (-1)^5 
\bigg[
\int \frac{d^3 q}{(2\pi)^3}
\big(\frac{v}{c} \big)^{2-\delta_{\nu 0}-\delta_{\beta 0}}
 \text{Tr} \big[
\gamma_\nu  \, S_F(q+p) \, S(q+Q) \gamma_\beta S_F(q) \big]
\bigg]
= - T^{\nu \beta}(Q,p).
\end{eqnarray}
Let
\begin{flalign}
&
\pi_1^{\mu \nu}
=
\int \frac{d^2 Q }{(2\pi)^2} \; \;
[ T^{\mu \alpha}(Q,p)] \; [ - T^{\nu \beta}(Q,p) ]
\times 
\bigg[ g_m \, \big( \frac{-i g}{\sqrt{N_f}} \big)^{4}   \, N_f^2  \,
D_{\alpha \beta}^{dis}(\bm{Q},Q_0=0)\, \delta(p_0=0)
 \bigg]
&
\end{flalign}
and
\begin{eqnarray}
&&
T_2(Q,p) \equiv T^{\mu \alpha} T^{\nu \beta} \\
&&
T_2(Q,p) = T_2(Q,0) + 
\frac{\partial T_2}{\partial p_x} p_x
+
\frac{\partial T_2}{\partial p_y} p_y
+ \frac{1}{2}
\bigg[
\frac{\partial^2 T_2}{\partial p_x^2} p^2_x
+ \frac{\partial^2 T_2}{\partial p_y^2} p^2_y
+ 2 \frac{\partial^2 T_2}{\partial p_x p_y} p_x \, p_y
\bigg] + \mathcal{O}(p^3).  \\
&&
=
\bigg(   
\frac{\partial T^{\mu \alpha}}{\partial p_x} 
\frac{\partial T^{\nu \beta}}{\partial p_x} 
\bigg) \bigg{|}_{\bm{p}=0} p_x^2 
+
\bigg(   
\frac{\partial T^{\mu \alpha}}{\partial p_y} 
\frac{\partial T^{\nu \beta}}{\partial p_y} 
\bigg) \bigg{|}_{\bm{p}=0} p_y^2
+
 \bigg(   
\frac{\partial T^{\mu \alpha}}{\partial p_x} 
\frac{\partial T^{\nu \beta}}{\partial p_y} 
+
\frac{\partial T^{\mu \alpha}}{\partial p_y} 
\frac{\partial T^{\nu \beta}}{\partial p_x} 
\bigg) \bigg{|}_{\bm{p}=0} p_x \, p_y
+ \mathcal{O}(p^3)   \nn \\
&&
\end{eqnarray}
Straightforward calculation gives
\begin{equation}
T^{\mu \alpha}(Q,p=0) = 0.
\end{equation}
For first order derivatives, we can also obtain (after lengthy algebra)
\begin{eqnarray}
\frac{\partial T^{\mu \alpha}}{\partial p_j} 
=
\big(\frac{v}{c} \big)^{2-\delta_{\mu 0}-\delta_{\alpha 0} }
\, \frac{1}{v^2}
\frac{i (-i)^3  }{32 |\bar{Q} |}    
\bigg(
\epsilon^{\mu \alpha j}
+ \frac{1}{\bar{Q}^2}
\big[
\epsilon^{\mu \alpha \tau} \bar{Q}_\tau \bar{Q}_j
+ \epsilon^{\alpha j \tau}  \bar{Q}_\tau \bar{Q}_\mu 
-\epsilon^{j \mu \tau} \bar{Q}_\tau  \bar{Q}_\alpha
\big]
\bigg).
\end{eqnarray}
Notice that this result is true in $3$d with general temporal component $Q_0$\\
~\\
Plugging into Eq.~(\ref{pi-munu}) and taking the four diagrams into consideration (each triangle has either clockwise or counterclockwise flowing momenta), the total result is 
\begin{equation}
\pi^{tot}_{\mu \nu} = 4 \pi_{\mu \nu},
\end{equation}
where
\begin{eqnarray}
&& 
\pi_{11}
= (-1)
\frac{  \; g_m(g_1^2 \Delta_0 + v^2 \Delta_j \kappa^2)}
{1024   \pi v^4 \epsilon (g_1^2+   g_1\,\frac{w_x}{v}+  \kappa^2)^2}
\; (\bar{p}_y^2+\bar{p}_x^2) ,
\\
&&  
\pi_{ij}
= (+1) 
\frac{    
g_m( g_1^2 v^2 \Delta_j+ 2 g_1 v \,w_x \Delta_j +w_x^2\Delta_j+ \Delta_0 \kappa^2  )}
{512 \pi v^4 \epsilon (g_1^2   +  g_1\,\frac{w_x}{v} +  \kappa^2)^2}
\;
(\delta_{ij} \bar{\bm{p}}^2- \bar{p}_i \bar{p}_j ),
\end{eqnarray}
and $\bar{p}_i = v \, p_i$. $\pi_{11}$ renormalizes $\Delta_j$ and $\pi_{ij}$ renormalizes $\Delta_0$.
This diagram scales as $1/N_f^2$ if $g_m, \Delta_0, \Delta_j$ scale as $1/N_f$.

\subsection*{With $W_{\alpha \beta}$ Propagator}

\begin{tabular}{lll}
 & \includegraphics[scale=0.55]{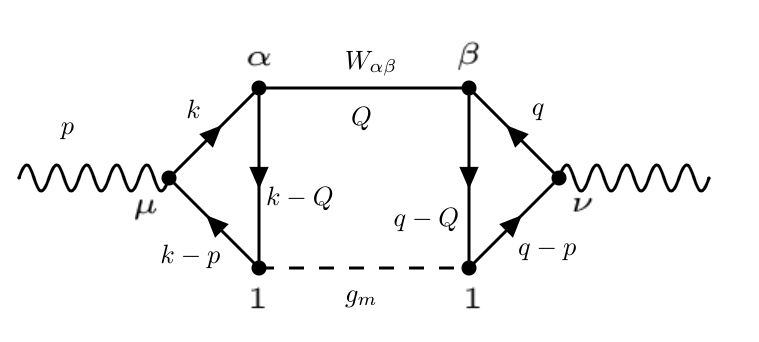}  &
\end{tabular}

Replace the internal propagator with $W_{\alpha \beta}$, the remaining calculations are the same:
\begin{equation}
\tilde{\pi}_{\mu \nu}^{tot} =4 \tilde{\pi}_{\mu \nu},
\end{equation}
where
\begin{eqnarray}
&& 
\tilde{\pi}_{11}
=  (-1) \,
\frac{  g_m  \; g_j \, N_f  }
{1024   \pi v^4 \epsilon}
 (\bar{p}_y^2+\bar{p}_x^2)
+ \phi_1 \;
\frac{ g_1^2 g_m  N_f \,(g_1^2 g_j + 2g_1 g_j \frac{w_x}{v} + (2g_j-g_0)\kappa^2 \;)  }
{1024 \pi v^4 \, \epsilon \, (g_1^2+ g_1 \, \frac{w_x}{v}+\kappa^2 )^2}
 (\bar{p}_y^2+\bar{p}_x^2)     \qquad\;   \\
&&
\tilde{\pi}_{ij}
=  (+1)\;
\frac{    g_0 g_m N_f}{512    \pi v^4 \epsilon }
\; (\delta_{ij} \bar{\bm{p}}^2- \bar{p}_i \bar{p}_j )  \nn \\
&& \qquad \qquad
+ \phi_1 \;
\frac{ g_1 g_m N_f \; \Big[
 g_1 g_j \kappa^2 
 - g_0 ( g_1 + \frac{w_x}{v} )\,(g_1^2 + g_1 \frac{w_x}{v} +2\kappa^2)
  \; 
\Big]   
}
{512 \pi v^4 \, \epsilon \, (g_1^2+ g_1 \, \frac{w_x}{v}+\kappa^2 )^2 }
\; (\delta_{ij} \bar{\bm{p}}^2- \bar{p}_i \bar{p}_j )  \nn   
\\
&&  
\end{eqnarray}
and $\bar{p}_i= v \, p_i$.
$\tilde{\pi}_{11}$ renormalizes $\Delta_j$, and $\tilde{\pi}_{ij}$ renormalizes $\Delta_0$.
This diagram scales as $1/N_f$ if $g_m, g_j$ scale as $1/N_f$.


\subsection*{With Gauge Propagator $G_{\alpha \beta} $ }
By dimensional analysis, this term should be UV finite,
\begin{equation*}
\sim \int d^2Q \,\frac{1}{Q} \frac{1}{Q} p^2  \frac{1}{Q} \bigg{|}_{Q_0=p_0}. 
\end{equation*}

\section{Summary}

\begin{eqnarray}
&& 
\bar{\delta}_1  \, p_0 \gamma^0  + \bar{\delta}_2 \, v \,  p_j \gamma^j 
= \Sigma_d + \Sigma_g + \Sigma_b   
 \end{eqnarray}
\begin{eqnarray}
&&  \overline{\delta}_{g_m} \;  (\bar{\psi} \bm{1} \psi) \; (\bar{\psi} \bm{1} \psi)
= \text{BOX}_{11}+ 
2 \Big( \Gamma_1^m + \Gamma_3^m+ \Gamma_4^m + 4 \Gamma_{X_1} + 2 \Gamma_{Z_1}  
 \Big) + 2 \Gamma^m_2
\end{eqnarray}

\begin{eqnarray}
\overline{\delta}_{g_0} \;  (\bar{\psi} \gamma^0 \psi) \; (\bar{\psi}  \gamma^0 \psi)
= \text{BOX}_{\gamma_0 \gamma_0}
 + 2\Big( 
  \Gamma^\mu_1 +\Gamma^\mu_3 +\Gamma^\mu_4
 \Big)_{\mu =1}   (\frac{ -i g }{ \sqrt{N_f} })^{-1}
 + 2 \Gamma^{\mu=1}_2  (\frac{ -i g }{ \sqrt{N_f} })^{-1}
\end{eqnarray}
\begin{eqnarray}
\overline{\delta}_{g_j} \;  (\bar{\psi}i \gamma^j \psi) \; (\bar{\psi} i \gamma^j \psi)
= \text{BOX}_{\gamma_j \gamma_j}
 + 2\Big( 
  \Gamma^\mu_1 +\Gamma^\mu_3 +\Gamma^\mu_4
 \Big)_{\mu =j}   \; (\frac{ -i g }{ \sqrt{N_f} } \, v )^{-1}
 + 2 \Gamma^{\mu=j}_2  \; (\frac{ -i g }{ \sqrt{N_f} } \, v )^{-1}  \;\;\;\;
\end{eqnarray}
where $\Sigma_b,\, \Gamma^m_2,\, \Gamma^{\mu=1}_2,\, \Gamma^{\mu=j}_2$ 
are the subleading order terms in the above expressions.
\begin{eqnarray}
\bar{\delta}_{\Delta_0}
  (\delta_{ij} \bm{p}^2 -p_i p_j)      
 = 4 \pi_{ij} +  4\tilde{\pi}_{ij}
\end{eqnarray}
\begin{eqnarray}
\bar{\delta}_{\Delta_j}  (p_x^2+p_y^2)      
 = 4 \pi_{11} + 4\tilde{\pi}_{11}
\end{eqnarray}


\bibliography{bigbib}{}
\bibliographystyle{utphys}

\end{document}